%% file: ITER_report03_DIIIDValidation.tex
\documentclass[12pt,letterpaper]{article}
\usepackage[latin1]{inputenc}

\usepackage{color}
\usepackage{graphics}
\usepackage{graphicx}
\usepackage{times}
\usepackage[T1]{fontenc}
\usepackage{amsmath}
\usepackage{amssymb}
\usepackage{amsthm}
\usepackage{tcolorbox}
\usepackage{ae,aecompl}
\usepackage{textcomp}   
\usepackage{hyperref}
\hypersetup{colorlinks=true}

\usepackage{float}
\usepackage{caption}    
\usepackage{subcaption} 

\usepackage{transparent}

\usepackage{tikz}

\usepackage{tabularx}
\newcolumntype{L}[1]{>{\hsize=#1\hsize\raggedright\arraybackslash}X}%
\newcolumntype{R}[1]{>{\hsize=#1\hsize\raggedleft\arraybackslash}X}%
\newcolumntype{C}[1]{>{\hsize=#1\hsize\centering\arraybackslash}X}%

\input macros.tex

\definecolor{clr1}{RGB}{0,255,255}
\definecolor{clr3}{RGB}{255,0,255}

\begin{document}

\title{
\vspace{1.25in}
\Huge{
\textbf{ITER-IA 3D MHD Simulations of \\Shattered Pellet Injection(SPI)}\\
}
\vspace{0.1in}
\huge{
\textbf{D1.3 Code Validation (DIII-D)} \\
}
\vspace{0.2in}
\large{
in partial fulfillment of ITER Agreement Ref: IO/IA/20/4300002130\\
to the Agreement on Scientific Cooperation Ref: LGA-2019-A-73}
\vspace{0.25in}
}

\author{Charlson.~C.~Kim\\
SLS2 Consulting\\ San Diego USA\\
Email: kimcc@fusion.gat.com\\[0.05in]
T.~Bechtel, J.~L.~Herfindal, B.~C.~Lyons,\\
Y.~Q.~Liu, P.~B.~Parks, L.~Lao\\
General Atomics\\ San Diego USA\\[0.05in]
}
\date{}

\clearpage
\maketitle
\thispagestyle{empty}

\newpage
\tableofcontents
\newpage
\listoffigures

\graphicspath{
    {./Figures/}
	}

\newpage
\section*{Executive Summary}
\noindent
This report is in partial fulfillment of deliverable \textbf{D1.3 Code Validation (DIII-D)}.  These simulations focus on
thermal quench phase of the SPI mitigation and are not typically carried beyond it to the current spike and subsequent
current quench.\footnote{It must be emphasized that our simulations (in contrast to engineering simulations)
do not set out to reproduce any particular experimental shot.  Many experimental features are simply ignored in the
numeric simulations due to limits of the model, computational cost, and expediency.  We do not painstakingly adjust
parameters to match experimental data.  Instead, we look for common trends and features that may give further insight 
into the experimental results.  In this way, experiments,  analytic theory, and numeric simulations form a strong 
and stable foundation for physical understanding and insight. } 
\\ \\
\noindent
NIMROD SPI simulations\cite{kimc:2019} are validated against DIII-D experiments for the following cases:
\begin{enumerate}
	\item Single pre-thermal quench injection SPI with a pure Ne pellet at high thermal energy.
	\item[*] \textit{Pure neon (instead of mixed) pellets were chosen to allow comparison with available DIII-D
		data.  Insufficient dual SPI mixed SPI DIII-D data is available.}
	\item Dual injection of pure Ne pellets from same and different toroidal locations.
		\begin{itemize}
			\item DIII-D dual SPI are separated by 120$^{\circ}$\ref{sec:explayo}
			\item dt=[0.0,$\pm$0.2,$\pm$0.4,$\pm$0.8]ms delays in pellet arrival time have been simulated
		\begin{itemize}
			\item positive values indicate that the delayed injector is at $\displaystyle \frac{2\pi}{3}$ 
			\item negative values indicate that the delayed injector is at $\displaystyle \frac{4\pi}{3}$
		\end{itemize}
		\end{itemize}
	\item[$\dagger$] \textit{We change the position of the second injector with the sign of the delay because it is more
		convenient and more readable in the input file.  This set up is equivalent to 0$^{\circ}$ injector
		leading is a positive delay and the 120$^{\circ}$ injector leading is a negative delay.}
\end{enumerate}

The target plasma for these simulations is DIII-D 160606@02990ms; detailed experimental parameters are reported in Ref.
\cite{Shiraki2016}.  Dual SPI simulations are compared with DIII-D experiments reported recently at IAEA-TM on Disruptions\cite{herfindal2022}.

\begin{table}[h]
\small{
\centerline{
	\begin{tabular}{|l||c|c|c|}\hline
		  	    &  thermal quench time $\tau_{TQ}$(ms)  &  peak radiation $\times10^8$(W)  &  radiated/thermal energy\\ \hline\hline
	  \clrlg{single load}    &       2.39        &     3.58                  &     0.46       \\ \hline
	  \clrlg{double load}    &       1.87        &     4.80                  &     0.57       \\ \hline\hline
	  dt=-0.8ms         &       2.39        &     3.55                  &     0.48       \\ \hline
	  dt=-0.4ms         &       2.13        &     7.33                  &     0.67       \\ \hline
	  dt=-0.2ms         &       1.96        &     5.93                  &     0.61       \\ \hline
	  dt= 0.0ms         &       1.97        &     5.05                  &     0.58       \\ \hline
	  \clrr{dt=+0.2ms}  &  \clrr{1.73}      &  \clrr{2.90}              & \clrr{0.35}    \\ \hline
	  dt=+0.4ms         &       2.18        &     3.58                  &     0.46       \\ \hline
	  dt=+0.8ms         &       2.37        &     4.44                  &     0.47       \\ \hline
	\end{tabular}
}
}
	\caption{Summary of NIMROD dual SPI delay: single SPI single load and single SPI double load and dual SPI with
	delays dt=[-0.8,+0.8]ms. Thermal quench times display a symmetry about dt=0 but radiation peak and thermal quench 
	efficiency(=radiated/thermal energy) with injector delay. Single injector single load and single injector double 
	load bracket thermal quench times but not radiation peak and radiation efficiency. (\textit{\clrr{dt=+0.2ms} case
	prematurely terminated due to numeric instabilities.})
	}
	\label{table:dsum}
\end{table}

The emphasis of the multiple injection simulations is on the impact of injection timing on the
pellet penetration and assimilation, the thermal quench efficiency (= radiated energy/thermal energy),
the radiation distribution, the electron density rise and distribution.  During the
SPI driven thermal quench, the primary loss is to the thermal energy while magnetic energy loss is marginal;
bulk magnetic energy
loss and current decay occur after the thermal quench once the temperature has dropped and the Spitzer-like temperature
dependent resistivity is significant.  Again, these simulations focus only on the thermal quench and not the current
quench.

\begin{figure}[H]
		\hspace{-0.5cm}
		\begin{subfigure}[b]{0.35\textwidth}
			\includegraphics[width=\textwidth]{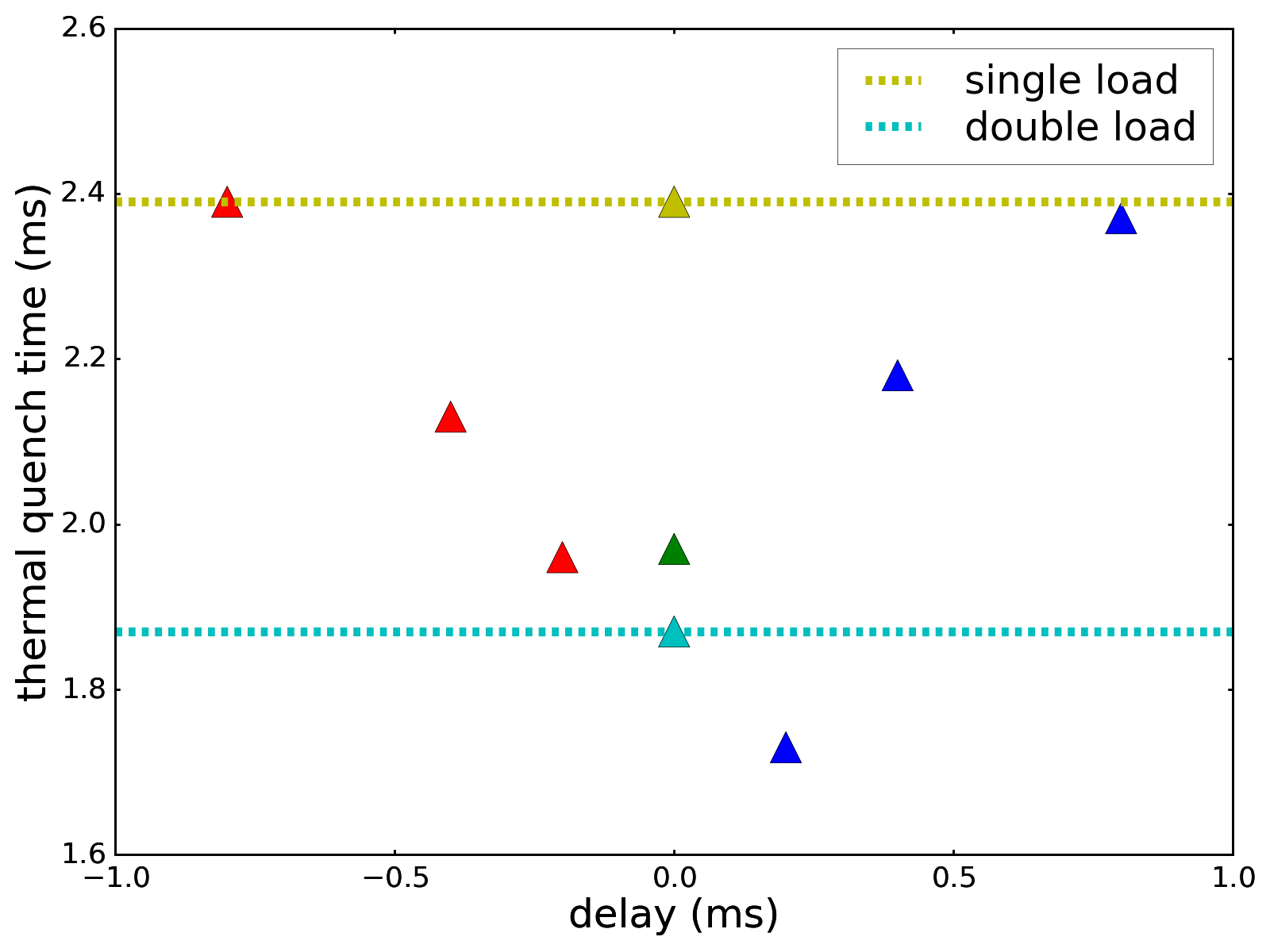}
			\vspace{-0.40cm}
			\caption{thermal quench time}
			\label{fig:dsumtauscat}
		\end{subfigure}
		\hspace{-0.5cm}
		\begin{subfigure}[b]{0.35\textwidth}
			\includegraphics[width=\textwidth]{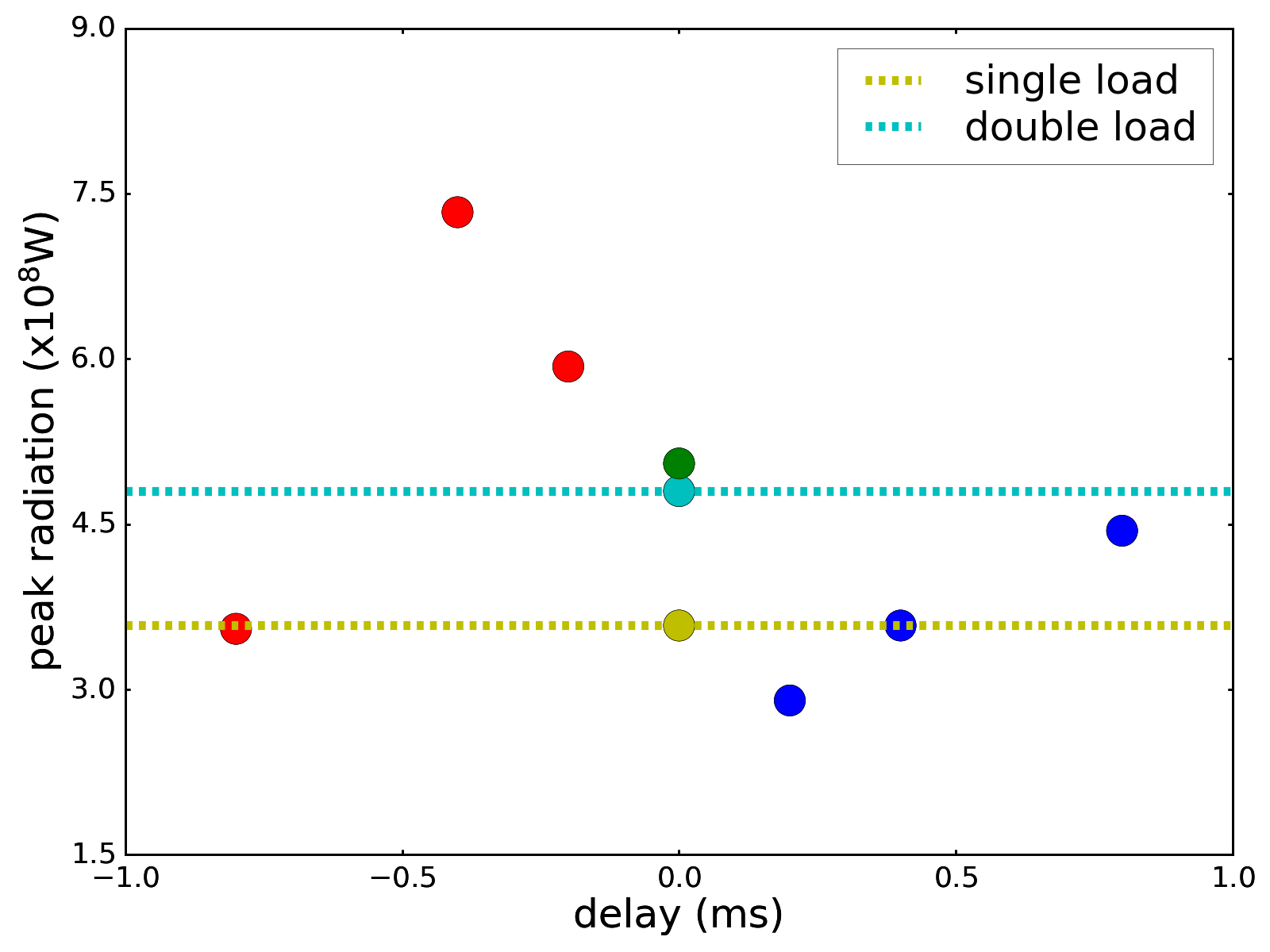}
			\vspace{-0.40cm}
			\caption{radiated power}
		\end{subfigure}
		\hspace{-0.5cm}
		\begin{subfigure}[b]{0.35\textwidth}
			\includegraphics[width=\textwidth]{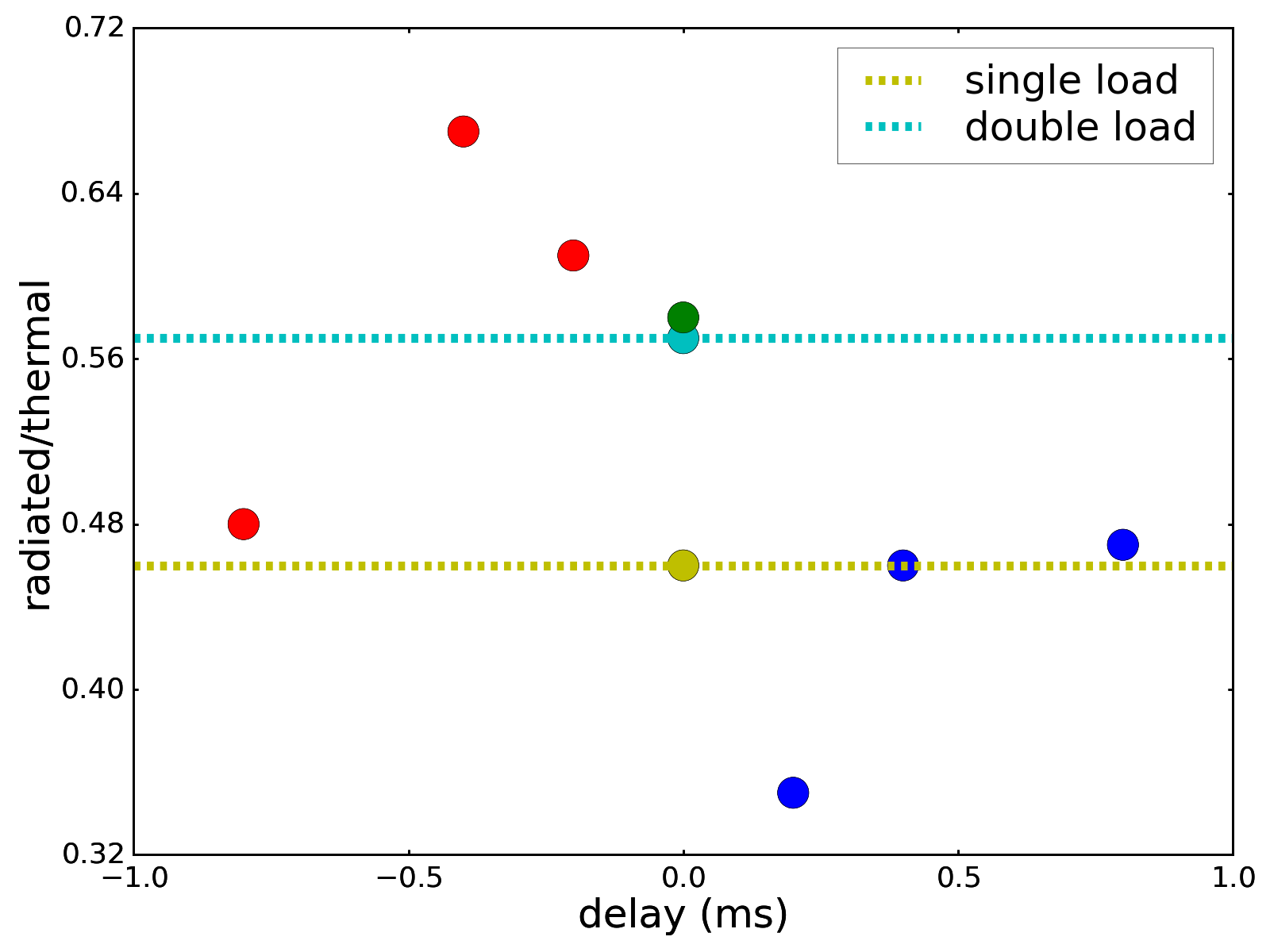}
			\vspace{-0.40cm}
			\caption{radiated/thermal energy}
		\end{subfigure}
		\caption{Thermal quench times, peak radiation, quench efficiency(=radiated/thermal energy) vs injector
		delay: red is negative delay, blue is positive delay and green is simultaneous dt=0.0ms.  Single
		SPI single load(yellow) and single SPI double load(cyan) bracket the thermal quench times of
		the dual SPI simulations but radiated power and quench efficiency defies expectations.}
		\label{fig:dsum}
	\end{figure}

Simulations results are summarized in Table \ref{table:dsum} and Figure \ref{fig:dsum} and show a surprising trend 
in efficacy of DIII-D dual SPI.  As might be expected, the thermal quench time increases as the absolute value of 
delay time increases.  However, the peak radiation and thermal quench efficiency 
show a clear increasing trend as dt is decreased from dt=+0.4ms to dt=-0.4ms.

Unfortunately, these simulations also show that the acceptable delay between the injectors is relatively small; at
dt=$\pm$0.8ms, the thermal quench behavior returns to the single SPI single load.

Figure \ref{fig:dsum} shows a near linear trend in radiation peak and quench efficiency between the injector delays
dt=[-0.4,+0.4]ms (ignoring the dt=+0.2ms case).  Naively, one might expect the simultaneous dual SPI where
dt=0.0ms to produce the optimal thermal quench.  This set of simulations clearly indicates this is not the case.

Comparison to experiment shows fair agreement and provides some reassurance that the simulations are reproducing many of
the important features of the thermal quench.  Thermal quench times and magnetic probe activity are comparable and share
many similar features.  Comparison of density profiles and interferometer line integrated densities show adequate
agreement. Follow up simulations are underway that include more refined fragment plume parameters and
additional physics such as temperature dependent thermal conduction. 

\newpage
\section{DIII-D Target Plasma and Simulation Parameters}
\vspace{-0.5cm}
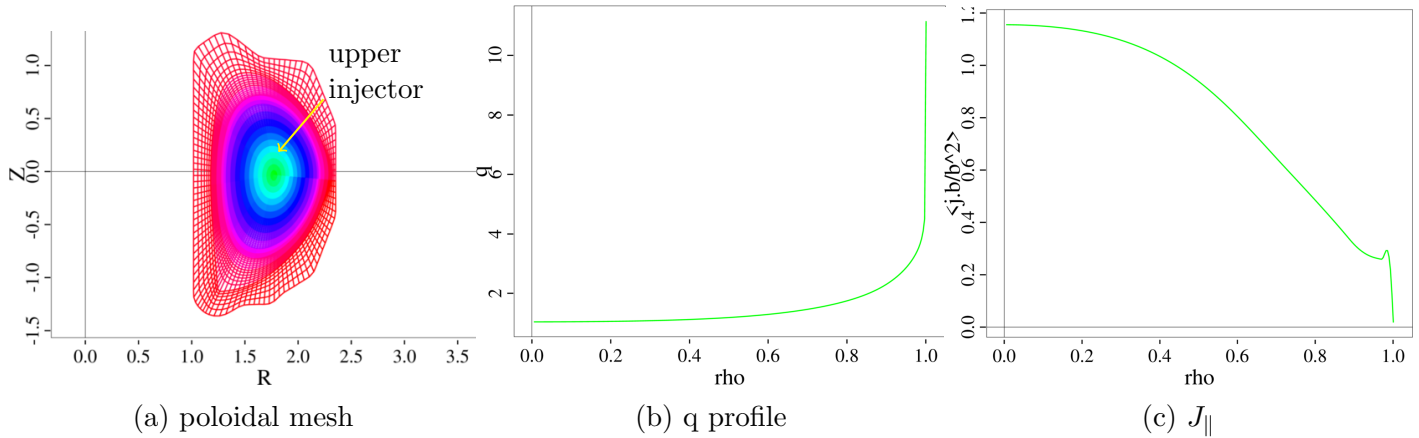
\begin{figure}[H]
	\hspace{-1.0cm}
	\begin{subfigure}[b]{0.35\textwidth}
		\includegraphics[width=\textwidth,trim={0.0cm 0.0cm 0.0cm 2.0cm},clip]{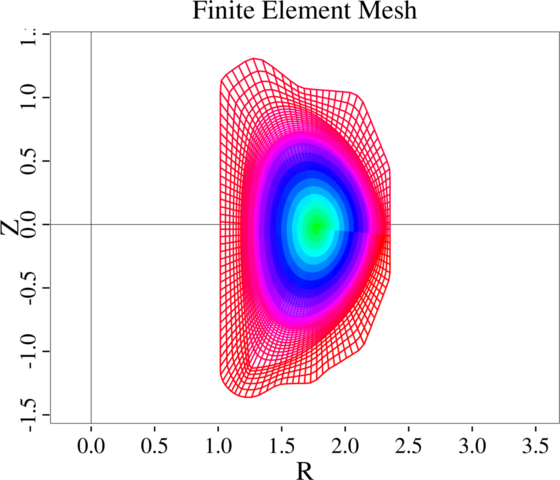}
		\caption{poloidal mesh}
		\label{fig:grid}
	\end{subfigure}
	\hspace{-0.25cm}
	\begin{subfigure}[b]{0.35\textwidth}
		\includegraphics[width=\textwidth,trim={0.0cm 0.0cm 0.0cm 2.0cm},clip]{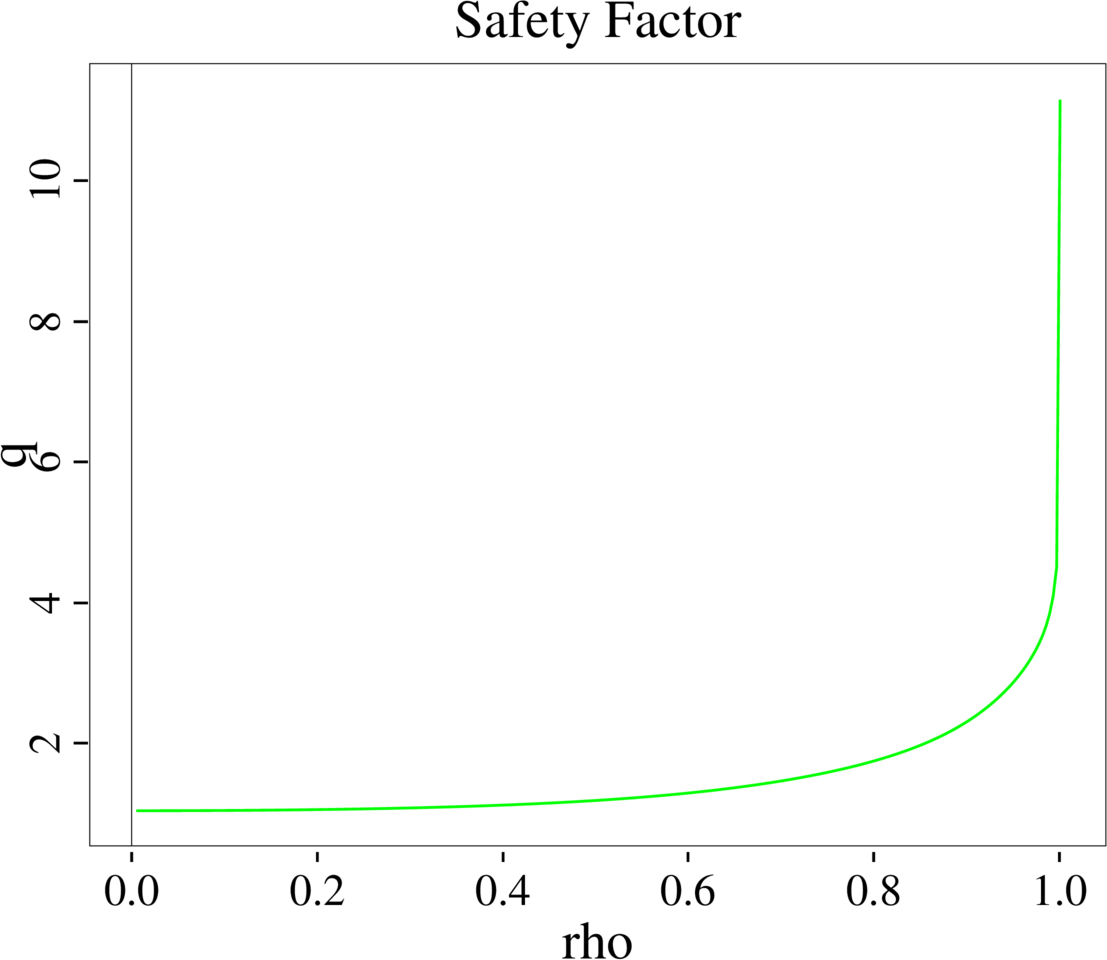}
		\caption{q profile}
		\label{fig:qprof}
	\end{subfigure}
	\hspace{-0.25cm}
	\begin{subfigure}[b]{0.35\textwidth}
		\includegraphics[width=\textwidth,trim={0.0cm 0.0cm 0.0cm 2.0cm},clip]{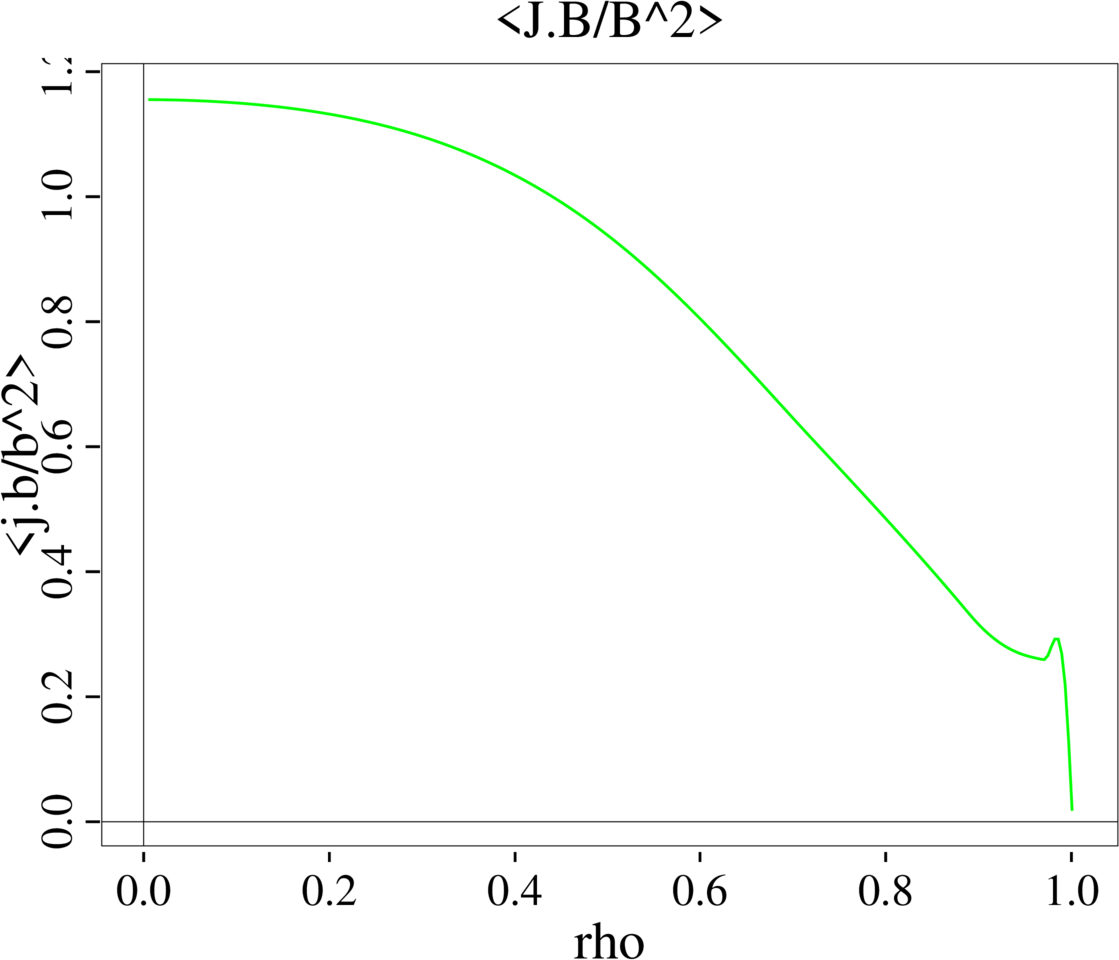}
		\caption{$J_\parallel$}
	\end{subfigure}
\begin{tikzpicture}[overlay]
	\draw[<-,thick,yellow] (2.7,4.30) -- (3.3,5.00);
	\node[align=left,font=\small,below,black] at (4.0,5.80){upper\\ injector};
\end{tikzpicture}
	\vspace{-0.5cm}
	\caption{Finite element poloidal mesh (colors denote domain decomposition), equilibrium q-profile and parallel current profile for DIII-D
	160606@02990.  The q-profile has been raised above q>1 to avoid (1,1) internal kink.}
	\label{fig:grideq}
\end{figure}

The NIMROD equilibrium is based on DIII-D 160606@02990ms with a thermal energy of 0.7MJ and total plasma current of 1.28MA.
The q profile has been raised to q$_0$=1.11 and q$_{min}$=1.05 to prevent (1,1) internal kink modes.  The q profile 
(Figure \ref{fig:qprof}) has been raised above q>1 to avoid a (1,1) internal kink.  As a single fluid
resistive MHD simulation, this NIMROD model does not contain addition physics (e.g. energetic particle effects) to
provide the stability observed in experiment.  

The simulation
uses a poloidal mesh of 96x112 elements (Figure \ref{fig:grid}) with polynomial degree 3 and 22 Fourier modes (n=[0,21])
in the toroidal direction.  Density diffusion is 3.0m$^2$/s and viscosity is 250.0m$^2$/s.  These simulations use a
modified temperature dependent Spitzer resistivity\cite{Sauter} that includes neoclassical corrections from trapped
electrons but a constant anisotropic thermal diffusion model with $\chi_\perp$=0.2m$^2$/s(equivalent to T$\sim$1eV) and
$\chi_\parallel$=1.0$\times$10$^9$m$^2$/s (equivalent to T$\sim$500eV).  This avoids the high computational cost of an
anisotropic temperature dependent thermal conduction model.  The parallel timescales are sufficiently fast enough that
the thermal loss is reasonably modeled by the lower value of the parallel thermal conduction.  Late in the thermal
quench and more so during the current quench, the constant parallel thermal conduction likely overestimates the 
thermal loss. Ongoing follow-up simulations address this limitations of constant thermal conduction with the use 
of a temperature dependent thermal conduction model and the development of a two temperature 
(separate T$_e$ and T$_i$) model for the impurities.

Simulations were performed on Cori-Cray XC40, at the National Energy Research Scientific Computing Center (NERSC)
located at Lawrence Berkeley National Laboratory.  These SPI simulation utilize 22 nodes, each node consists of 32 Haswell
Intel Xeon Processors (704 total processors).

Simulations begin with an initial timestep of dt=1.0$\times$10$^{-7}$s and may decrease it (e.g. to satisfy CFL constraints
or solver tolerance) by a factor of $\times$100
before NIMROD terminate.  The different shades of colors for the plots in Figures \ref{fig:svdouble} and
\ref{fig:svdual} denote 48hour compute cycles on Cori. 48hours is the limit in wall clock run time for each 
queue submission to the
Cori-NERSC system.  Wait times in the queue are typically a few days.  Simulations require 4-6 runs but 
some are run out
to 8 or more to include the current spike and the beginning of the current quench phase.  The shorter lengths
of color segments denote periods where the time step has decreased and/or the solver iterations have increased due to
increasing nonlinearity of the  system, typically near the end of the thermal quench and especially during the 
current spike.

\subsection{NIMROD SPI Fragment Plume Parameters}
A single SPI load is a pure neon pellet approximated as a sphere of radius r$_p$=2.0mm, with a
shatter parameter of S=10, where S=$\displaystyle\frac{r_{pellet}}{r_{fragment}}$ is the ratio of pellet radius to 
fragment radius; the resulting number of fragments 
is then S$^3$. Of the nominal 1000 fragments, we assume 80\% are lost, resulting in an injected plume
composition of 200 fragments, each fragment with a radius of r$_f$=0.2mm and nominal velocity of v=120.0m/s.  
We reduce the inventory so that enough of the vanguard fragments of the leading injector plume ablate and allow 
the trailing injector fragments to catch up.  More details regarding the reasoning for this reduction are provides
in Sections \ref{sec:dualSPI} and \ref{sec:follow}. 

The fragments have a linear velocity dispersion with $\frac{\Delta v}{v}$=0.5 such that fragment velocities v$_i$ 
can range from
v$_i$ = [60.0m/s,180.0m/s].  The fragments have a nominal trajectory along the upper port injector directed at the
magnetic axis as illustrated in Figure \ref{fig:grid} with a linear poloidal spread with $\Delta\theta_{hw}=20^{\circ}$.
All fragments remain in the poloidal plane of their respective injector, i.e. no toroidal component to the fragments'
velocity.  

The ablated neutral deposition uses a Gaussian shape function with r$^{Gaussian}_{hw}$=3.0cm in the poloidal plane and a
von Mises distribution\cite{vonMises} with half-width of $0.03\times2\pi$ in the toroidal direction.  These simulations
represent the smallest deposition sizes to date.  Future updates will change the deposition function to a finite
cylindrical puck to eliminate `leaking' deposited neutrals into the plasma.

\subsubsection{Single Injector Single Load vs. Single Injector Double Load}
\vspace{-0.5cm}
	\begin{figure}[H]
		\hspace{-1.0cm}
		\begin{subfigure}[b]{0.35\textwidth}
			\includegraphics[width=\textwidth,trim={0.0cm 0.0cm 0.0cm 2.0cm},clip]{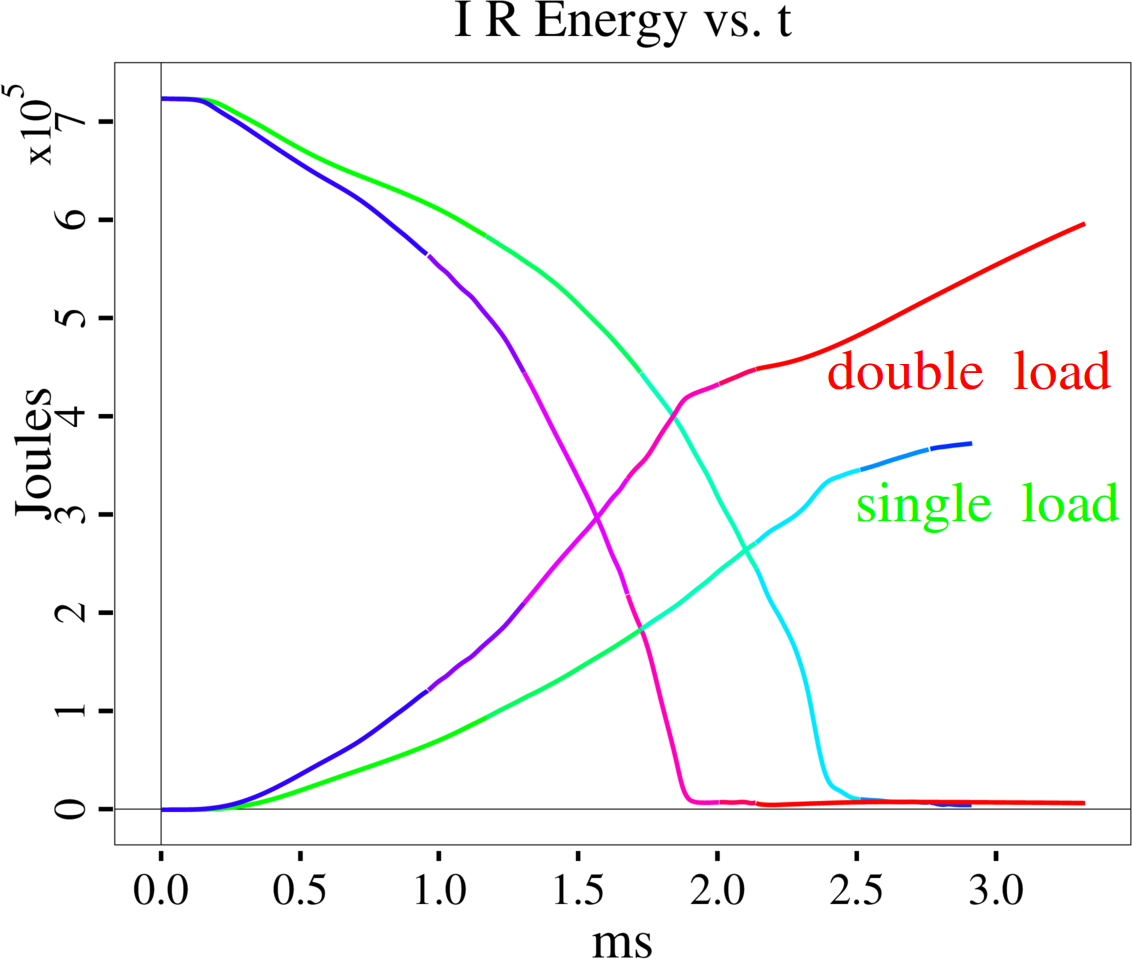}
			\vspace{-0.60cm}
			\caption{Thermal and Radiated Energy}
		\end{subfigure}
		\hspace{-0.25cm}
		\begin{subfigure}[b]{0.35\textwidth}
			\includegraphics[width=\textwidth,trim={0.0cm 0.0cm 0.0cm 2.0cm},clip]{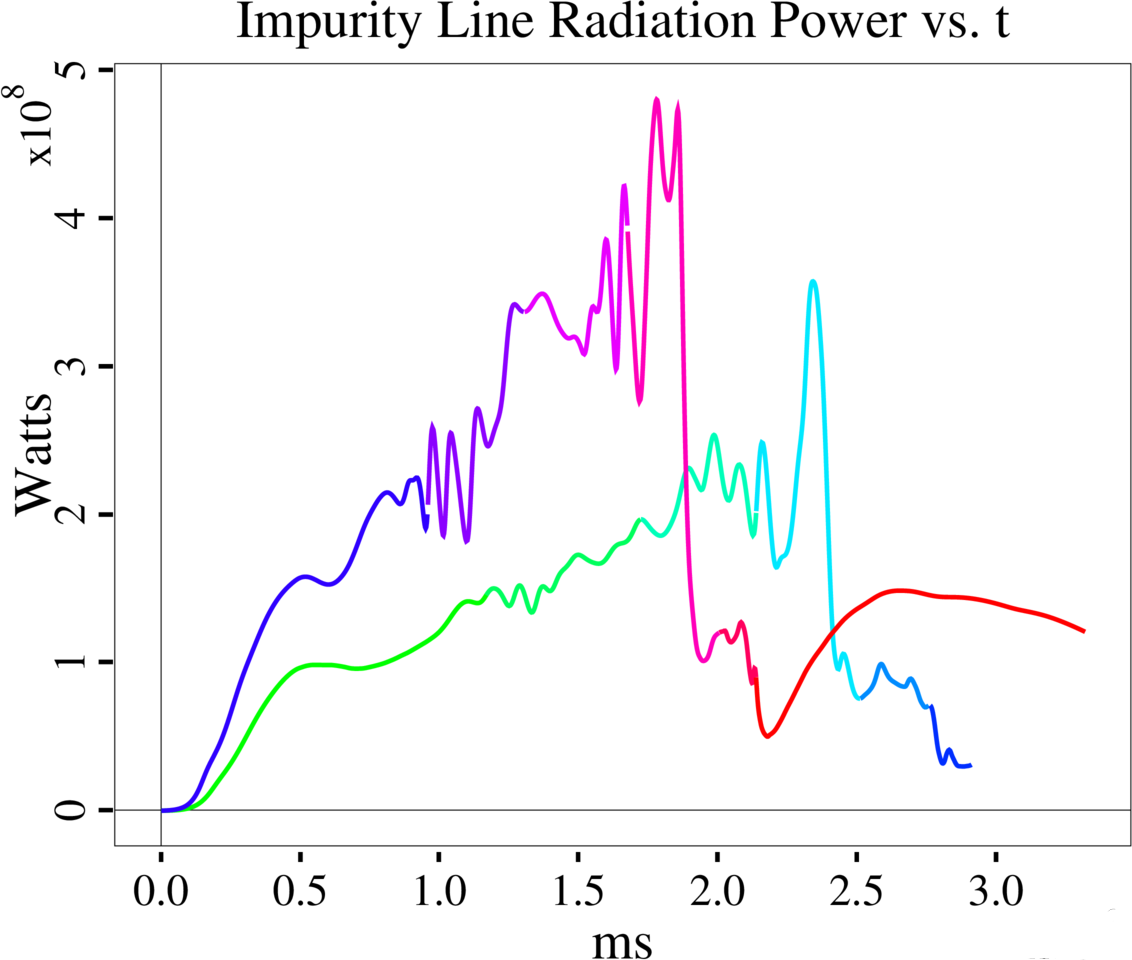}
			\vspace{-0.60cm}
			\caption{Total Radiated Power}
			\label{fig:radpow_single}
		\end{subfigure}
		\hspace{-0.25cm}
		\begin{subfigure}[b]{0.35\textwidth}
			\includegraphics[width=\textwidth,trim={0.0cm 0.0cm 0.0cm 2.0cm},clip]{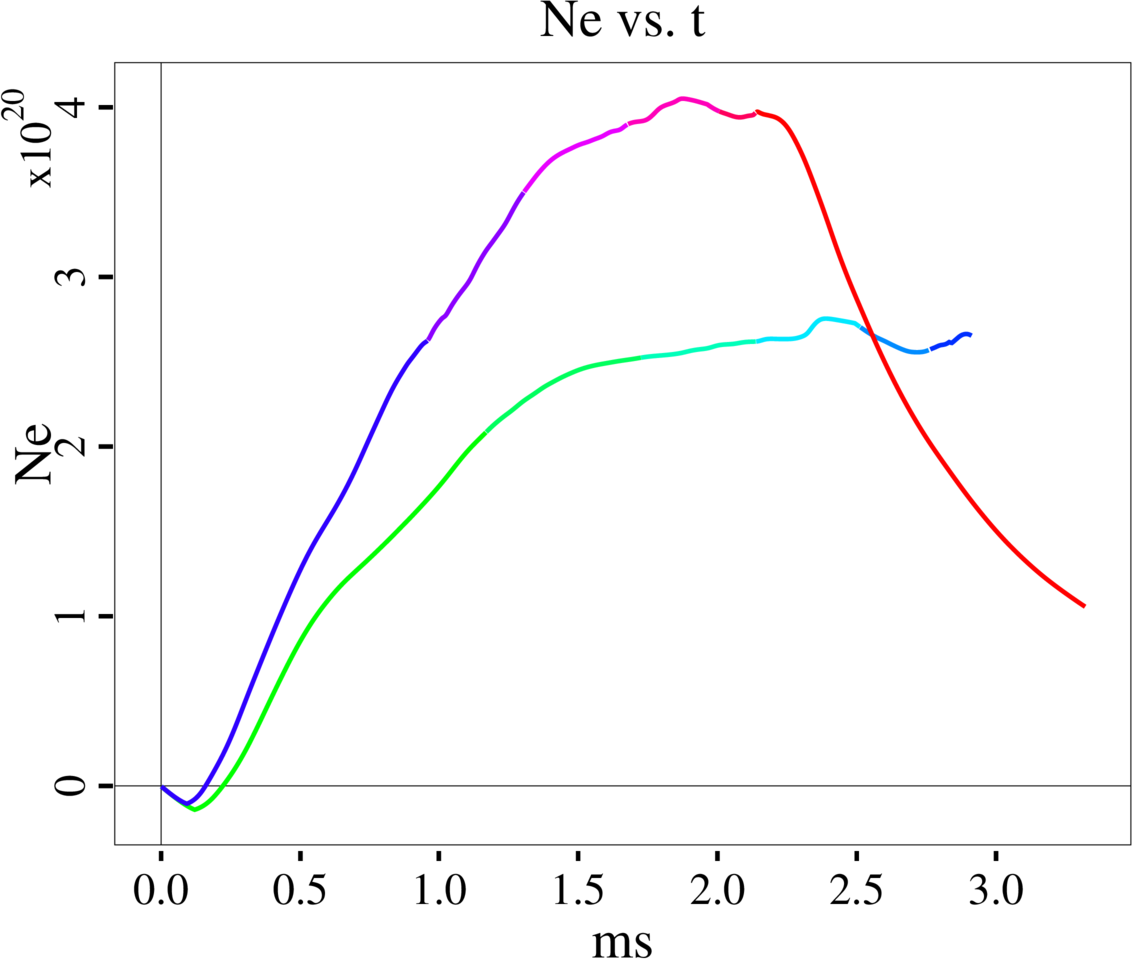}
			\vspace{-0.50cm}
			\caption{Total Electron Number}
			\label{fig:ne_single}
		\end{subfigure}
		\caption{Comparison of thermal energy, radiated energy, and radiated power shows the double load
		decreases thermal quench time, improves radiation efficiency, and increases radiation intensity
		and electron density. Different colors along the same curve indicate 48hrs intervals of
		wall time, giving an indication of the relative computing workload.}
		\label{fig:svdouble}
	\end{figure}

Figure \ref{fig:svdouble} shows a comparison of single SPI \textbf{single load} to a single SPI
\textbf{double load} (with 400 fragments) thermal quench simulation.  Plotted are time histories of the volume
integrated thermal and radiated energies, the radiate power, and electron population.  

Thermal quench simulations consist of two general phases: 1) the initial phase dominated by the fragment ablation where
the thermal loss and radiation and MHD activity are modest and primarily axisymmetric and 2) the late phase where
MHD is very active, radiation is
high and burstie, and thermal loss is more rapid.  The initial phase is seen in the linear rise in the electron 
population and the late phase is the plateau.  

For these NIMROD SPI simulations, we define the end of the thermal quench as the maximum in electron population; 
when most of the thermal energy is lost and recombination exceeds
ionization (ionization is the process of an atom/ion increasing in charge whereas recombination is the process of 
an ion decreasing in charge).
The use of MAX(electron population vs time) is a more convenient and precise definition in the simulations
(see figure \ref{fig:Ne_dt_comp}). 
Choosing a point along the curve of the thermal energy vs time is less 
convenient and less precise due to the ambiguity caused by the finite ``tail'' as the plasma reaches thermal quench.

A large spike in radiation (typically the largest amplitude) accompanies the end of the thermal quench (Figure \ref{fig:radpow_single}).

During the initial phase when the fragment ablation dominates, the thermal quench and radiation remains local,
concentrated around the ablating fragments.  During the late phase of the thermal quench, fragment ablation is a
subdominant process as indicated by the plateau in electron population.  MHD motion driven by the SPI
pushes the plasma to interact with the ambient impurities ablated during the initial phase. (see Section \ref{sec:viz})

Comparison of the \clrr{double load} to the \clrlg{single load} shows :
		\begin{itemize}
			\item faster thermal quench time : \clrr{1.87ms} vs. \clrlg{2.39ms}
			\item better radiation efficiency : \clrr{57\%} vs. \clrlg{46\%}
			\item higher peak radiation : \clrr{4.8$\times10^8$W} vs. \clrlg{3.58$\times10^8$W}
			\item more ionized electrons : \clrr{$\sim4.0\times10^{20}$} vs. \clrlg{$\sim2.5\times10^{20}$}
		\end{itemize}

These results are not surprising; we expect that doubling the plume inventory should decrease the quench time and
improve the thermal quench efficiency.  These results bracket the expectations for the dual SPI simulations
since naively, the delay between injectors should span the continuum between single load and double load.  However,
the separation of the injectors allows for MHD dynamics to confound expectations.

\section{Dual Injector SPI Simulations}
\label{sec:dualSPI}
The DIII-D dual SPI system is composed of 2 upper injectors separated by 120$^{\circ}$ (see Section \ref{sec:explayo}).
In the simulation, it is easier to change the position of the second injector, either +120$^{\circ}$ or -120$^{\circ}$,
rather than changing the timing between the two injectors.  Negative and positive dt, refers to the sign of the toroidal
displacement in the simulation.  Due to the axisymmetry in the toroidal direction, this is equivalent to the experiment
where they change the relative timings between injectors.  However, some of the synthetic diagnostics do not
line up with the experiment.  In those case, multiple angles are provided to demonstrate the toroidal variation.  

In the dual SPI simulations, a delay of dt=0.1ms is equivalent to dx=dt*v(120m/s)$\sim$1cm difference in distance
between the two plumes' trajectory, i.e. the leading plume needs to ablate away $\sim$1cm of vanguard
fragments before the second injector (delayed by dt=0.1ms) fragments can close the gap and catch up to the same 
flux surface
(but different toroidal location).  The approximate plume length is 20cm; the ablated 1cm represents about 
5\% of the plume. 
At the end of the thermal quench, the total ablation typical does not exceed 25\% of the injected simulation
inventory.  A significant fraction of
the total ablation (5\% of the total 25\%) needs to occur before the second SPI participates in the thermal quench.
For a delay of dt=0.1ms, the first injector must complete $\sim$20\% of the thermal quench before the
second injector fragments reach the plasma.  This increases in proportion to the delay between injectors.
(This estimate assumes a stationary quenching plasma column that does not move or distort.)

\subsection{Dual Injector dt=0.0ms vs Single Injector Double Load}
\vspace{-0.5cm}
	\begin{figure}[H]
		\hspace{-1.0cm}
		\begin{subfigure}[b]{0.35\textwidth}
			\includegraphics[width=\textwidth,trim={0.0cm 0.0cm 0.0cm 2.0cm},clip]{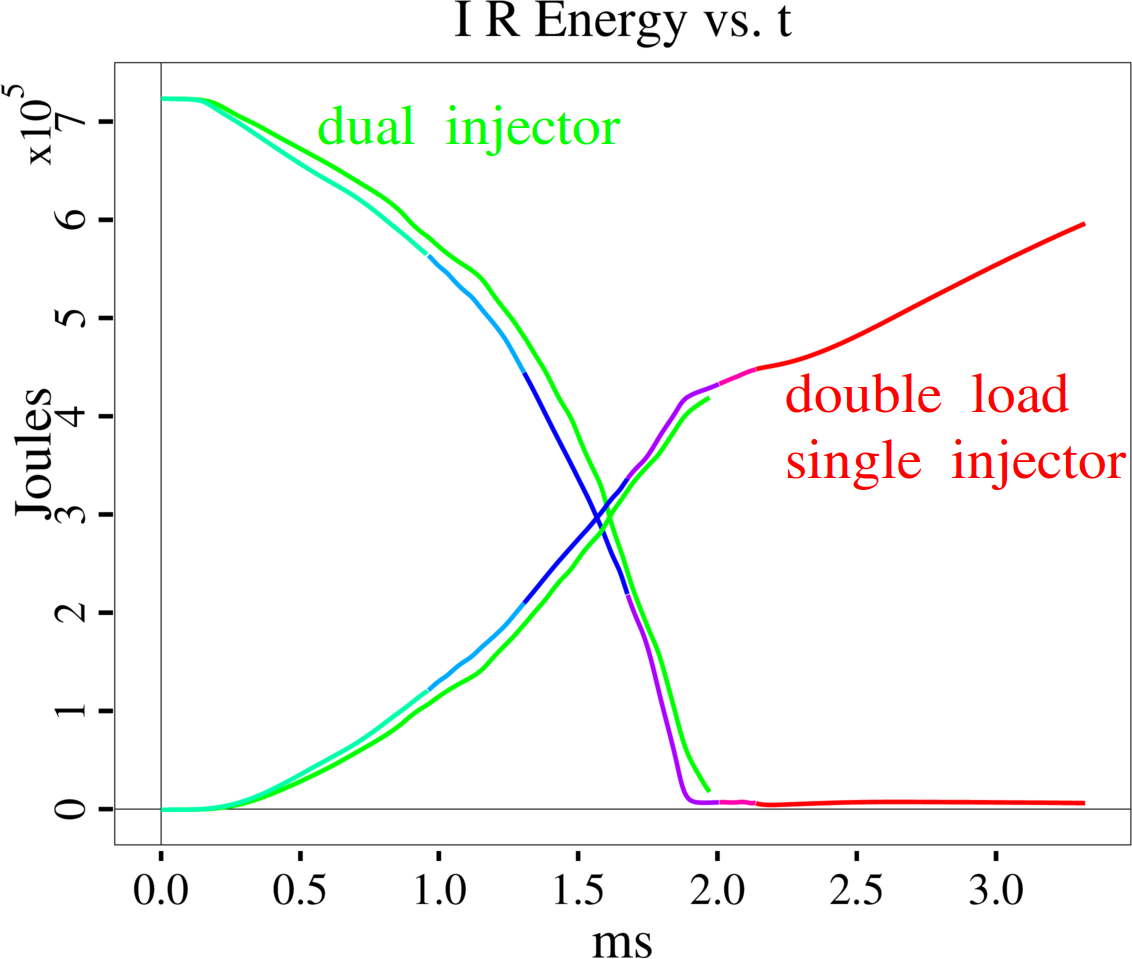}
			\vspace{-0.60cm}
			\caption{Thermal and Radiated Energy}
		\end{subfigure}
		\hspace{-0.25cm}
		\begin{subfigure}[b]{0.35\textwidth}
			\includegraphics[width=\textwidth,trim={0.0cm 0.0cm 0.0cm 2.0cm},clip]{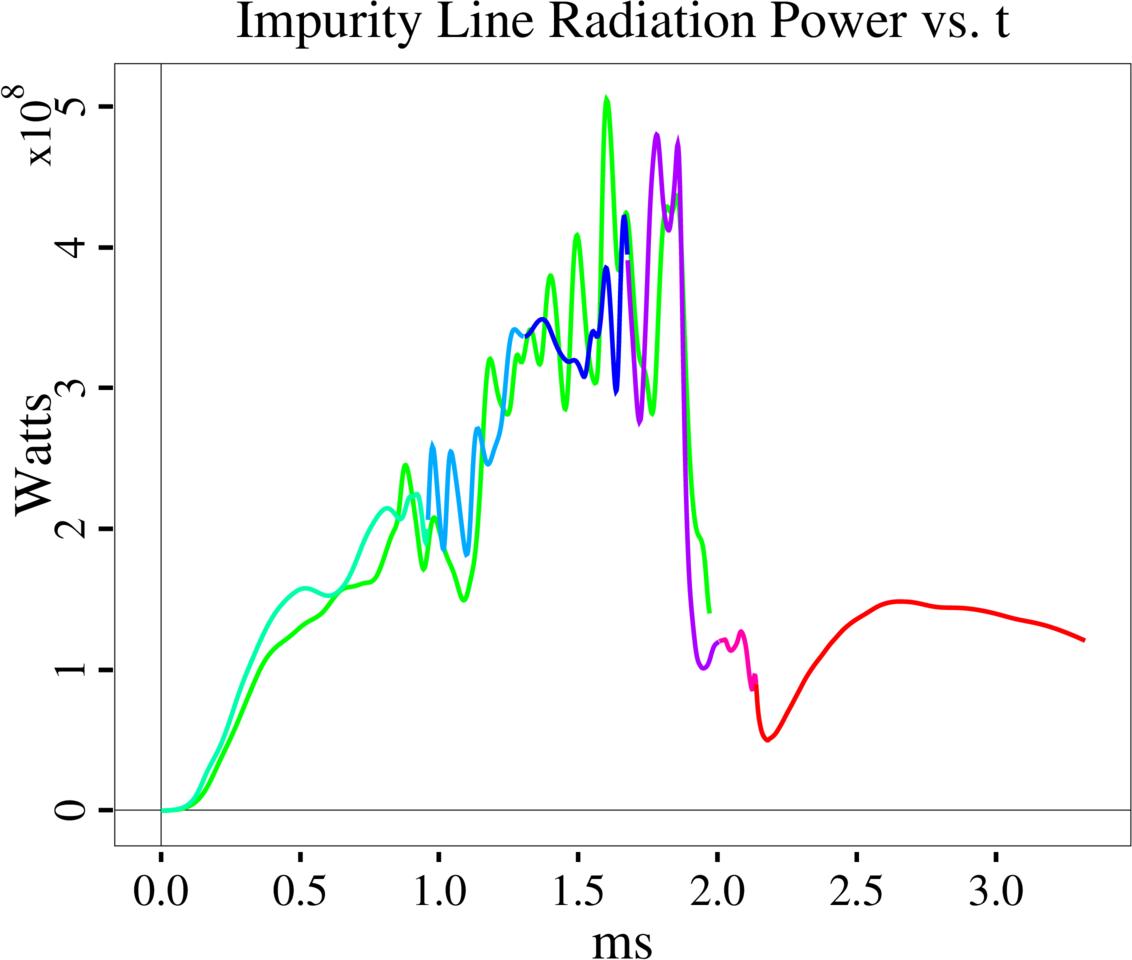}
			\vspace{-0.60cm}
			\caption{Total Radiated Power}
		\end{subfigure}
		\hspace{-0.25cm}
		\begin{subfigure}[b]{0.35\textwidth}
			\includegraphics[width=\textwidth,trim={0.0cm 0.0cm 0.0cm 2.0cm},clip]{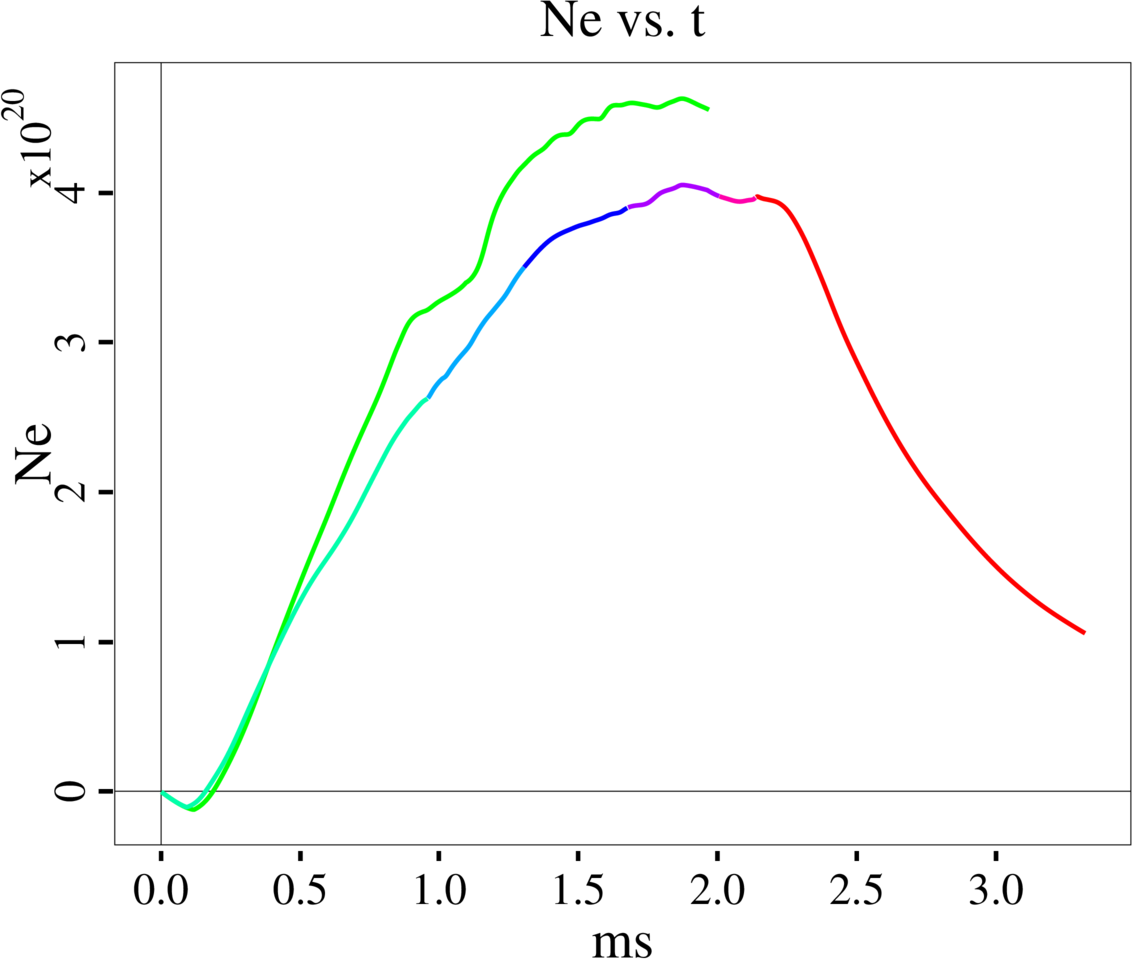}
			\vspace{-0.50cm}
			\caption{Total Electron Number}
			\label{fig:ne_dual}
		\end{subfigure}
		\caption{Comparison of single SPI double load to simultaneous dt=0 dual SPI shows
		similar thermal and radiated energy traces and radiated power.  It can be inferred
		from the similarity in radiated power, that the radiation intensity and toroidal peaking is lower for the dual
		SPI compared to the single SPI. A modest relative increase in electron population is observed
		in the dual SPI case.}
		\label{fig:svdual}
	\end{figure}
Figure \ref{fig:svdual} shows that the energies and radiated power of the simultaneous (dt=0.0) dual SPI and
single SPI double load are very similar, almost overlaying each other.  These two SPI simulations inject the same
impurity inventory but with different toroidal distributions.  It can be inferred that the dual SPI has reduced
radiation intensity and toroidal peaking since the total radiated powers are the same.  

There is a modest relative increase in electron population observed in the dual SPI case.  We also note a step
in the linear rise of the electron population for
the dual SPI case that is absent in the single SPI double load case; a display of difference in dynamics
due to the separation of the dual SPI.

For dual SPI with a finite time delay, the spatial separation must be closed in the initial phase of the 
thermal quench since it is during the initial phase when most of the fragment ablation occurs.  
With this considerations, it is not surprising that the acceptable delay is a narrow window.

\subsection{Dual Injector: dt=$\pm$0.2ms}
\begin{figure}[H]
		\hspace{-0.50cm}
		\begin{subfigure}[b]{0.50\textwidth}
			\includegraphics[width=\textwidth,trim={0.0cm 0.0cm 0.0cm 2.0cm},clip]{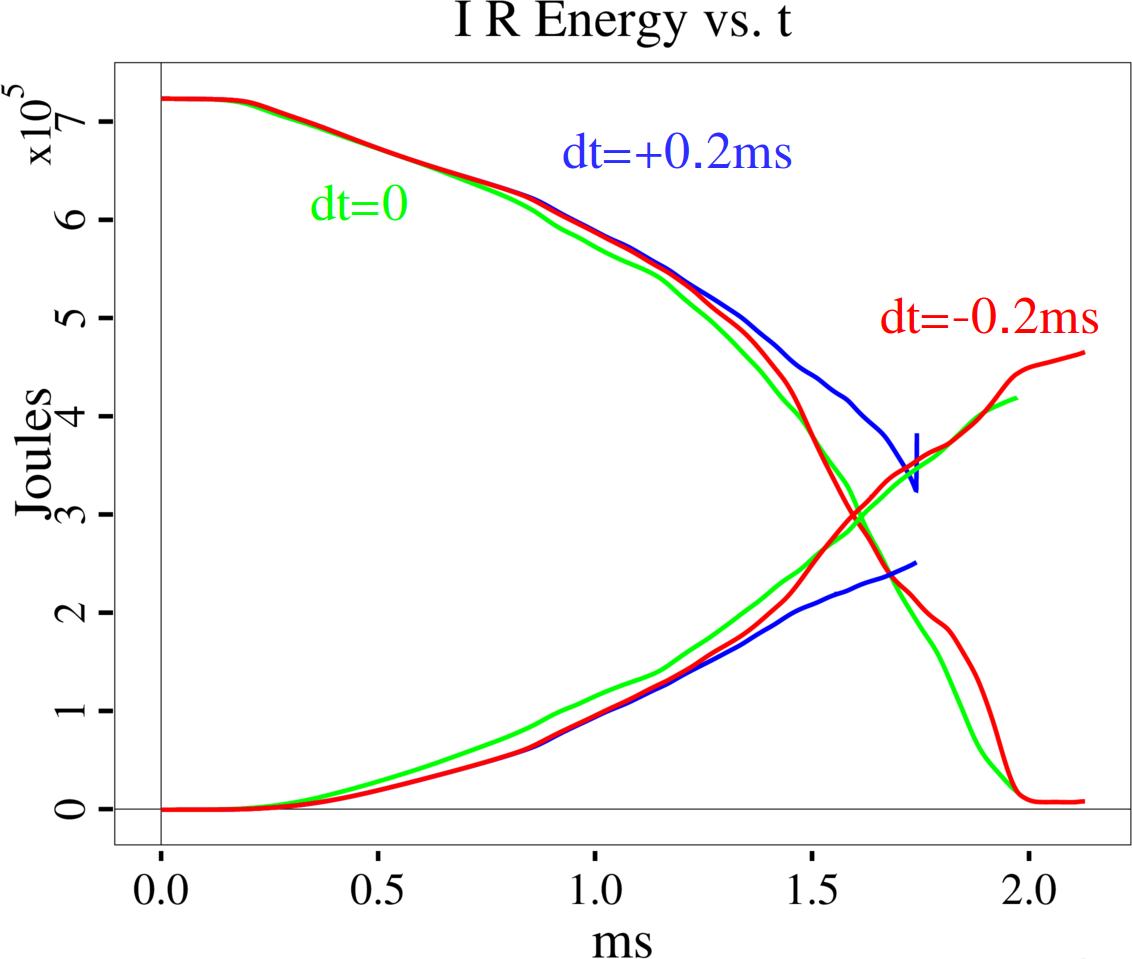}
			\vspace{-0.60cm}
			\caption{Thermal and Radiated Energy}
		\end{subfigure}
		\hspace{0.00cm}
		\begin{subfigure}[b]{0.50\textwidth}
			\includegraphics[width=\textwidth,trim={0.0cm 0.0cm 0.0cm 2.0cm},clip]{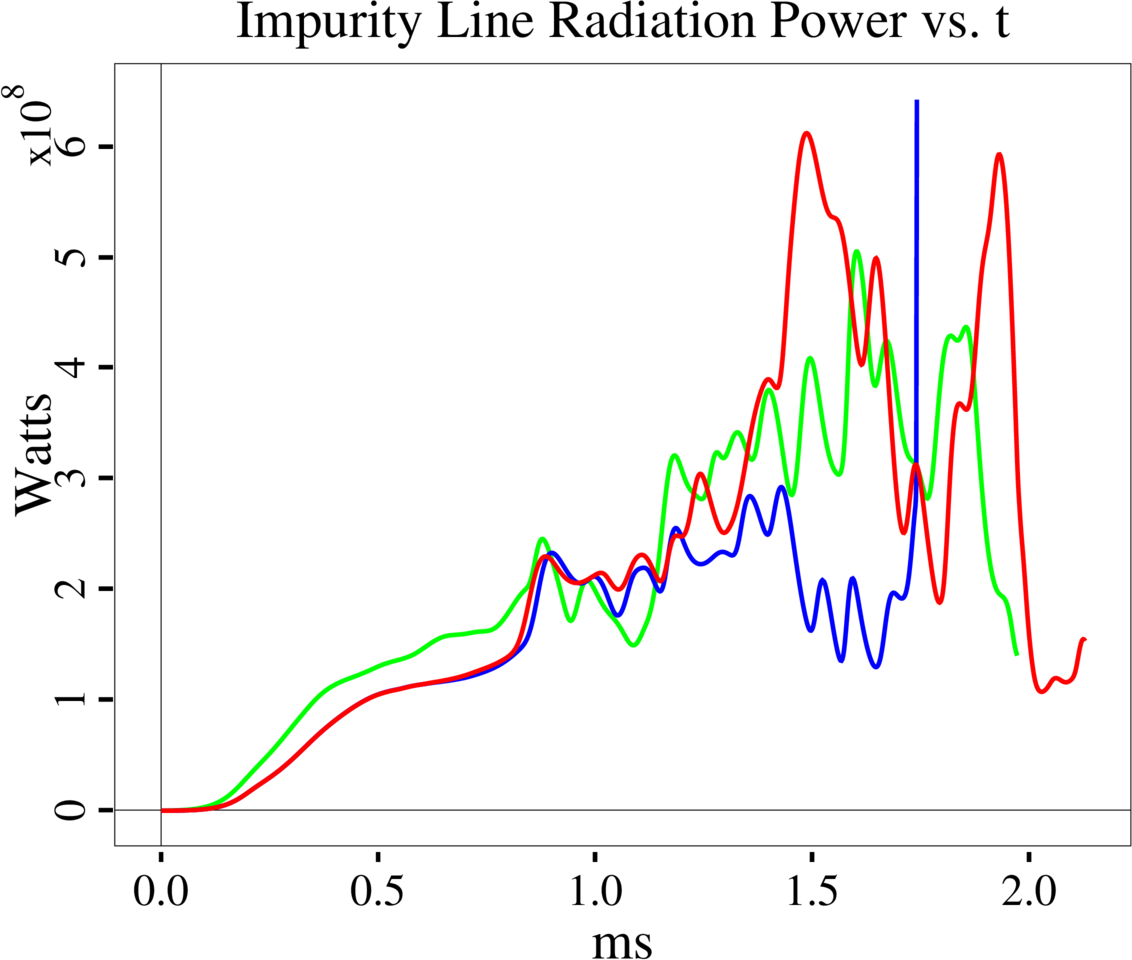}
			\vspace{-0.60cm}
			\caption{Total Radiated Power}
		\end{subfigure}
		\caption{Comparison of dt=$\pm$0.2ms injector delay indicates differences in positive versus
		negative delay times.  Unfortunately, the \clrb{dt=+0.2ms} simulation did not complete due to
		numeric instability.}
		\label{fig:dt2}
	\end{figure}

Figure \ref{fig:dt2} shows a comparison of thermal and radiated energy and total radiated power for the dt=$\pm$0.2ms
injector delays, indicating a differences in positive versus negative delay times.  The plots show that the two
delays \clrb{dt=+0.2ms} and \clrr{dt=-0.2ms} proceed identically until about t$\simeq$0.6ms when the radiated power plots 
begin to diverge.  Up
until t$\simeq$0.6ms, the thermal quench is driven the the single leading injector.  Even though the delay between
injectors is only 0.2ms, it required 0.6ms to burn through the vanguard fragments of the first injector fragments
and allow the second injector fragments to catch up to the plasma surface, which is receding due to the quenching 
by the leading
injector fragments.  By t=1.2ms, the \clrr{dt=-0.2ms} case is clearly radiating more.  Unfortunately, the \clrb{dt=+0.2ms}
simulation did not complete due to numeric instabilities making further conclusions difficult.

Due to this incomplete simulation, minimal analysis is applied to the dt=$\pm$0.2ms cases.  We will instead focus
our attention on analysis and comparison of the other more viable cases.

\subsection{Dual Injector: dt=$\pm$0.4ms}
\label{sec:dtn04}
\vspace{-0.5cm}
	\begin{figure}[H]
		\hspace{-0.50cm}
		\begin{subfigure}[b]{0.50\textwidth}
			\includegraphics[width=\textwidth,trim={0.0cm 0.0cm 0.0cm 2.0cm},clip]{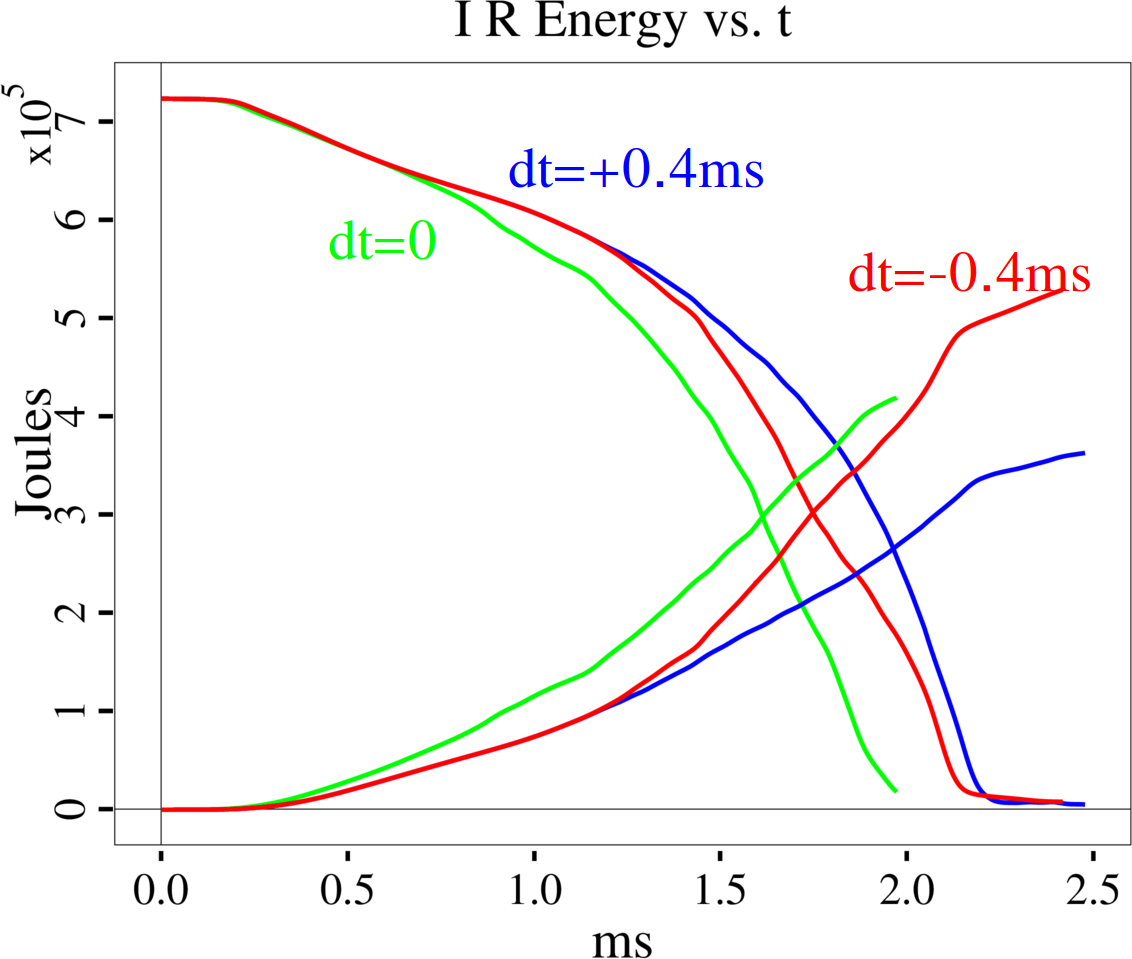}
			\vspace{-0.60cm}
			\caption{Thermal and Radiated Energy}
		\end{subfigure}
		\hspace{0.00cm}
		\begin{subfigure}[b]{0.50\textwidth}
			\includegraphics[width=\textwidth,trim={0.0cm 0.0cm 0.0cm 2.0cm},clip]{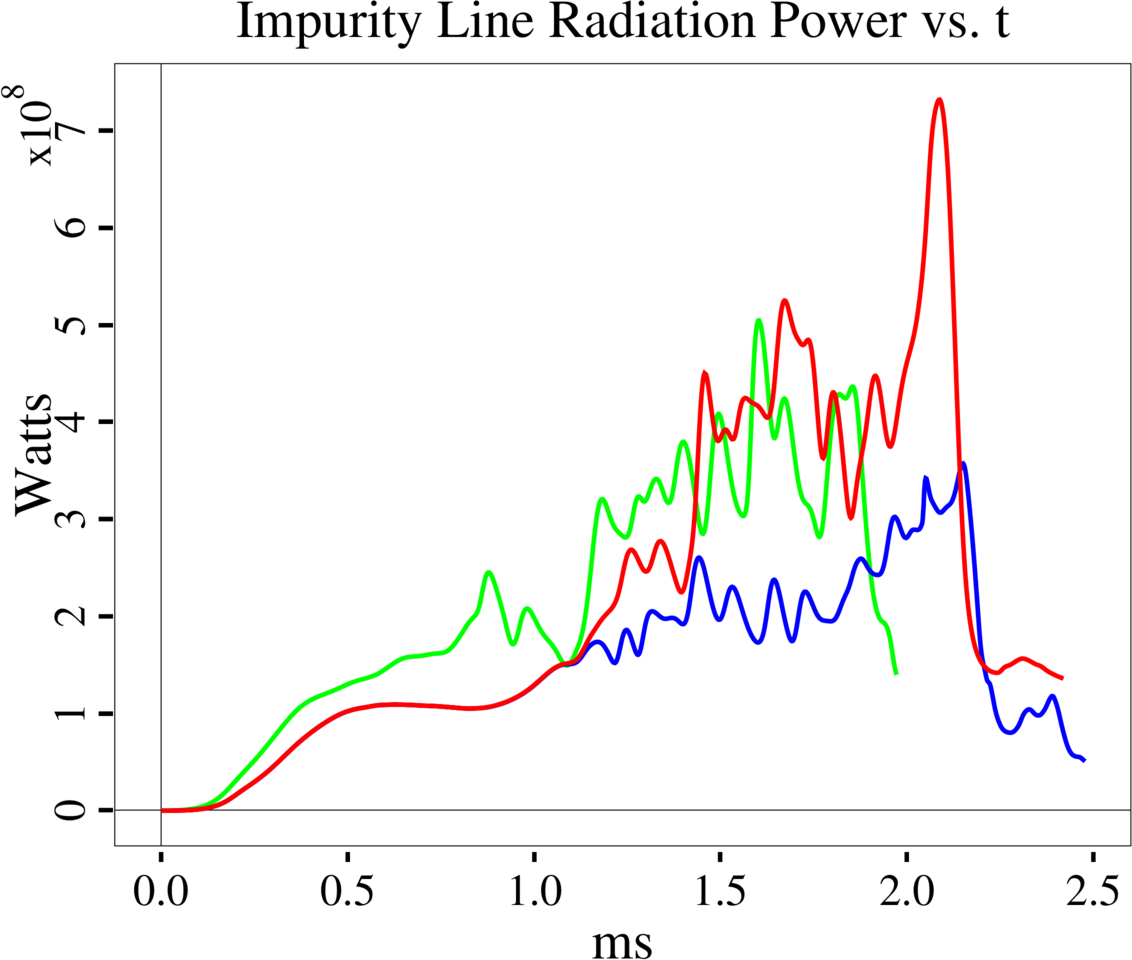}
			\vspace{-0.60cm}
			\caption{Total Radiated Power}
		\end{subfigure}
		\caption{Comparison of dt=$\pm$0.4ms injector delay shows clear differences in positive versus
		negative delay times. Negative delay \clrr{dt=-0.4ms} shows improved thermal quench whereas
		\clrb{dt=+0.4ms} shows degraded thermal quench compared to \clrlg{dt=0ms}. 
		}
		\label{fig:dt4}
	\end{figure}
Figure \ref{fig:dt4} shows a comparison of thermal and radiated energy and total radiated power for the \clrb{dt=+0.4ms}
and \clrr{dt=-0.4ms} injector delays.  The two dt=$\pm$0.4ms cases are identical up to about t$\simeq$1.1ms,
almost twice longer
than the two dt=$\pm$0.2ms cases.  Again, as in the dt=$\pm$0.2ms cases, the leading injector dominates and the 
thermal quench begins as a single SPI single load quench.  At t$\simeq$1.1ms, the dt=-0.4ms case begins to produces
more radiation as the second fragment plume catches up to the plasma (see Section \ref{sec:vizdtn04}) and the
two cases diverge.

Increasing the time delay to dt=$\pm$0.4ms shows a clear difference in positive versus negative delay:
\begin{itemize}
	\item radiation efficiency improves for \clrr{dt=-0.4ms} from \clrlg{0.58}(\clrlg{dt=0}) to \clrr{67\%}!
	\item radiation efficiency decreases for \clrb{dt=+0.4ms} to \clrb{46\%} 
	\item but \clrr{dt=-0.4ms}, \clrb{dt=+0.4ms} have similar $\tau_{TQ}$
	\item \clrr{dt=-0.4ms} radiation peak increases to \clrr{7.33$\times10^8$W} (from \clrlg{dt=0 5.05$\times10^8$W})
	\item \clrb{dt=+0.4ms} radiation peak decreases to \clrb{3.58$\times10^8$W}
\end{itemize}

Comparing the dt=$\pm$0.2ms to the dt=$\pm$0.4ms we noted already that the dt=$\pm$0.4ms proceeds almost twice as long
on a single SPI single load thermal quench trajectory than the dt=$\pm$0.2ms.  During the interval t=[0.6,1.1],  
both the dt=$\pm$0.2ms cases track relatively closely.  It is not until
t$\simeq$1.2ms that the dt=$\pm$0.2ms cases significantly diverge.  For the dt=$\pm$0.4ms cases, there is a clear
divergence at t$\simeq$1.1ms with the \clrr{dt=-0.4ms} radiation exceeding the \clrlg{simultaneous case} while the
\clrb{dt=+0.4ms} case produces less radiation than the simultaneous case.

\subsubsection{Magnetic Helicity Important in Symmetry Breaking}\label{maghel}
\vspace{-0.5cm}
	\begin{figure}[H]
	\centerline{
		\begin{subfigure}[b]{0.550\textwidth}
			\includegraphics[width=0.95\textwidth,trim={0.0cm 0.0cm 0.0cm 0.0cm},clip]{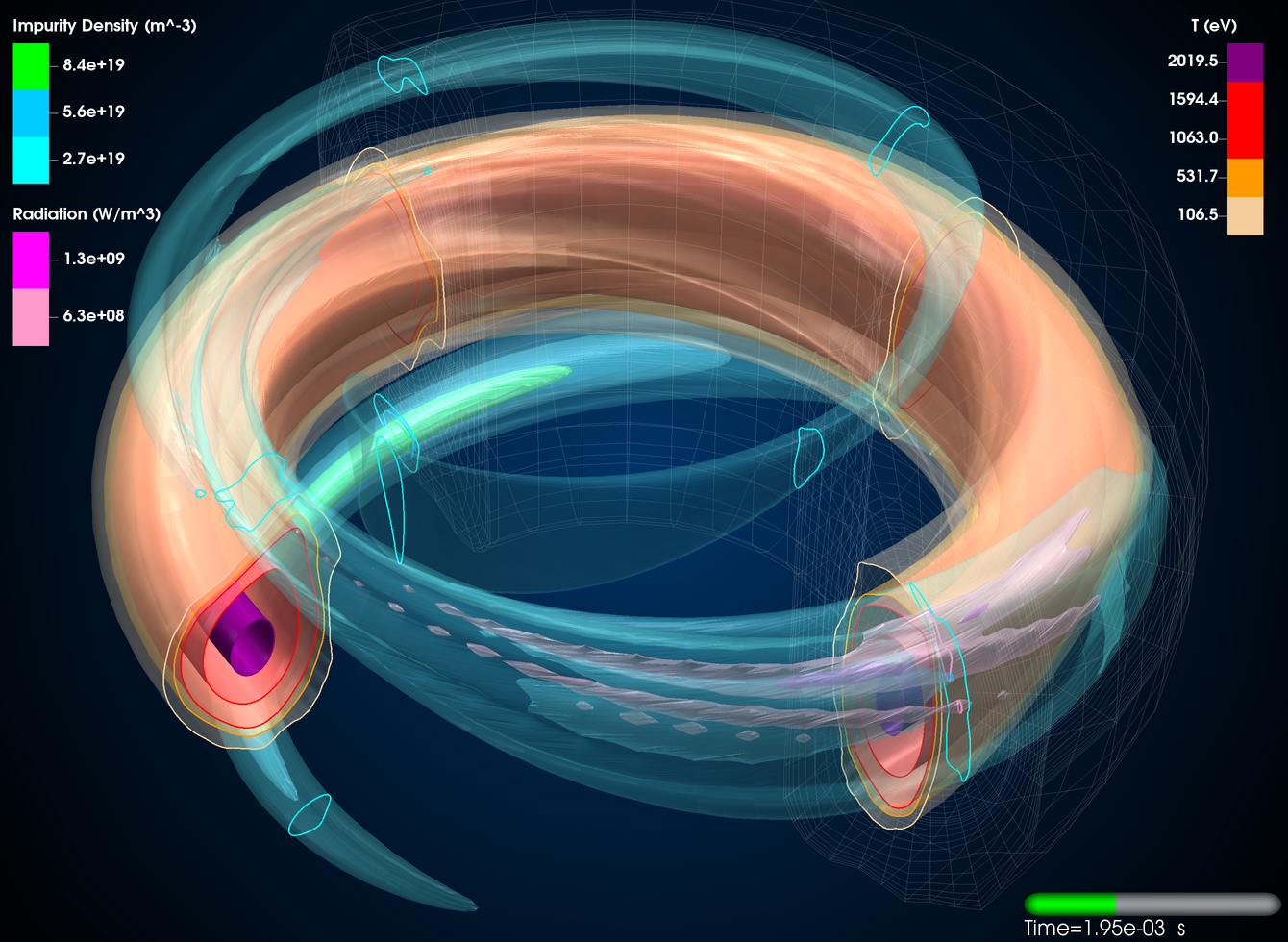}
		\end{subfigure}
		\hspace{-0.50cm}
		\begin{subfigure}[b]{0.250\textwidth}
			\includegraphics[width=0.95\textwidth,trim={0.0cm 1.5cm 0.0cm 1.0cm},clip]{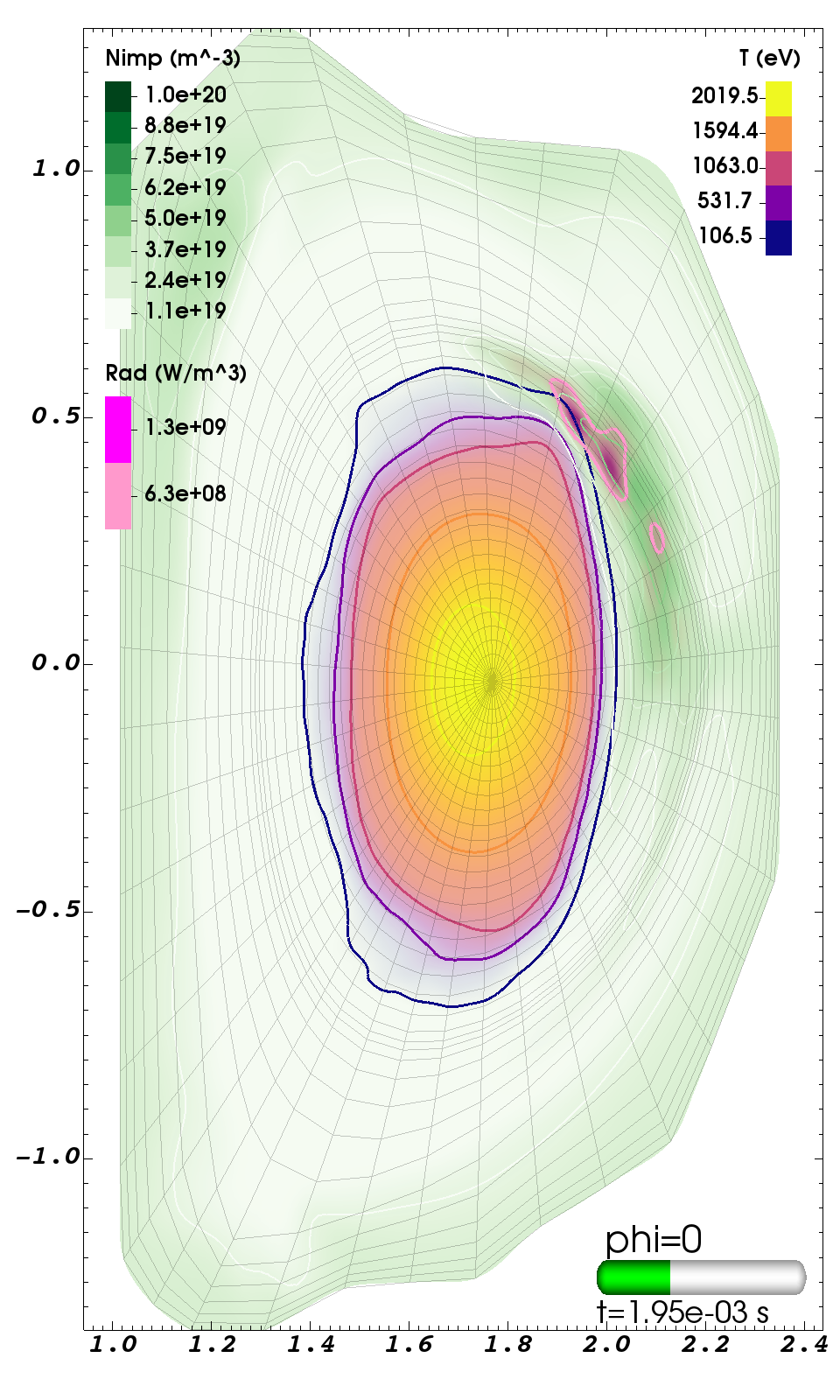}
		\end{subfigure}
		\hspace{-0.50cm}
		\begin{subfigure}[b]{0.250\textwidth}
			\includegraphics[width=0.95\textwidth,trim={0.0cm 1.5cm 0.0cm 1.0cm},clip]{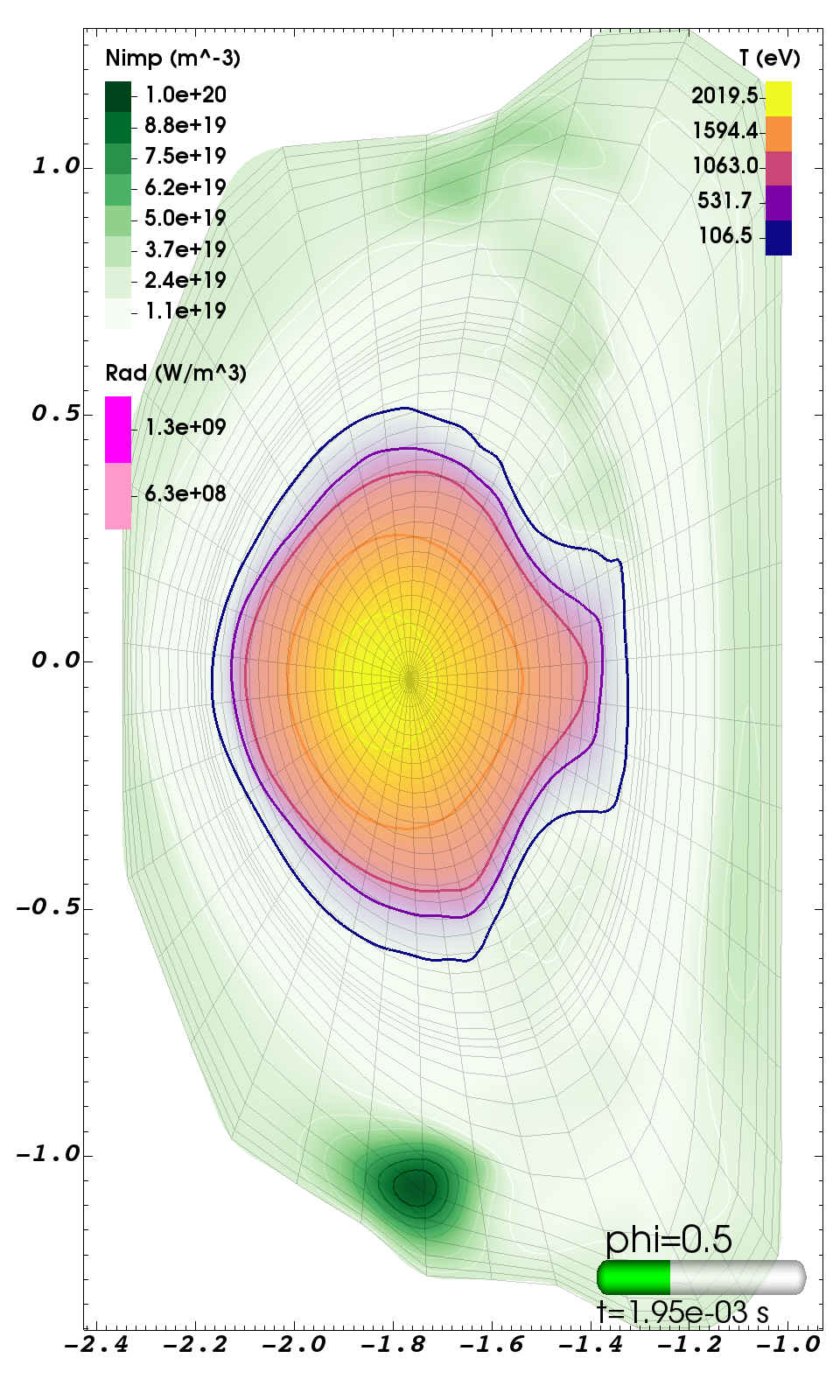}
		\end{subfigure}
		}
		\caption{DIII-D single upper injector, single load SPI 3D visualization and poloidal cross sections at
		$\phi=0^\circ$ and $\phi=180^\circ$ of temperature, impurity density, and radiation at t=1.95ms
		(see legend for values).
		The impurity density has flowed primarily along the fields winding around the plasma.  Close
		observation reveals that the impurity density plume has lifted away from its originating flux surface as
		it streams along the field lines due to a flux normal drift.  This drift lifts the
		downstream impurities away from the hot plasma surface, limiting interaction and contributes to the
		localized radiation peaking.  The radiation filament seen in the 3D image is much shorter than the
		impurity density filament.  The hot spot seen in the $\phi=0^\circ$ cross section show the localized
		nature of the radiation.  Radiation hot spots are >1.0GW/m$^3$.
		}
		\label{fig:3dviz}
	\end{figure}

The magnetic helicity (handedness of the plasma) and the connection path between injectors is essential in breaking
the symmetry in injector delays.  

Figure \ref{fig:3dviz} shows Visit\cite{VisIt2012} 3D visualization and poloidal cross sections at
$\phi=0^\circ$ and $\phi=180^\circ$ of temperature, impurity density, and radiation at t=1.95ms(see legend for values).
The impurity density has flowed primarily along the fields winding around the plasma.  Close
observation reveals that the impurity density plume has lifted away from its originating flux surface as
it streams along the field lines due to a flux normal drift\cite{Pegouri2006}.  This drift lifts the
downstream impurities away from the hot plasma surface, limiting interaction and contributes to the
localized radiation peaking.  Note that the radiation filament seen in the 3D image is much shorter than the
impurity density filament.  The hot spot seen in the $\phi=0^\circ$ cross section show the poloidally localized
nature of the radiation.  Radiation hot spots are >1.0GW/m$^3$.
The 3D visualization show
strong distortions of the confining flux surfaces (temperature isosurfaces are proxies for the flux surface)
as the plasma is quenched.  Portions of the plasma column bulge
outboard while other parts retreat in response to the deposited impurities from the ablating fragments.  
It is easy to imagine the ample
opportunities for increased or decreased interaction with a second toroidally separated SPI plume.

We have seen that the evolution of the ablated and ionizing impurities is dominantly along a parallel flow.
Along a toroidally helical path, the upstream trajectory from the injector plane at $\phi=0^\circ$ to the injector 
plane at $\phi=120^\circ$
is not the same as the the upstream trajectory from the injector at $\phi=120^\circ$ to the injector at $\phi=0^\circ$.
This difference in trajectories and the non-axisymmetric plasma disturbances accounts for the asymmetry
observed in the dual SPI delay scan.

Section \ref{sec:viz} and the animation links within give further insight into the role of the driven MHD and
interaction with the dual SPI fragments.

\subsubsection{Improved Thermal Quench Correlates with Reduction in MHD}
\begin{figure}[H]
		\begin{subfigure}[b]{0.5\textwidth}
			\includegraphics[width=\textwidth]{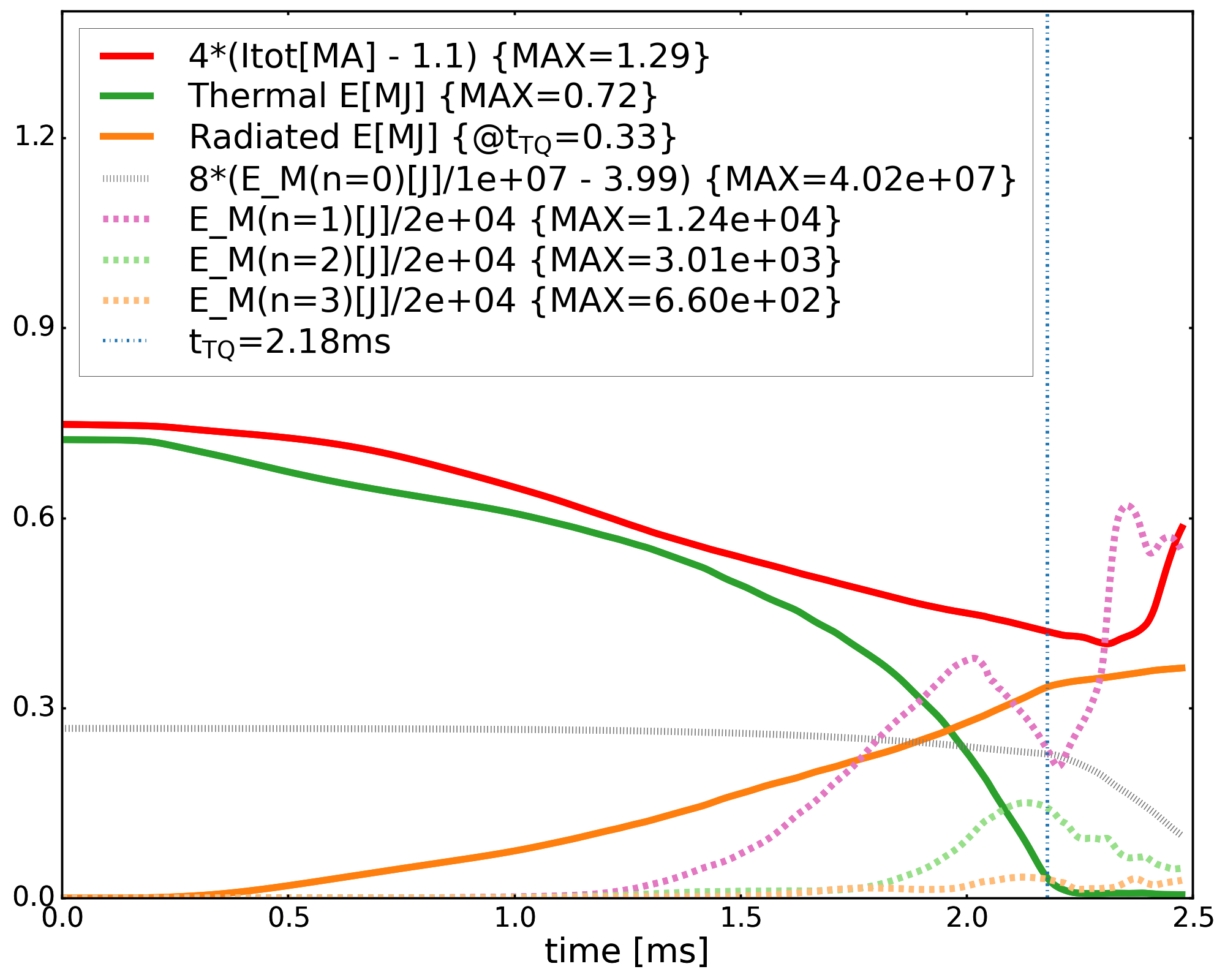}
			\vspace{-0.60cm}
			\caption{dt=+0.4ms}
			\label{fig:allddtp04}
		\end{subfigure}
		\hspace{-0.25cm}
		\begin{subfigure}[b]{0.5\textwidth}
			\includegraphics[width=\textwidth]{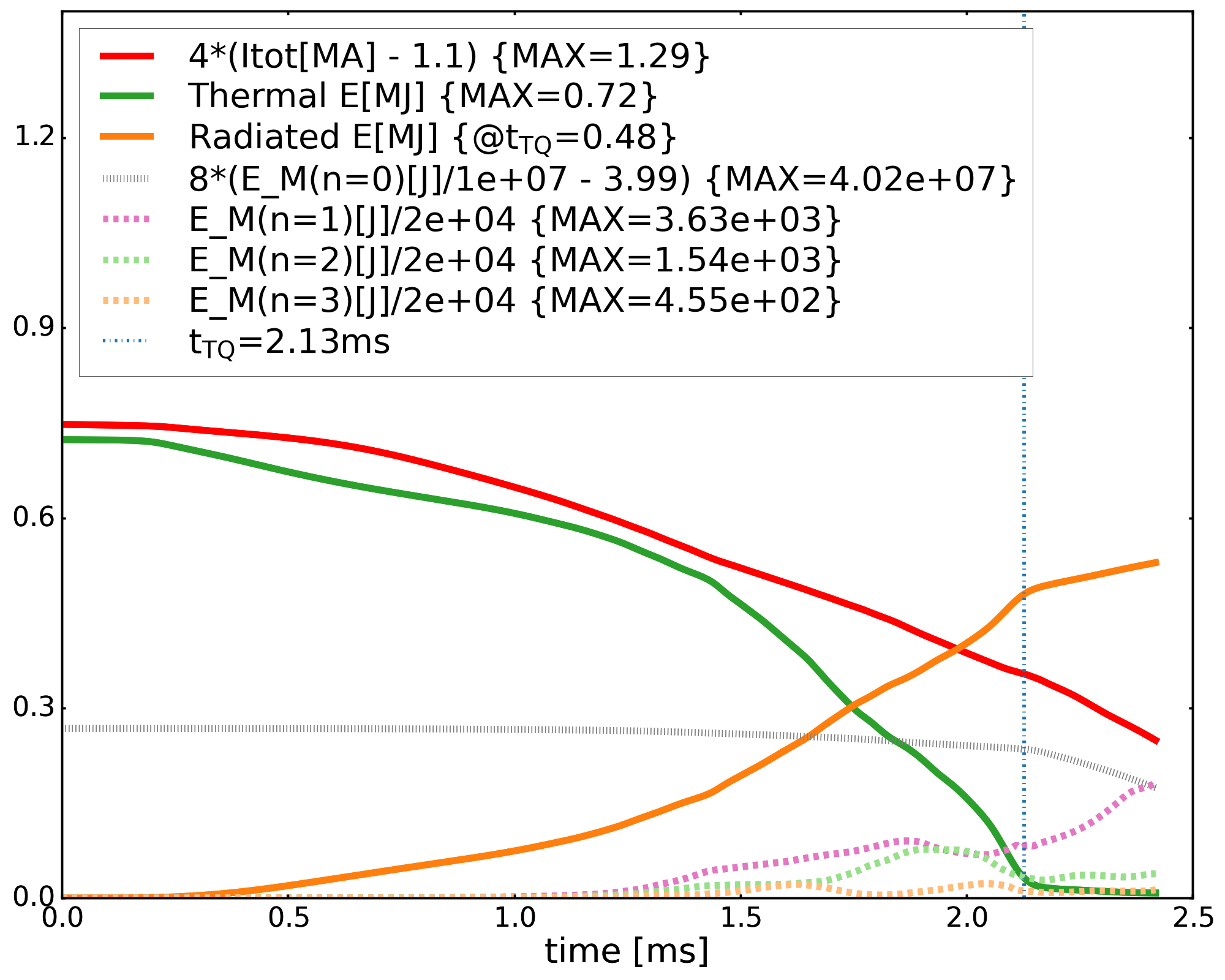}
			\vspace{-0.60cm}
			\caption{dt=-0.4ms}
			\label{fig:allddtn04}
		\end{subfigure}
		\\
		\begin{subfigure}[b]{0.5\textwidth}
			\includegraphics[width=\textwidth]{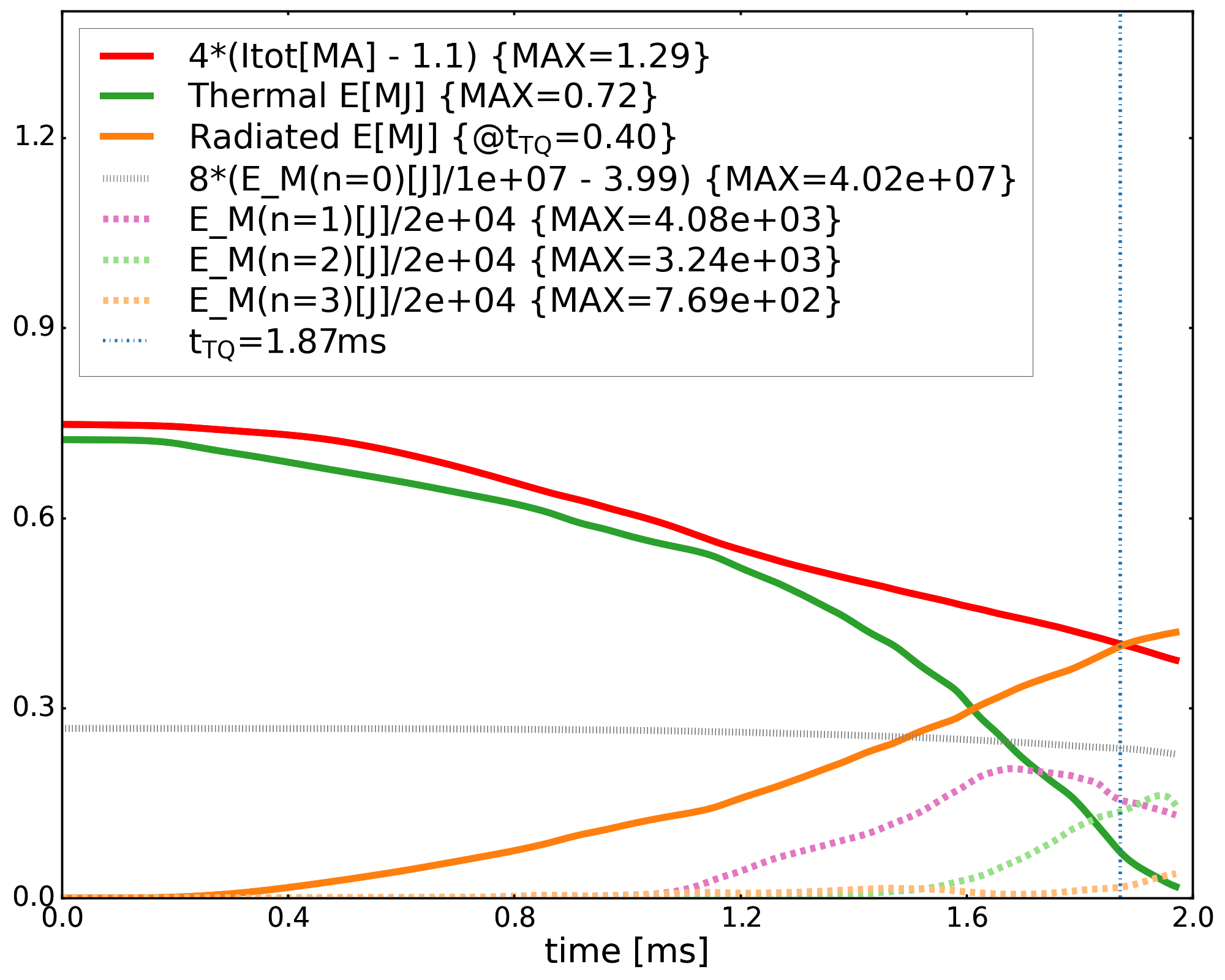}
			\vspace{-0.60cm}
			\caption{dt=0}
			\label{fig:allddt0}
		\end{subfigure}
		\hspace{-0.25cm}
		\begin{subfigure}[b]{0.5\textwidth}
			\includegraphics[width=\textwidth]{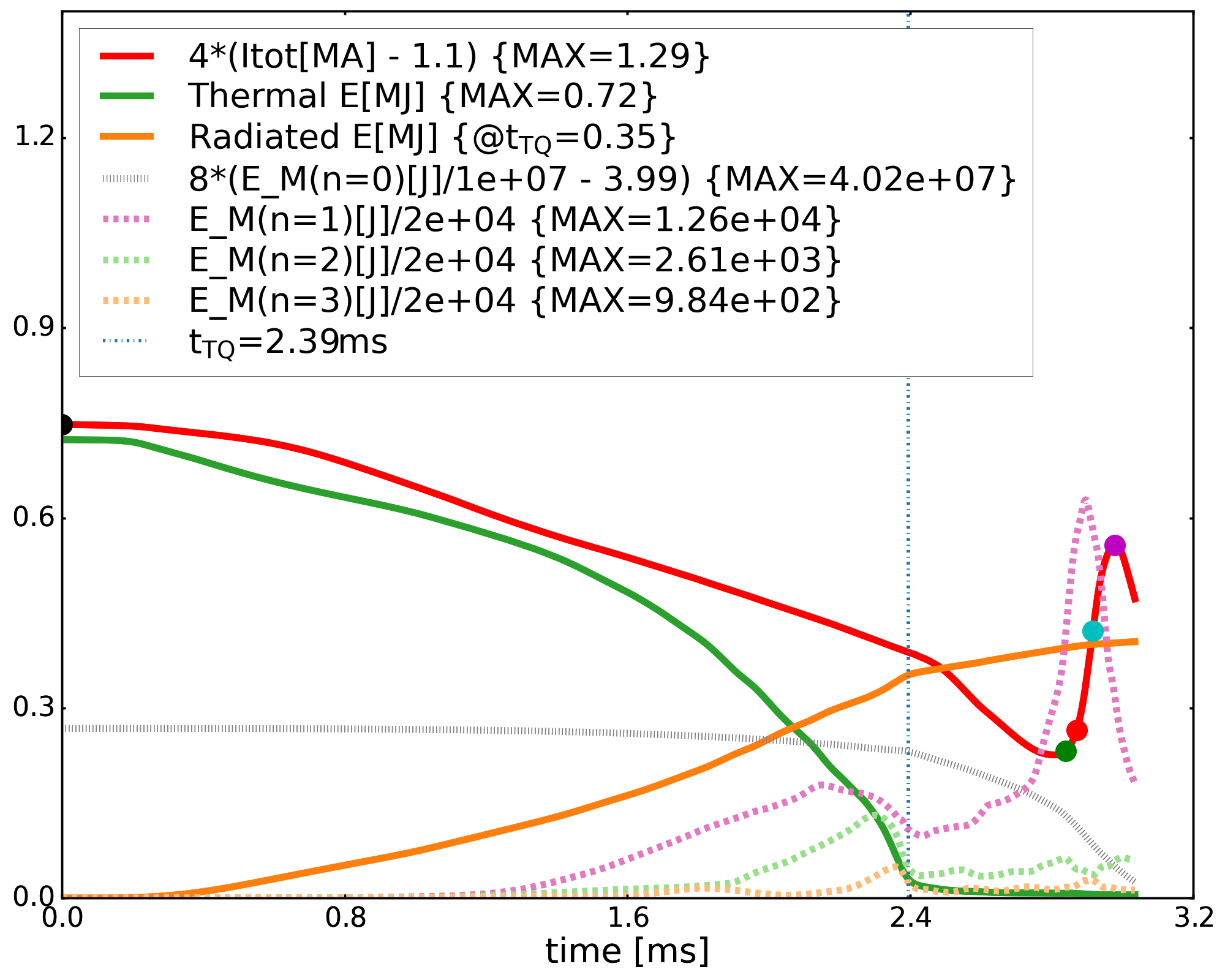}
			\vspace{-0.60cm}
			\caption{dt=-0.8ms}
			\label{fig:allddt08}
		\end{subfigure}
		\caption{Summary plot of the thermal quench for dt=[$\pm$0.4,0.0,-0.8]ms showing plasma current, thermal
		energy, radiated energy, and magnetic energies for n=[0,1,2,3].  The legend includes the maximum values
		of each trace, except the radiated energy which indicates the radiated energy at the thermal quench
		shown by the dot-dash vertical line (note the scaling factors for the traces) .  A more efficient
		thermal quench correlates with a reduction in mode energies.  The relative amplitudes and phases of the
		the dominant n=[1,2,3] modes shift as the delay is altered.  The dt=-0.8ms is a proxy for a single
		SPI single load case and runs through the current spike and provides additional comparison to experiment.
		The colored dots along the plasma current trace correspond to the last closed flux surfaces in Figure
		\ref{fig:lcfsdt08} .
		}
		\label{fig:MEdt4}
	\end{figure}

Figure \ref{fig:MEdt4} shows summary plots of the thermal quench for dt=[+0.4,-0.4,0.0,-0.8]ms showing plasma current,
thermal energy, radiated energy, and magnetic energies for n=[0,1,2,3], where some traces have been scaled to fit on
the same plot.  The legend includes the maximum values of
each trace, except the radiated energy which includes the radiated energy at the end of the thermal quench shown
by the dot-dash
vertical line.  The legend also includes the scaling and shift factors for each of the traces.  All plots use identical
scaling and shift factors for ease of comparison.  Units are labeled in the legend in the square brackets.

We include the dt=-0.8ms case to complete the comparison and stand as a proxy for a single SPI case.  In addition,
dt=-0.8ms case runs through the current spike and provides an additional comparison to experiment.  The colored dots
along the plasma current trace on the current spike correspond to the last closed flux surfaces in 
Figure \ref{fig:lcfsdt08} .

Examination of the magnetic mode energies in Figure \ref{fig:MEdt4} clearly show that the delays 
between the injectors alters the MHD evolution, particularly prominent among the dominant n=[1,2,3] modes. 
The dominant n=1 magnetic energy is smaller by
almost a factor of $\sim5$ from dt=+0.4ms to dt=-0.4ms as the radiated energies increase from .33MJ to .48MJ .  

The n=1 mode is suppressed by the dt=-0.4ms delayed second plume fragments (compared to simultaneous and single case)
resulting in less mode activity and less stochasticity and more radiation.  For the dt=+0.4ms case, the n=1 mode is
promoted, increasing mode activity and stochasticity and thermal loss.  The n=2 mode is also larger amplitude.  

The improvement in radiated energy for delay dt=-0.4ms over dt=0.0ms is an instance of ``being in the right
place at the right time''.  As mentioned in Section \ref{maghel}, the helical distortion/motion of the quenching plasma
intercepts the dt=-0.4ms delayed plume fragments and dampens the distortion resulting in a more quiescent and less
stochastic plasma through the thermal quench.  

The lower performance of the dt=+0.4ms case is disappointing but we note that it is no worse than the single SPI
single load case.

    \begin{figure}
        \centering
        \includegraphics[width=0.75\textwidth]{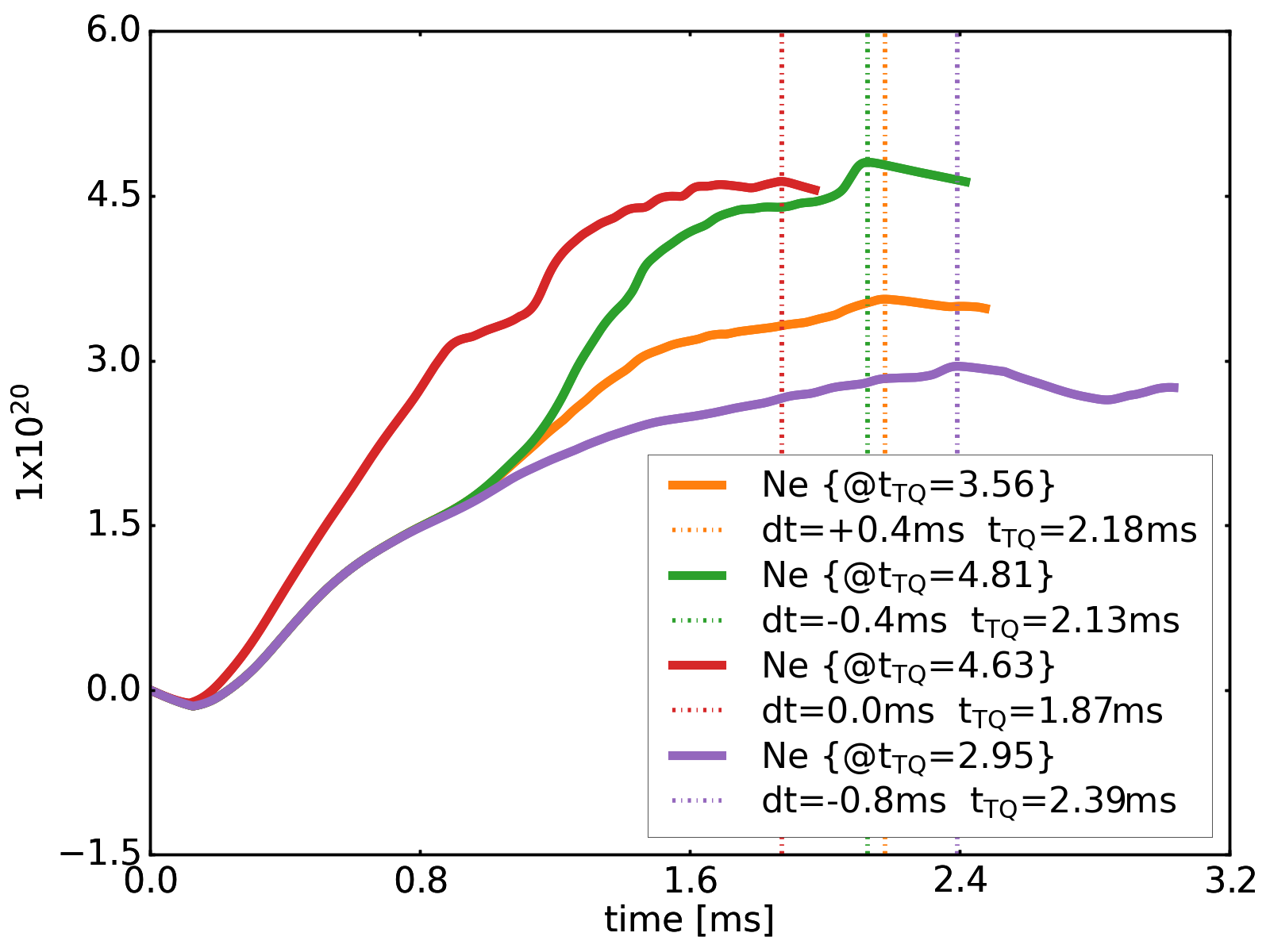}
        \caption{Electron Population for dual SPI delays dt=$[\pm$0.4,0.0,-0.8]ms.  The dotted vertical lines
        indicate the thermal quench time for each case.  The legend includes the MAX(N$_e$).  The electron count
        is $\sim$50-60\% higher in the more optimal dual SPI case than the single SPI case (dt=-0.8). }
        \label{fig:Ne_dt_comp}
    \end{figure}
    
\subsection{Dual Injector: dt=$\pm$0.8ms}
\label{sec:dt08}
\vspace{-0.5cm}
	\begin{figure}[H]
		\hspace{-0.50cm}
		\begin{subfigure}[b]{0.50\textwidth}
			\includegraphics[width=\textwidth,trim={0.0cm 0.0cm 0.0cm 2.0cm},clip]{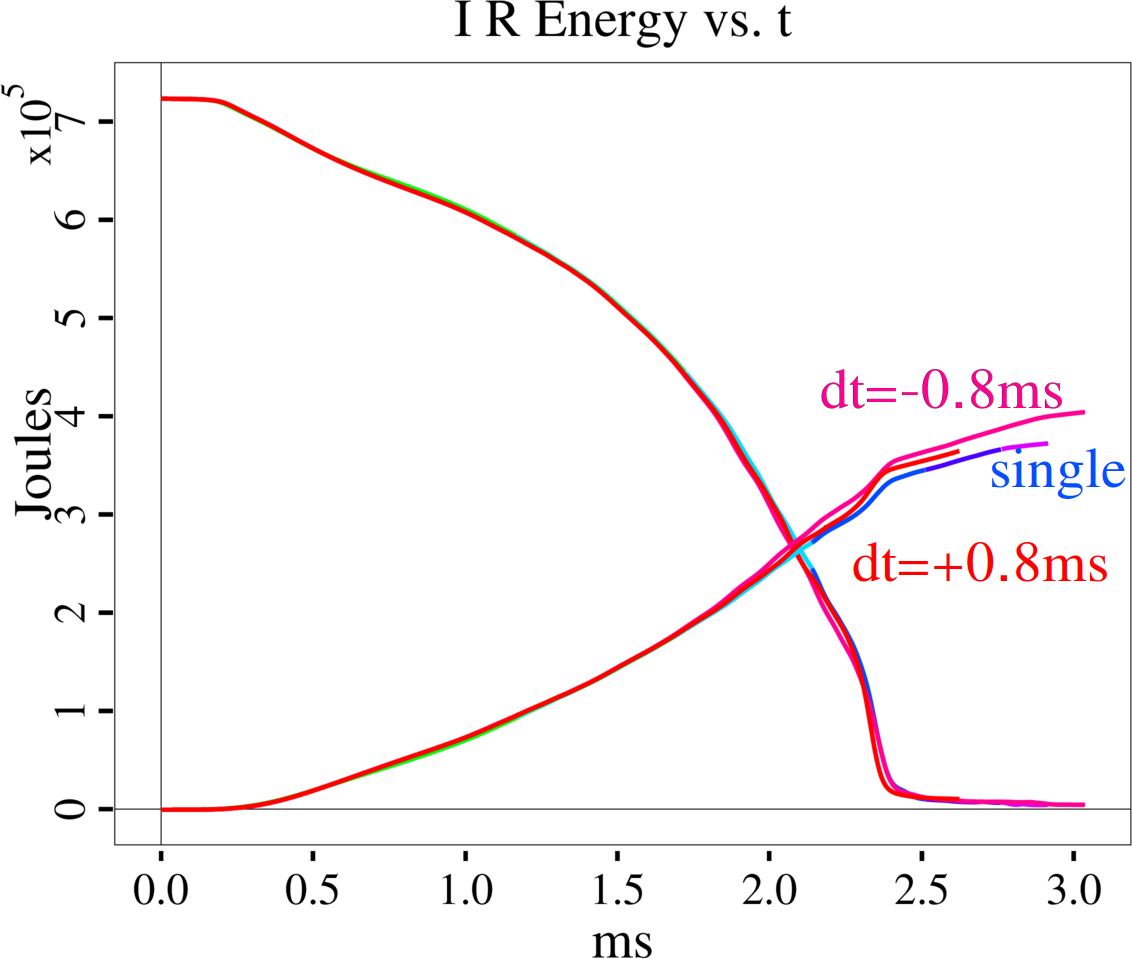}
			\vspace{-0.60cm}
			\caption{Thermal and Radiated Energy}
		\end{subfigure}
		\hspace{0.00cm}
		\begin{subfigure}[b]{0.50\textwidth}
			\includegraphics[width=\textwidth,trim={0.0cm 0.0cm 0.0cm 2.0cm},clip]{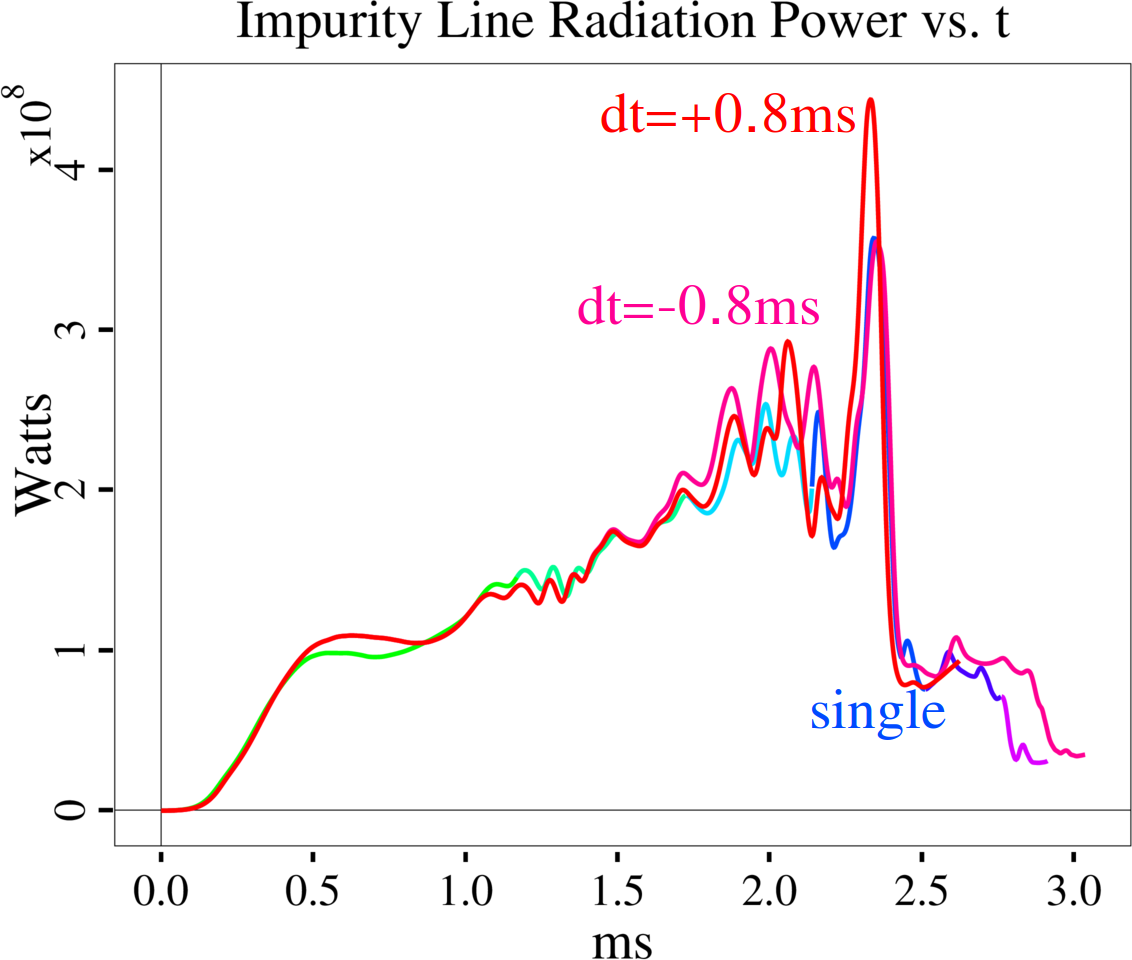}
			\vspace{-0.60cm}
			\caption{Total Radiated Power}
		\end{subfigure}
		\caption{Comparison of single SPI single load with dual SPI delays dt=$\pm$0.8ms shows a return to
		single SPI single load quench behavior.  The injector delay is too large for the second injector plume
		to reach the plasma and contribute to the thermal quench.}
		\label{fig:dt8}
	\end{figure}
Figure \ref{fig:dt8} shows a comparison of dt=$\pm$0.8ms to the single SPI single load case.  The two
dt=$\pm$0.8ms case are identical up until t$\simeq$1.6ms.  Examination of the single SPI single load trace shows
that t$\simeq$1.6ms is the transition from the initial to the late thermal quench phase.  Interaction with the second
injector fragments, inferred from the separation in traces, is very late.

Increasing the delay to 0.8ms between injectors returns the thermal quench behavior to the single SPI single load.  The 
delay is too large for the second injector fragments to reach the receding plasma in time and the thermal quench is
dominated by the fragments of the injector arriving first.  Later in the thermal quench (t>1.6ms), there is some 
minor variability suggesting late interaction with the second fragment plume.  In particular, the dt=+0.8ms case shows 
a larger late radiation spike at the end of the thermal quench.  But we see from the total radiated energy that 
these variations do not contribute much to the overall thermal quench efficiency.

These simulations suggest that the acceptable delay between injectors is relatively small.  For the fragment
parameters - 200 fragments of radius r$_f$=0.2mm, velocity v=120.0m/s($\frac{\Delta v}{v}$=0.5) and
poloidal spread with $\Delta\theta_{hw}=20^{\circ}$ - there is already a return to single SPI behavior at
dt=$\pm$0.8ms.  The acceptable time window decreases with increasing velocity.

Consider that for a nominal velocity of 200m/s, a delay of 0.5ms translates to a 10cm separation between the two
injector fragment plumes.  10cm is too large a gap for the second injector fragments to catch up over the thermal quench
time.  This requires 10cm of burn through of the first injector plume.  With a velocity spread of $\frac{\Delta
v}{v}$=0.5, the plume is 20cm long at t=1.0ms.  10cm burn through means half(50\%) of the first fragment plume
inventory has been
ablated.  These DIII-D SPI simulations, with reduced fragment inventory, result in a total assimilation of less than
20\%.  

10cm is a large radial excursion for the plasma distortion, described in Section \ref{maghel}, to intercept the second
fragment plume.  A combination of the two is possible, as demonstrated by the dt=$\pm$0.8ms simulations, but the
interaction is minimal.

\section{Comparison to Experiment}
As an initial value code, NIMROD is constrained by finite resolution in time and space; necessitating dissipation
parameters that are typically beyond experimental values.  As a numeric model, the simulations make a number of
assumptions and approximations; of most concern are the constant thermal conduction and lack of rotation.  Both of 
these will be improved upon in future simulations.  The constant parallel thermal conduction of 
1$\times$10$^9$m$^2$/s is too 
low for the core and too high for the edge.  The lack of rotation can concentrate the deposition to a single
toroidal location and obscure comparison with the experiment.  Even modest rotation frequencies of a few 100Hz
can rotate the plasma sufficiently to smear the impurity deposition across a significant toroidal fraction 
and mix the toroidal mode interaction.  Further a modest rotation can bring features into and out of the view
of toroidally localized diagnostics.  

The repeatability of the Shattered Pellet Injection experiments are difficult.   Formation and injection of the
cryopellet has it variability.  Often, the pellet will break in flight before the shatter.  The shattering of the pellet
produces variability in the distribution of fragment sizes and velocities\cite{Gebhart21}.  For these simulations 
we ignore the variation in fragment sizes.  The difficulty is compounded in the dual
SPI experiments, since necessary control of the timing for comparison suggested by the simulations is absent.

\begin{figure}[H]
	\hspace{-1.0cm}
	\includegraphics[width=1.1\textwidth]{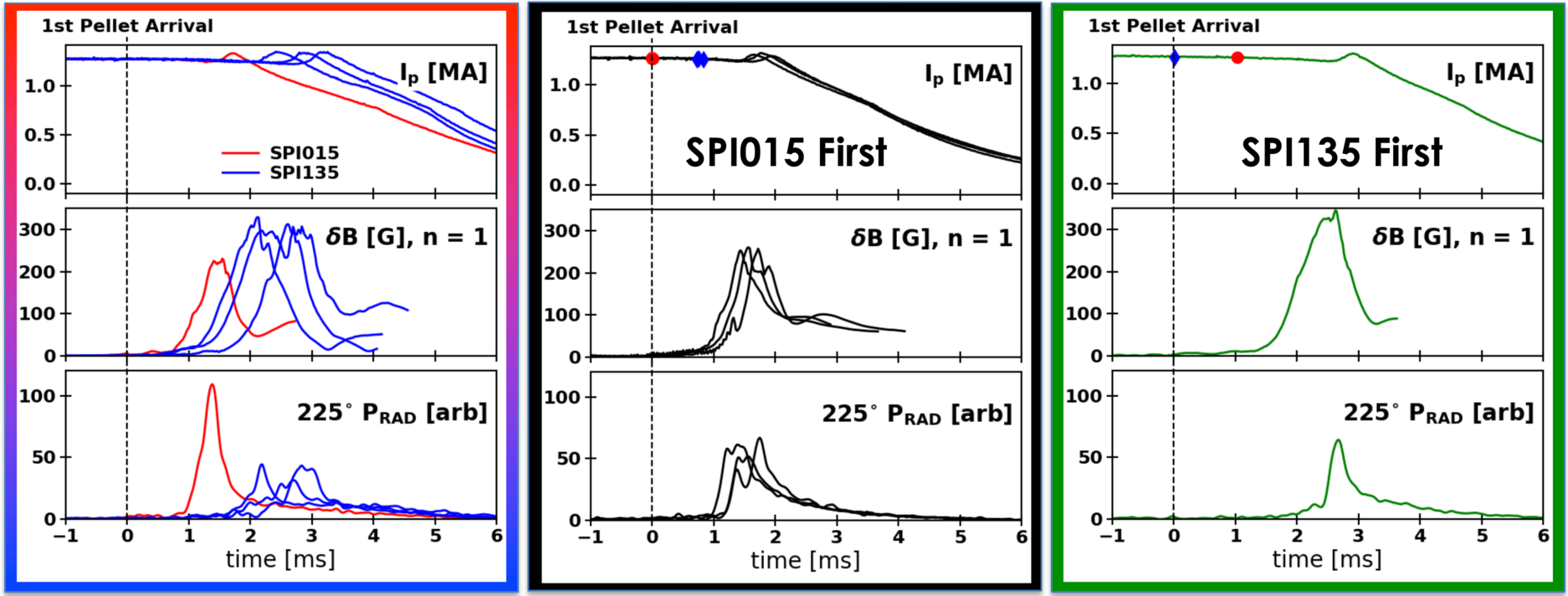}
	\caption{Four types of shots considered in DIII-D dual SPI experiments: single SPI at
	15$^\circ$ (1 case), single SPI at 135$^\circ$ (3 cases), dual SPI where 15$^\circ$ is first (3
	cases), and dual SPI where 135$^\circ$ is first (1 case).  The red circles (SPI015) and blue diamonds (SPI135)
	indicate the estimated arrival of the fragments. }
	\label{fig:d3dexsmallsum}
\end{figure}

Figure \ref{fig:d3dexsmallsum} shows four types of shots from the most recent DIII-D pure neon dual SPI 
experiment\cite{herfindal2022}: single
SPI at 15$^\circ$ (1 case [184410]), single SPI at 135$^\circ$ (3 cases [184411, 184412, 184413]), dual SPI where 
15$^\circ$ is first (3 cases [184414, 184416, 184417]), and dual SPI where 135$^\circ$ is first (1 case [184421]).
Plotted are the plasma current showing the
characteristic current spike (that denotes the transition from the end of the thermal quench to the beginning of the
current quench), the n=1 $\delta$B, and integrated radiation from the soft x-ray array at 225$^\circ$.  All cases 
show similar levels of $\delta$B$\sim$200-300G and similar levels of radiation at 225$^\circ$, except the single
SPI SPI015 case which shows almost twice the radiation of the other cases.  The same case also has one of the
smallest, if not the smallest, $\delta$B amplitude.

For comparison, the experiment uses a 150Torr-L pure neon pellet, with a diameter D=5.3mm and a L/D=1 (equivalent sphere
of r$_p$=3.03mm).  The experimental pellet inventory is about 20 times larger than NIMROD's single SPI plume of
200 r$_f$=0.2mm fragments.  Recall that this significantly reduced inventory is used in the simulation plume to 
allow enough of the leading SPI fragments to ablate such that the second injector plume can close the gap.  Past single SPI
simulations have demonstrated that only a small fraction (<5\%) of the nominal full experimental inventory is typically
ablated; the majority of the inventory merely passing through the simulation domain.

The factor of 20 discrepancy suggests perhaps a too low ablation rate in the simulation and/or an overestimate of
expected/assumed assimilation of the SPI experiment.  Other differences may be details missing from the 
fragment plume model: size distribution of fragments, smaller fragments are faster, pulverized pellet
gas, non-uniform distribution of fragments in the plume.  Research is ongoing to rectify this difference.

\subsection{Cooling Duration and Thermal Quench Time}
\begin{figure}[H]
		\hspace{-0.5cm}
		\begin{subfigure}[c]{0.35\textwidth}
			\includegraphics[width=\textwidth]{scat_tauTQ}
			\vspace{-0.10cm}
			\caption{NIMROD}
		\end{subfigure}
		\hspace{-0.5cm}
		\begin{subfigure}[c]{0.35\textwidth}
			\includegraphics[width=\textwidth]{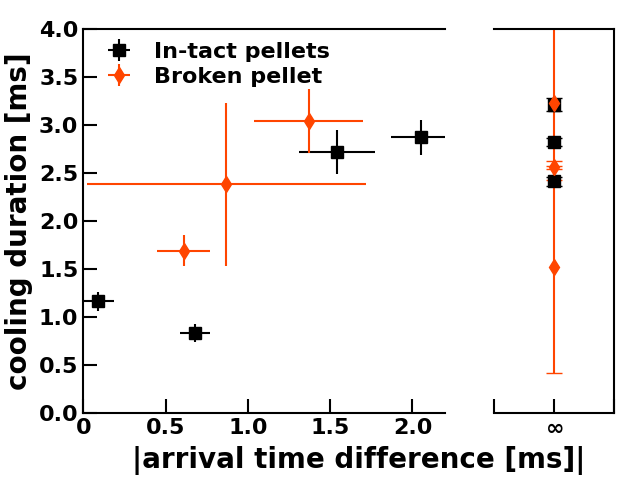}
			\vspace{-0.10cm}
			\caption{DIII-D old}
		\end{subfigure}
		\hspace{-0.5cm}
		\begin{subfigure}[c]{0.35\textwidth}
			\includegraphics[width=\textwidth]{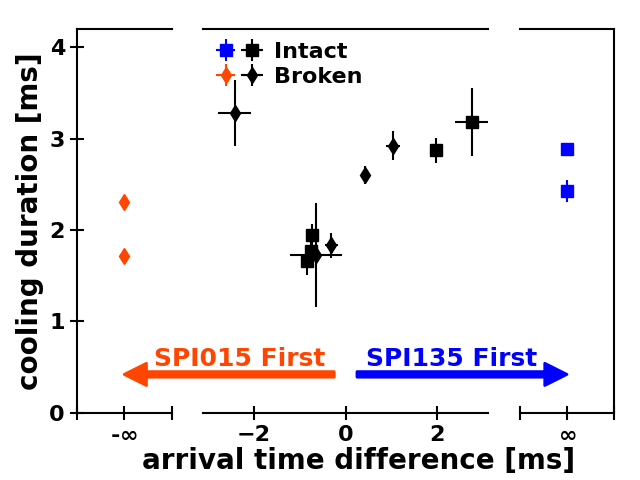}
			\vspace{-0.10cm}
			\caption{DIII-D new}
		\end{subfigure}
		\vspace{-0.20cm}
		\caption{Thermal quench time vs delay from NIMROD (from Figure \ref{fig:dsumtauscat}) and DIII-D cooling
		duration versus injector delay from two sets of dual SPI run days.  The older experiment shows a `V'
		like structure similar to the NIMROD results.  The newer experimental data looks to divide into two sets
		: SPI015 FIRST shows cooling duration t$\simeq$1.9ms and SPI135 FIRST t$\simeq$3.0ms.
		}
		\label{fig:cooldur}
	\end{figure}

Figure \ref{fig:cooldur} show plots of the thermal quench times from NIMROD and the cooling duration from two DIII-D
dual SPI run day\cite{herfindal2021}\cite{herfindal2022}.  The cooling duration in DIII-D is measured as time 
between initial arrival of fragments until
the current spike.   The total plasma current diagnostic is temporally well resolved and the current spike is 
a clear sign post of the transition from the end of the thermal quench to the beginning of the current quench. 
In NIMROD, the thermal quench time is measured as the time between initial arrival of the fragments
until the maximum in volume integrated electron number (i.e.  when recombination exceeds ionization).  This is a 
convenient and objective definition for the end of the thermal quench.  NIMROD does not use the cooling duration
since we typically do not run out to the current spike due to the high computational cost.  NIMROD typically observes
$\Delta t\simeq0.3-0.5$ms between the end of the thermal quench and the current spike.

Although not an ``apples-to-apples'' comparison, these easily identifiable features from well resolved time traces
give an objective relative comparison in thermal quench time trends, even if absolute comparisons are a bit obscured.

The dual SPI delay scan from NIMROD shows a `V' like structure, with the positive and negative delays forming an
approximately symmetric linear trend in thermal quench times, with the minimum time set by the single SPI double 
load and the maximum time by the single SPI single load.  Note that the simultaneous dual SPI is a little slower 
than the single SPI double load.   As stated earlier(see sec.\ref{sec:dt08}), at injector delays dt=$\pm$0.8ms, 
the thermal quench behavior
returns to the single SPI single load.   From this single scan, we estimate that acceptable delays must be 
at most within dt$\le$0.5$\tau_{TQ}$ to see any affect from the delayed injector.  

The symmetry in thermal quench time is a little surprising in light of the asymmetry observed in the thermal quench
efficiency and maximum radiation and significant role played by the MHD dynamics.  However, it must be remembered that
dynamics and time scale is driven by the SPI and it is the fragments that are the primary driver of the quench even 
if the resulting 
efficiency (how much radiation is generated and heat is lost) is determined by the ensuing MHD.

Cooling duration versus injector delay from the older experiment shows a `V' like structure similar to the NIMROD results.

The newer experimental data looks to divide into two sets : SPI015 FIRST around t=1.9ms and SPI135 FIRST around t=3.0ms.  
The experimental delays for the dual SPI cases lie at the edge of the acceptable delay window.  

It is the interpretation of the first author (not the experimentalist) that in all dual SPI cases from the new run
day, the second injector has minimal impact on the thermal quench, similar to what is observed in NIMROD for the
injector delay of dt=-0.8ms.  The two sets of times in the newer data are attributed to a faster fragment velocities
for SPI015
compared to SPI135.  More evidence of the difficulty in the precise control of SPI experiments.  At the time of
this writing, experimental values for the velocities are not available.

It is somewhat reassuring that range of thermal quench times are similar between the NIMROD simulations and DIII-D
experiments.  This is an indication that the simulation fragment velocities are close to the experimental values.

\subsection{Line Integrated Density}
\begin{figure}[H]
		\hspace{-0.5cm}
		\begin{subfigure}[b]{0.5\textwidth}
			\includegraphics[width=\textwidth]{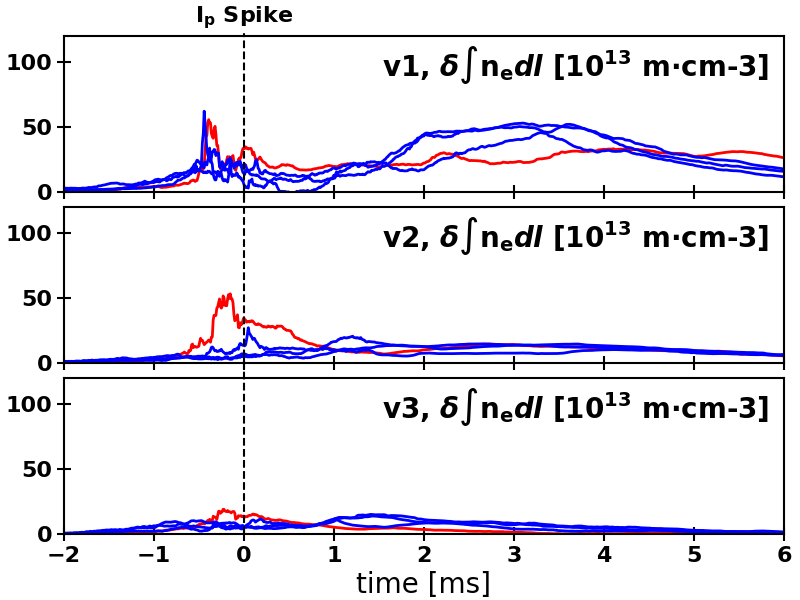}
			\vspace{-0.60cm}
			\caption{single SPI}
			\label{fig:d3dsingleinj}
		\end{subfigure}
		\hspace{0.0cm}
		\begin{subfigure}[b]{0.5\textwidth}
			\includegraphics[width=\textwidth]{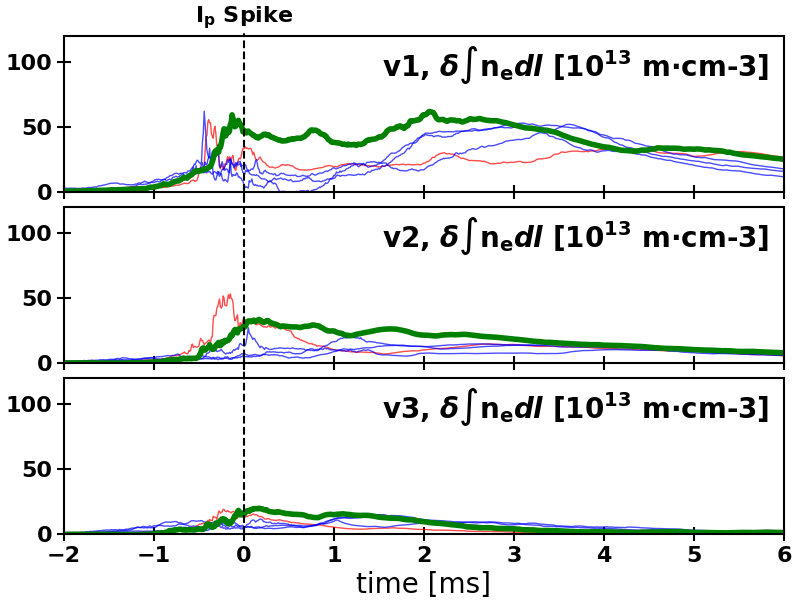}
			\vspace{-0.60cm}
			\caption{\textbf{SPI135 First}}
			\label{fig:spi135first}
		\end{subfigure}
		\caption{Line integrated density from 3 vertical chords in DIII-D SPI experiments.  
		Figure \ref{fig:d3dsingleinj} is from the \textbf{single SPI} shots where the red trace corresponds to
		SPI015 and blue corresponds to SPI135.  Figure \ref{fig:spi135first}
		shows the dual SPI \textbf{SPI135 First}(in green) overlaid on top of Figure \ref{fig:d3dsingleinj}.
	    Single SPI (\ref{fig:d3dsingleinj}) show roughly similar values.  SPI015 (red trace) shows more activity
		in \textbf{v2}.  The \textbf{SPI135 First} (\ref{fig:spi135first}) also shows similar values early 
		on in the thermal quench and some excess density late in the thermal quench, particularly in 
		\textbf{v1}, that persists into the current quench.
		}
		\label{fig:v1exp}
	\end{figure}

Figure \ref{fig:v1exp} shows the CO2 interferometer line integrated densities for the three vertical chords
(see Section \ref{sec:explayo}) located at $\phi$=230$^\circ$.  Figure \ref{fig:d3dsingleinj} is from the 
\textbf{single SPI} shots.  
Figure \ref{fig:spi135first} shows \textbf{SPI135 First}(in green) overlaid on top of Figure \ref{fig:d3dsingleinj}.

The interferometer line integrated electron densities show a steady rise over the initial phase of the thermal quench
(t=[-2.0,-0.5]ms), then a rapid burstie rise in the late phase of the thermal quench then settles during the current
quench.  The density is sustained well after the end of the thermal quench and persists during the current quench,
particularly in the inner vertical chord \textbf{v1}.  This is probably an indication that the plasma has collapsed
(after the thermal quench) and is leaning on the inner wall during the current quench.  As mentioned, this work focuses
on the thermal quench and synthetic integrated density do not include the current quench.  However, we will make note of
this persisting density as we transition our focus to simulating the current quench.

The experimental layout (Section \ref{sec:explayo}) shows that the interferometer is located on the opposite side of
the tokamak compared to the two SPI.  As mentioned, due to the magnetic helicity of the plasma, the path from
each injector to the interferometer plane differs and so it is not surprising to see differing signals for each of the
individual injectors.  

	\begin{figure}[H]
	    \centerline{
		\includegraphics[width=0.75\textwidth]{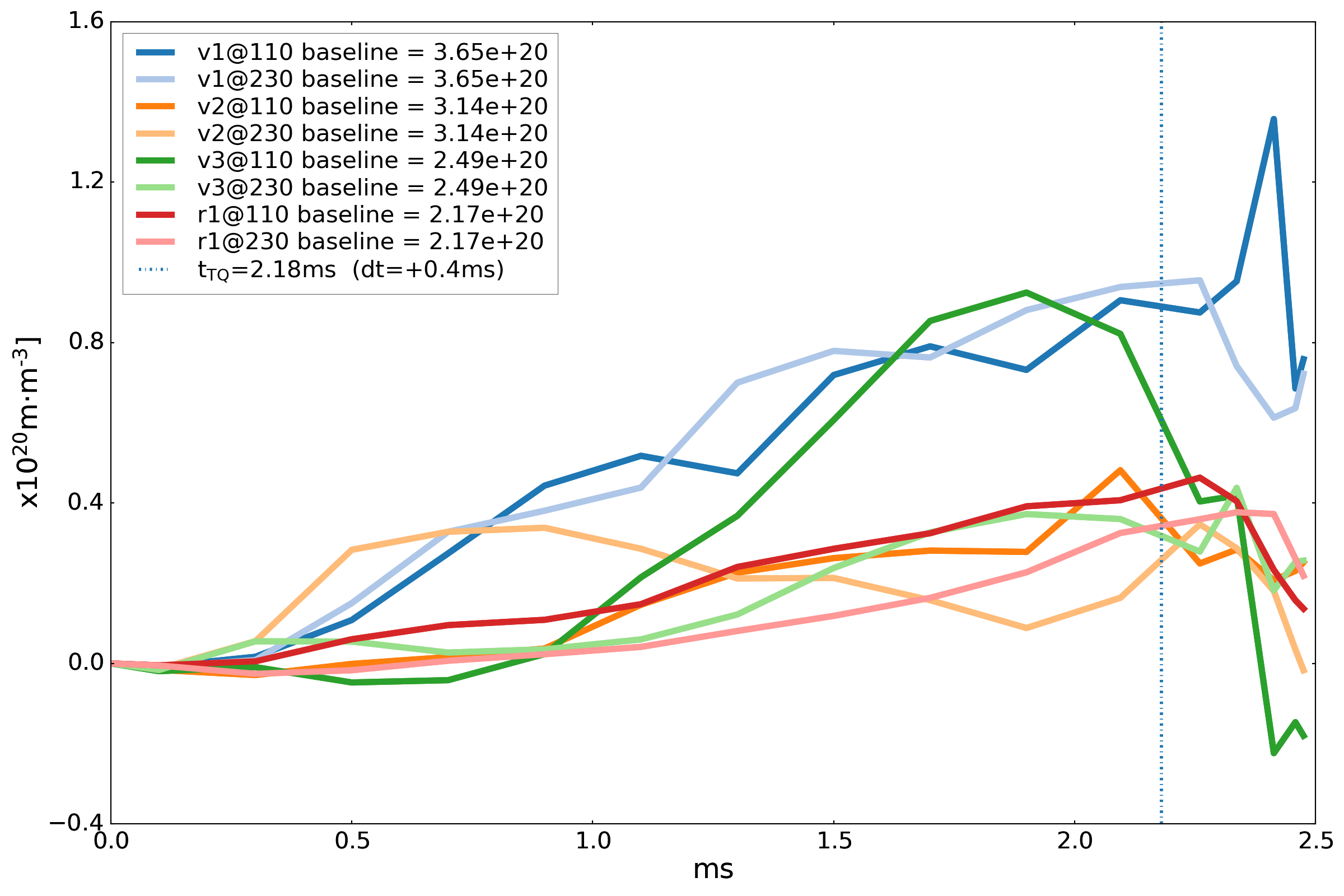}
		}
		\vspace{-0.40cm}
		\caption{Synthetic interferometer for dual SPI with delay of dt=+0.4ms.
		During the early phase of the thermal quench, the v2@230 chord shows increasing signal compared to
		v2@110.  During the late phase of the thermal quench, the outboard chord v3@110 shows a larger signal
		than v3@230.  Both inboard v1 chords show a steady increase throughout the thermal quench.
		The inboard v1@110 chord continues to increase a bit after the end of the thermal quench whereas 
		the other chords begin to trend downwards.
	    }
	    \label{fig:sintdtp04}
	\end{figure}
	\begin{figure}[H]
	    \centerline{
		\includegraphics[width=0.75\textwidth]{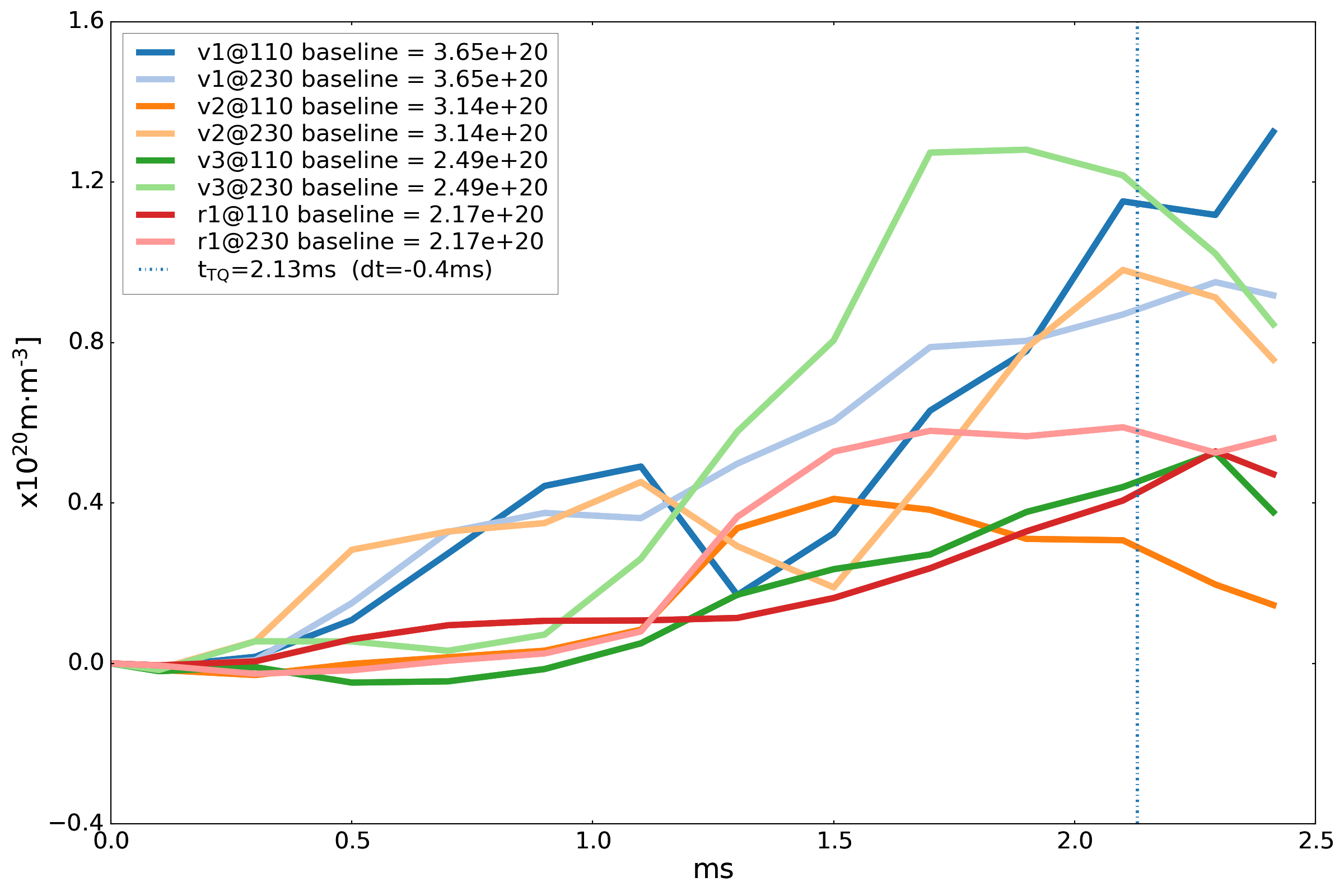}
		}
		\vspace{-0.40cm}
		\caption{Synthetic interferometer for delay of dt=-0.4ms.  Early in the thermal quench, the v2@230 chord shows a larger signal than v2@110, similar in behavior to the dt=+0.4ms case.  During the later phase of the
		thermal quench, the v3 chords show some difference, similar to the dt=+0.4ms delay, but here v3@230 is larger than v3@110.  v2@230 also increases compared to v2@110 in the later phase unlike the dt=+0.4ms case.}
		\label{fig:sintdtn04}
	\end{figure}
	\begin{figure}[H]
	    \centerline{
		\includegraphics[width=0.75\textwidth]{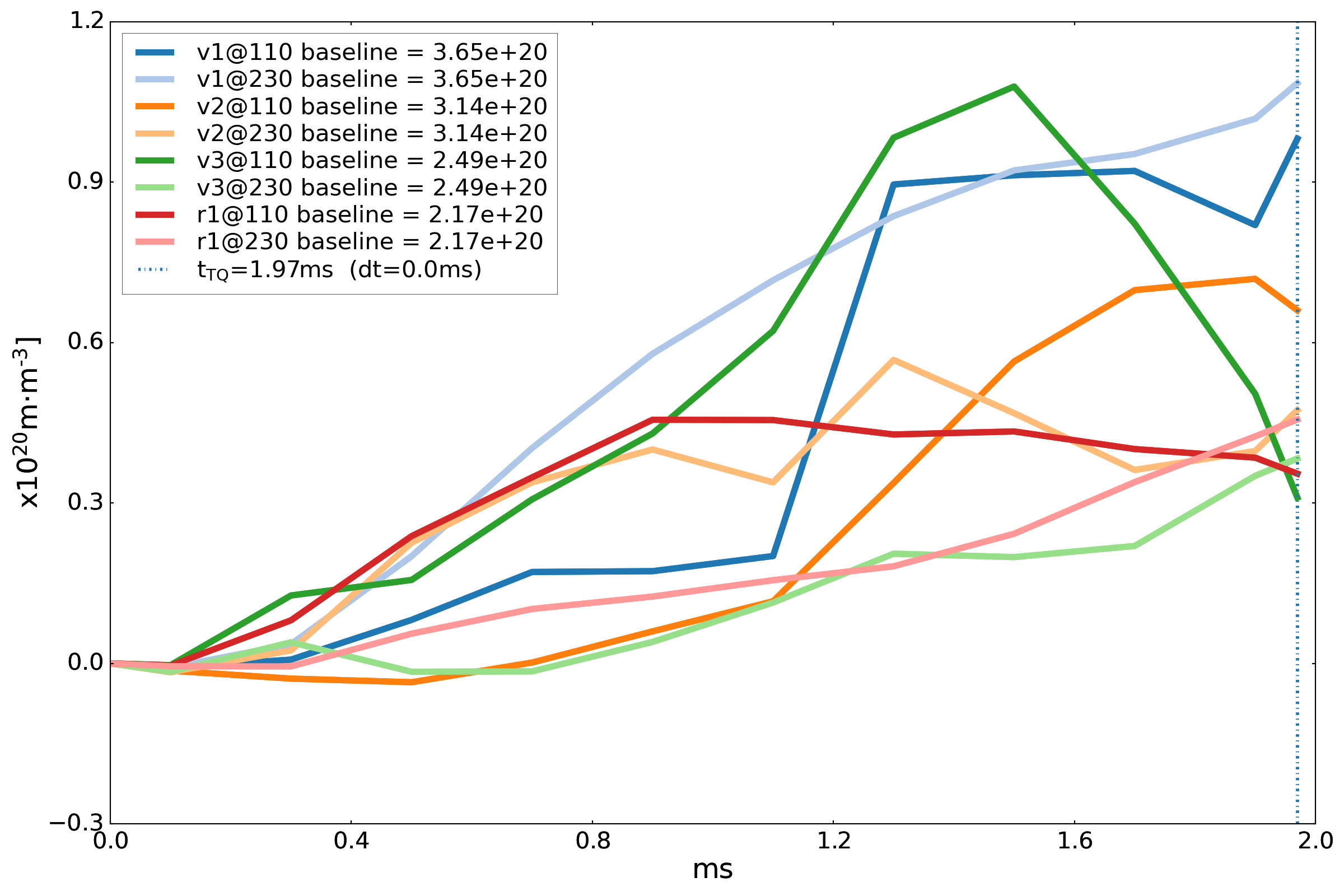}
		}
		\vspace{-0.60cm}
		\caption{Synthetic interferometer for simultaneous dual SPI dt=0.  Early in the thermal quench, we see
		all chords show a separation in the toroidal angels.  Later in the thermal quench, there is an increase in 
		all vertical chords @230.}
		\label{fig:sintdt0}
	\end{figure}
	\begin{figure}[H]
	    \centerline{
		\includegraphics[width=0.75\textwidth]{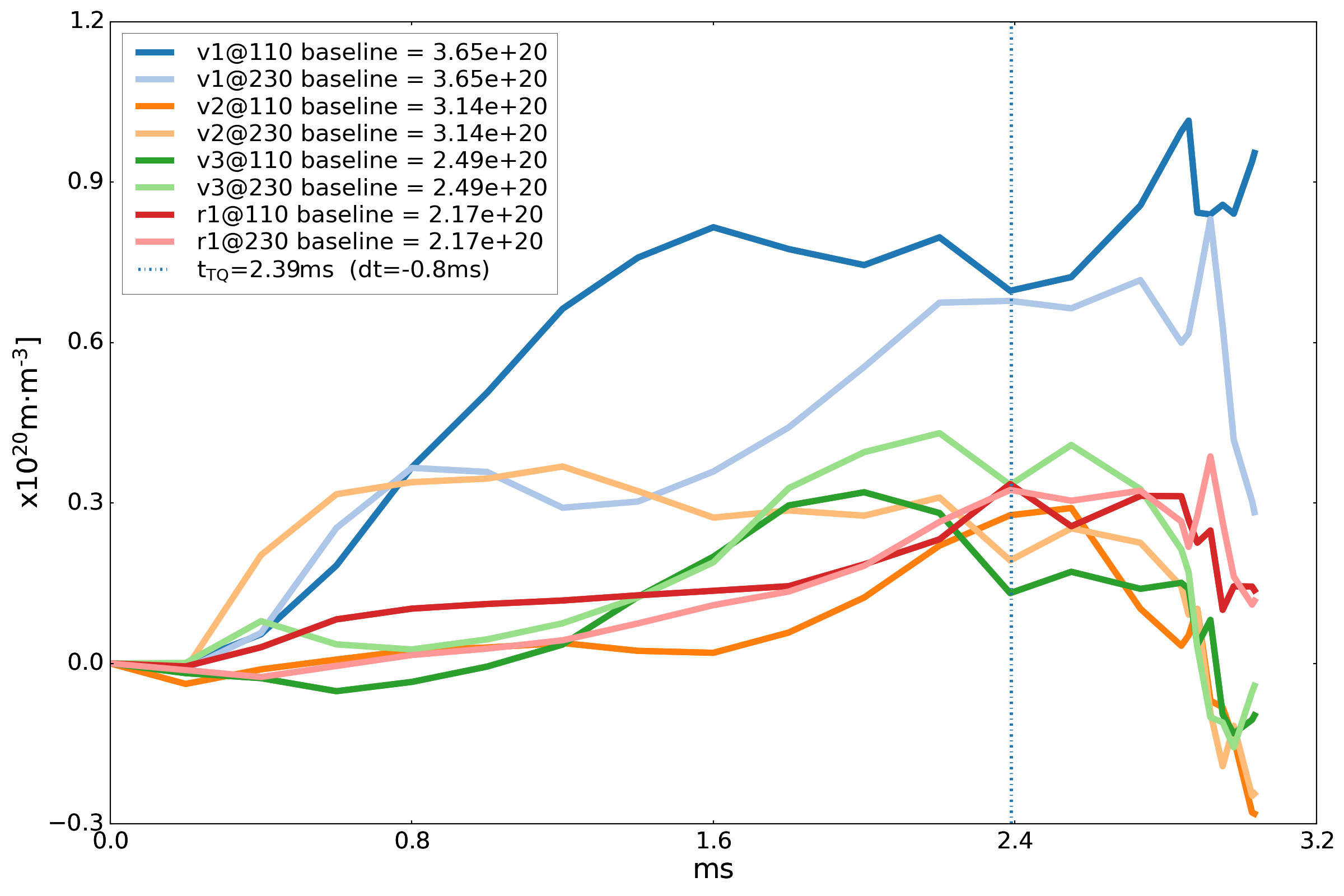}
		}
		\vspace{-0.60cm}
		\caption{Synthetic interferometer for dual SPI with delay dt=-0.8ms. Early in the thermal quench,
		the v2@230 signal increases more than the v2@110, similar to the other cases.  The late thermal quench phase
		differs from the other 3 cases shown, no significant increase in either v3 outboard chords.  Interferometer
		signals persist after the thermal quench, then begin to trend downward after the current spike.
		}
		\label{fig:sintdtn08}
	\end{figure}

Overall, these interferometer traces, both single and dual, are of comparable respective values.  The dual SPI case
shows some late increase in density, probably due to the late interaction of the second injector fragments.  However,
later in the current quench, as the plasma homogenizes, they settle to similar values.

Figures \ref{fig:sintdtp04}-\ref{fig:sintdtn08} show the synthetic interferometer measurements for the 3 vertical chords
and 1 radial midplane chord that approximate the experimental layout shown in Section \ref{sec:explayo} at toroidal angles
110$^\circ$ and 230$^\circ$ for time delay dt=[0.0,$\pm$0.4,-0.8]ms.  The vertical dotted line indicates the end of the
thermal quench.  The baseline values of each chord are included in the legend.

The synthetic interferometers show a steady increase in the integrated line density for all traces similar
to the experiment during the initial phase of the thermal quench.  There is also indication of burstie behavior 
during the late phase of the thermal quench.  

Figure \ref{fig:sintdtp04} shows the dual SPI with delay of dt=+0.4ms.
The early phase of the thermal quench (t=[0,1.0]ms) shows an increasing v2@230 signal while the v2@110 remains near 
the baseline.  During the later phase of the thermal quench, the outboard chord v3@110 shows an increasing signal
due to the ablation of the delayed SPI plume (located at 120$^\circ$).  This increase in ablation corresponds to 
the growth of the n=1 mode in figure \ref{fig:allddtp04}.  Both inboard v1 chords show a steady increase
throughout the thermal quench.  The inboard v1@110 chord continues to increase a bit after the end of the thermal 
quench whereas the other chords begin to trend downwards.

Figure \ref{fig:sintdtn04} shows the dual SPI for delay of dt=-0.4ms.  Early in the thermal quench, the v2@230 chord 
shows some increase while v2@110 remains near the baseline, similar in behavior to the dt=+0.4ms case.  During the 
later phase of the thermal quench, the v3 chords show some separation.   Here it is 
the v3@230 that increases rather than v3@110.  Recall for negative time delays, the delayed injector is at 240$^\circ$.
v2@230 also increases compared to v2@110 in the later phase, which differs from the dt=+0.4ms case.		

Figure \ref{fig:sintdt0} shows the simultaneous dual SPI dt=0.  Early in the thermal quench, we see
all chords show a difference in the toroidal angels.  Later in the thermal quench, there is an increase in 
all vertical chords @230 suggesting enhanced ablation due to plasma motion.

Figure \ref{fig:sintdtn08} shows dual SPI with delay dt=-0.8ms. Early in the thermal quench,
we again see the v2@230 signal increases more than the v2@110, similar to the other cases.
The late thermal quench phase differs from the other 3 cases shown; no significant
increase in either outboard v3 chords; i.e. no second fragment plume ablation (recall that the delay is too large
and this case is a proxy for the single SPI).   The interferometer signals persist after the thermal quench, 
then begin to trend downward after the current spike.	

All three finite delay cases, dt=[$\pm$0.4,-0.8]ms, show similar traces for the initial part of the thermal quench;
an increase in both inboard v1 chords and an increase in v2@230.  We can conclude that these early interferometer signals
are due to the ablation from the first SPI fragments at $\phi=$0$^\circ$. 
Late in the thermal quench we see the symmetric response of the dt=$\pm$0.4ms case in the outboard v3 chords, 
with an increase in v3@110 for dt=+0.4ms and an increase in v3@230 for dt=-0.4ms; a response to the delayed plume
fragments ablating.  The timing corresponds to those reported in Section \ref{sec:dtn04}.  Note that the
dt=-0.4ms case has a higher amplitude in its v3 signal and also show increases in additional chords
not seen in the dt=+0.4ms case.

Comparing to the simultaneous case, we see a clear difference due to the presence of the second injector fragment
ablation.  The outboard v3@110 chord increases while the inboard v1@110 is suppressed.

In general, the inner chord (\textbf{v1}) is the largest amplitude, similar to the experiment.  This is primarily due to 
the longer path length.  There is some indication that the inner chord signals persist longer that the outboard chords 
but the limited data beyond the thermal quench makes this conclusion a bit tenuous.  A persisting inner chord signal
while the outer chords drop could suggest that after the thermal quench, the plasma condenses a bit inboard. 
The total electron population does not increase after the thermal quench; for the integrated
density to increase, more electrons must be carried in from somewhere else.

	\begin{figure}[H]
		\centerline{
		\includegraphics[width=0.75\textwidth]{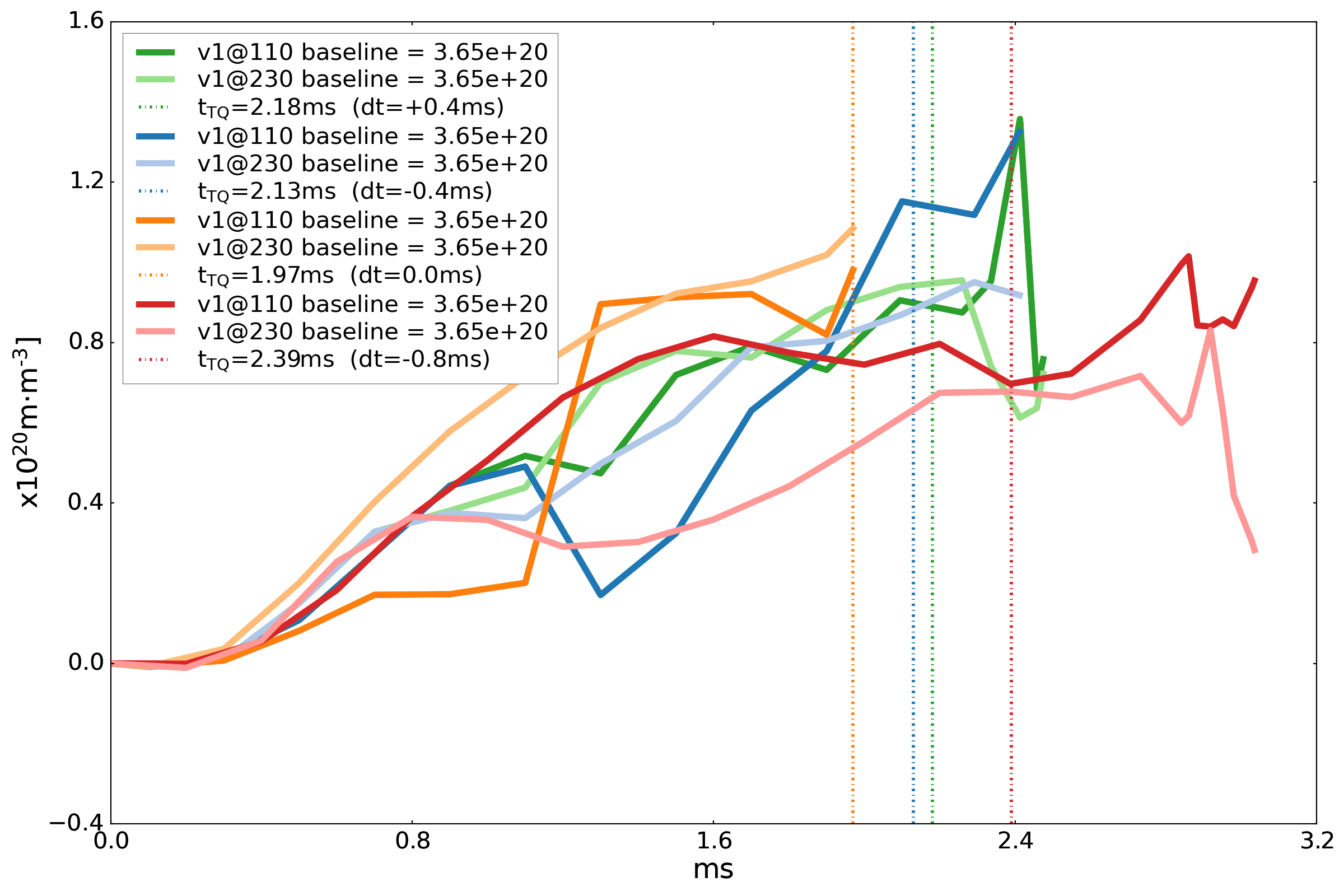}
		}
		\vspace{-0.20cm}
		\caption{Comparison of \textbf{v1} chords show similar behavior for dt=+0.4ms, dt=-0.4ms, and dt=-0.8ms
		up until t=0.9ms, since all three are dominated by the first injector early on.  At t=0.9ms the two sets
		of three traces separate as some of the second injector fragments begin to interact with the plasma.
		The \textbf{v1} chord increase even after the thermal quench, similar to the experiment.
		}
		\label{fig:v1comp}
	\end{figure}

Comparison of the 110$^\circ$ and 230$^\circ$ chords show toroidal variation of the impurity electron density, implying
that the impurities remain localized toroidally and poloidal as seen in Figure \ref{fig:3dviz}.  The largest
difference between the same chords at different toroidal angles appear in the outboard chord \textbf{v3}.  
The large values are likely due to the proximity to the injector and again suggest that the impurities are localized.

There is a general trend of greater increase in integrated density with increasing thermal quench efficiency.  The 
baseline integrated density at t=0 is denoted in the legend and ranges from 2.2-3.7e20; the increase in density does 
not exceed 30\%.

Figure \ref{fig:v1comp} shows a comparison of \textbf{v1} chord for dual SPI delays dt=[0.0,$\pm$0.4,-0.8]ms.
We do not see any particular correlated trend in the integrated density wrt the thermal quench efficiency.  
There are even periods during the thermal quench when the integrated density is larger for the single injector
(dt=-0.8ms) case than the simultaneous dual injector (dt=0.0ms).  

We see a factor $\times$2-3 larger values in the experiment compared to the simulations.  We find this discrepancy
acceptable, particularly considering the factor of 20 smaller nominal inventor in the simulated fragments, the additional
simplifying assumptions made in the fragment plume (e.g. uniform size fragments of r$_f$=0.2mm), and the single
temperature resistive MHD model with constant thermal conduction.   The inner chord interferometer signal for the
experiment persists much longer than is observed in simulations suggesting missing physics in the simulations.
However, we defer this until we commence with current quench simulations.

\begin{figure}[H]
		\centerline{
		\includegraphics[width=0.7\textwidth]{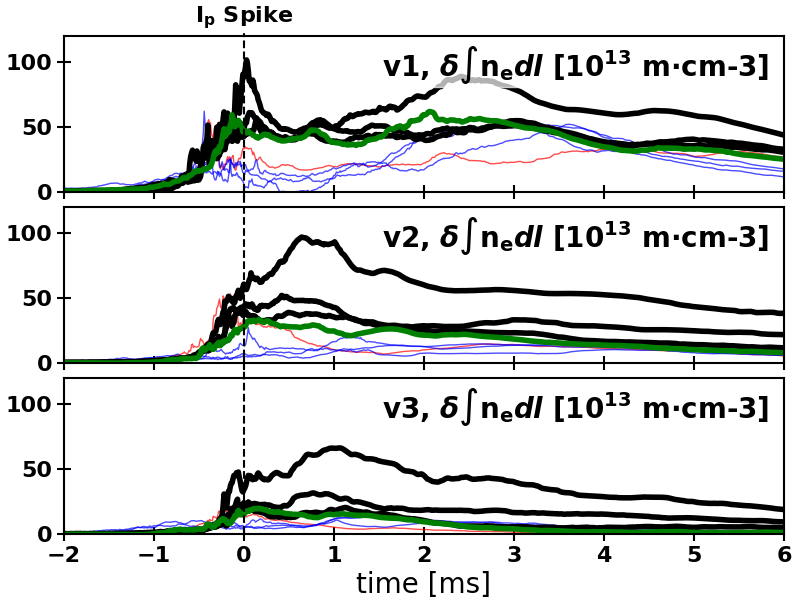}
		}
		\vspace{-0.20cm}
		\caption{SPI015 FIRST (from Figure \ref{fig:d3dexsmallsum}) interferometers (in black) overlaid on
		previous cases from Figure \ref{fig:v1exp}.  Two of the cases are comparable to SPI135 FIRST (green).  One case
		is an outlier producing almost twice as much density.
		}
		\label{fig:spi015first}
	\end{figure}

For completeness, we include a final interferometer plot from the experiment.  Figure \ref{fig:spi015first} shows the
interferometer measurements from SPI015 FIRST (from Figure \ref{fig:d3dexsmallsum}) (in black) overlaid
on previous cases from Figure \ref{fig:v1exp}.  Two of the cases are comparable to SPI135 FIRST (green).  One case is an
outlier producing $\times$2-3 as much density.  This outlier is difficult to reconcile with the simulations.

\subsection{Temperature and Density Profiles}
Temperature and density profiles are measured with Thompson Scattering System located at $\phi=$120$^\circ$ 
(see sec.\ref{sec:explayo}).  Timing and triggering complications for the SPI experiments result in limited
viable data sets for SPI run days.  

	\begin{figure}[H]
		\begin{subfigure}[c]{0.325\textwidth}
			\includegraphics[width=\textwidth]{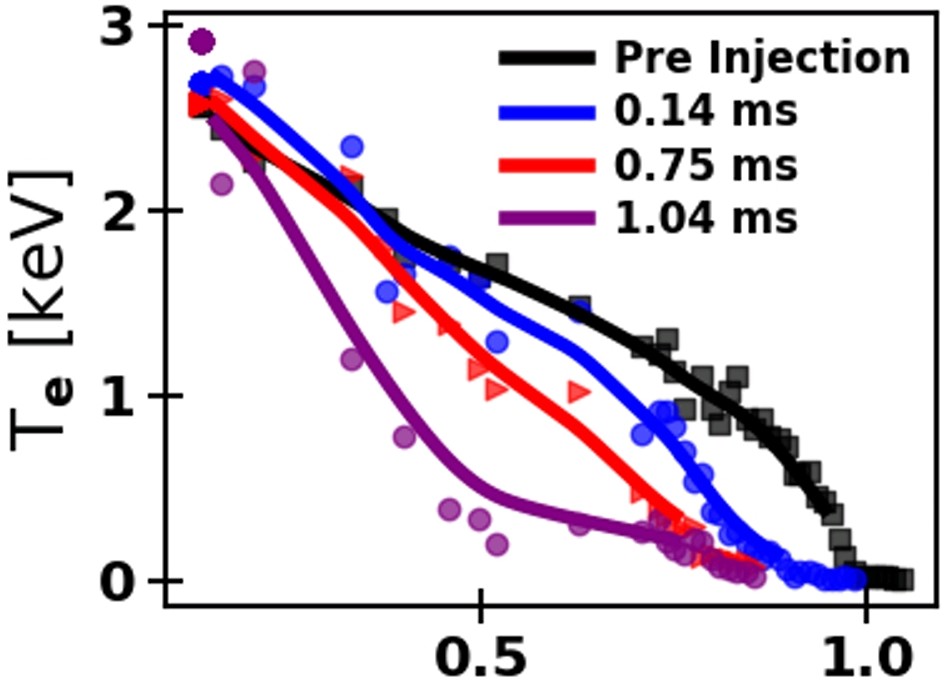}
			\vspace{-0.10cm}
			\caption{Te SPI015}
			\label{fig:te_p1}
		\end{subfigure}
		\hspace{-0.25cm}
		\begin{subfigure}[c]{0.325\textwidth}
			\includegraphics[width=\textwidth]{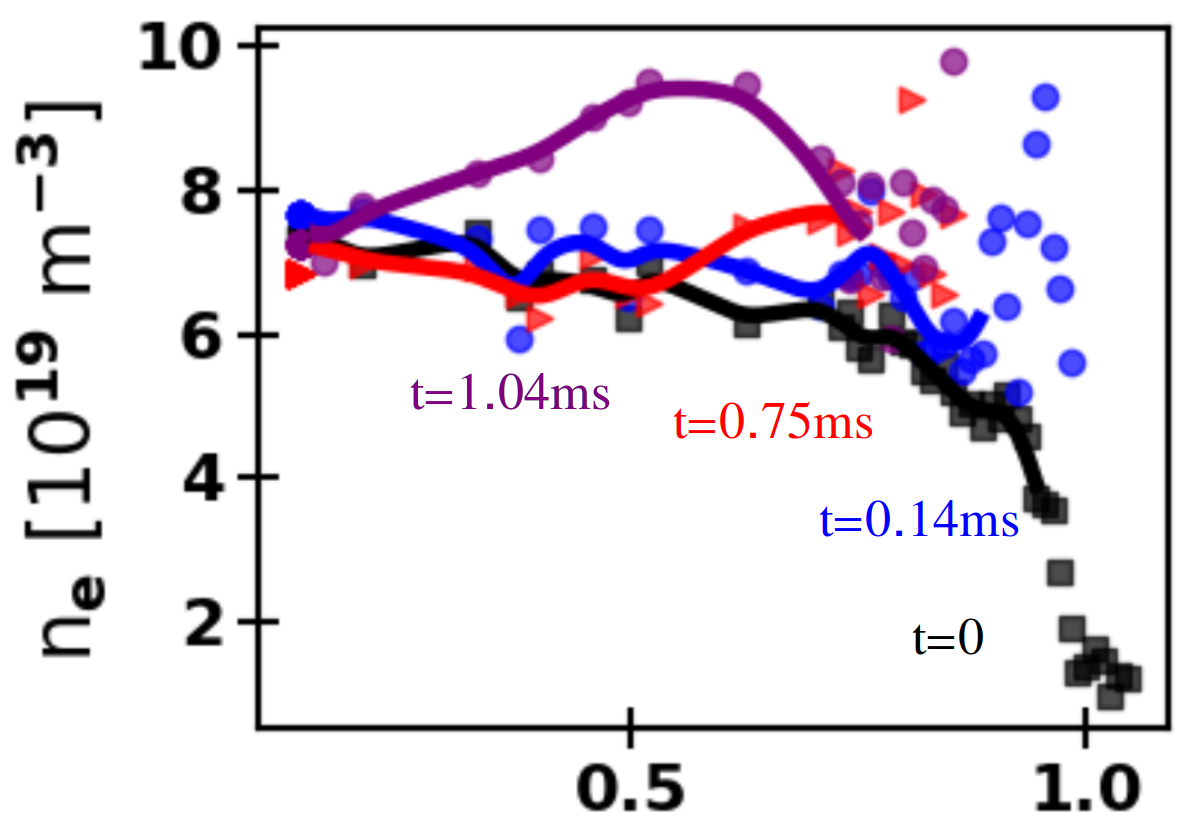}
			\vspace{-0.10cm}
			\caption{ne SPI015}
			\label{fig:ne_p1}
		\end{subfigure}
		\hspace{-0.25cm}
		\begin{subfigure}[c]{0.325\textwidth}
			\includegraphics[width=\textwidth]{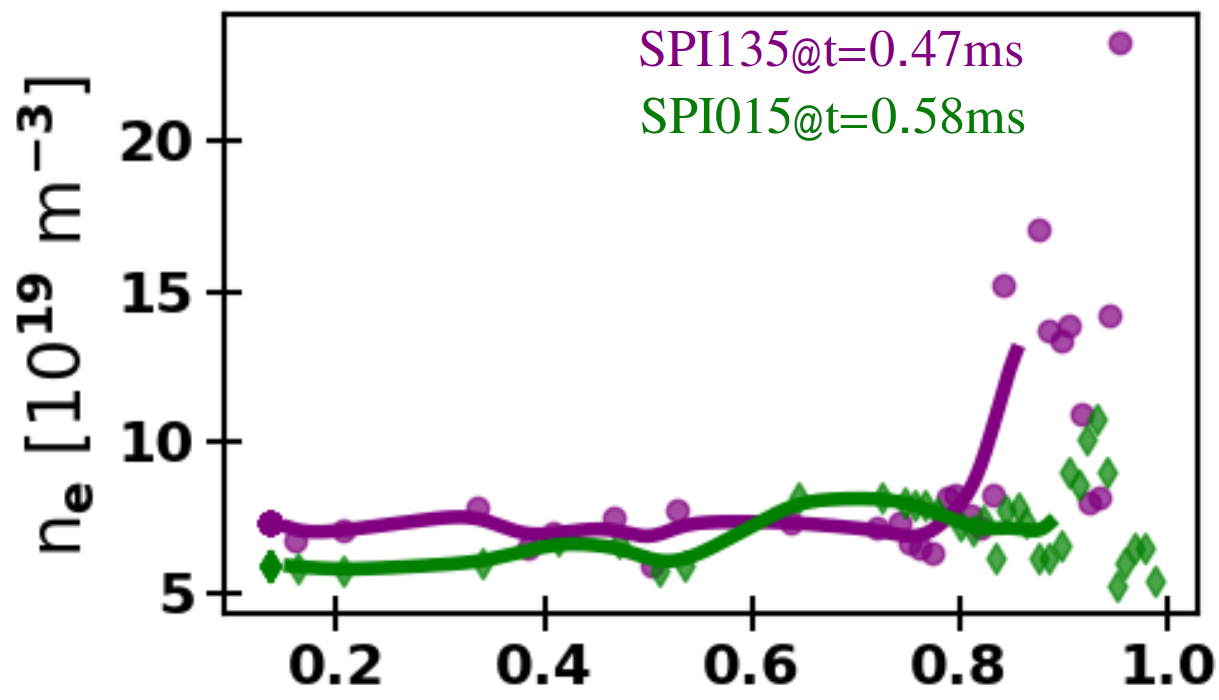}
			\vspace{0.50cm}
			\caption{SPI015 vs. SPI135}
			\label{fig:spinecomp}
		\end{subfigure}
		\caption{Electron temperature and density profiles vs normalized flux for several times during the
		thermal quench from the DIII-D dual SPI \textbf{SPI015 First} and a comparison of density profiles
		from SPI015 FIRST vs SPI135 FIRST (\ref{fig:spinecomp}) at comparable times, early in the thermal quench.
		The location of the diagnostic relative to the injectors account for the difference in figure \ref{fig:spinecomp} 
		and indicates localization and finite propagation time of the impurities.
		}
		\label{fig:neexprofs}
	\end{figure}

Figure \ref{fig:neexprofs} shows electron temperature and density profiles vs normalized flux for several times 
during the thermal quench
from the DIII-D shot 184416 (\ref{fig:te_p1}\&\ref{fig:ne_p1}), a dual SPI015 First case, and a comparison
of density profiles from SPI015 First (shot 184414)  vs  SPI135 First (shot 184421) (\ref{fig:spinecomp}) at comparable 
times.  The location of the diagnostic relative to the injector (see sec.\ref{sec:explayo}) accounts for the 
much of the  difference and demonstrates localization and finite propagation time of the impurities.

    \begin{figure}[H]
        \centering
        \begin{subfigure}[c]{0.325\textwidth}
            \includegraphics[width=\textwidth,trim={0.0cm 0.0cm 0.0cm 2.8cm},clip]{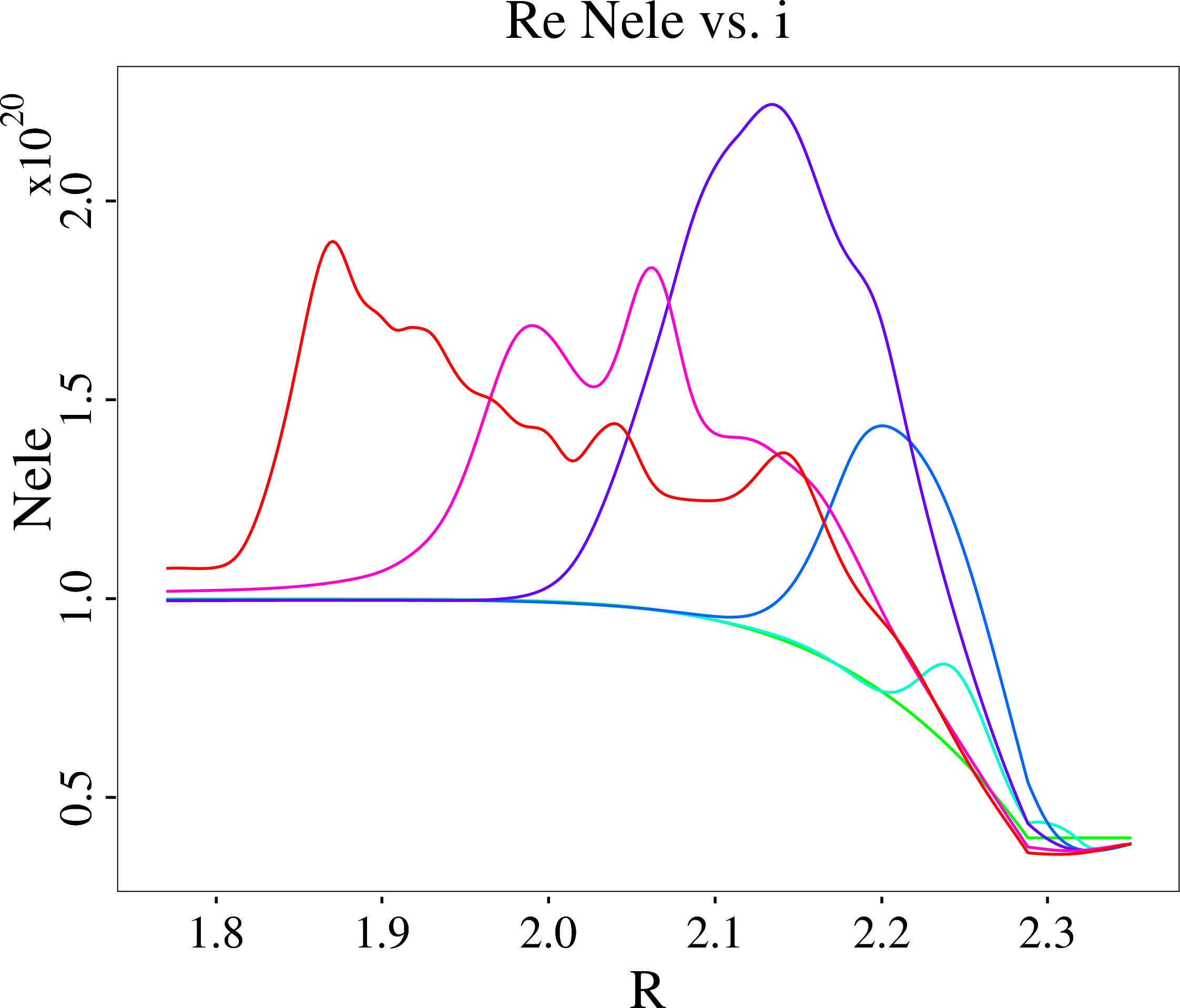}\\
            \includegraphics[width=\textwidth,trim={0.0cm 0.0cm 0.0cm 2.8cm},clip]{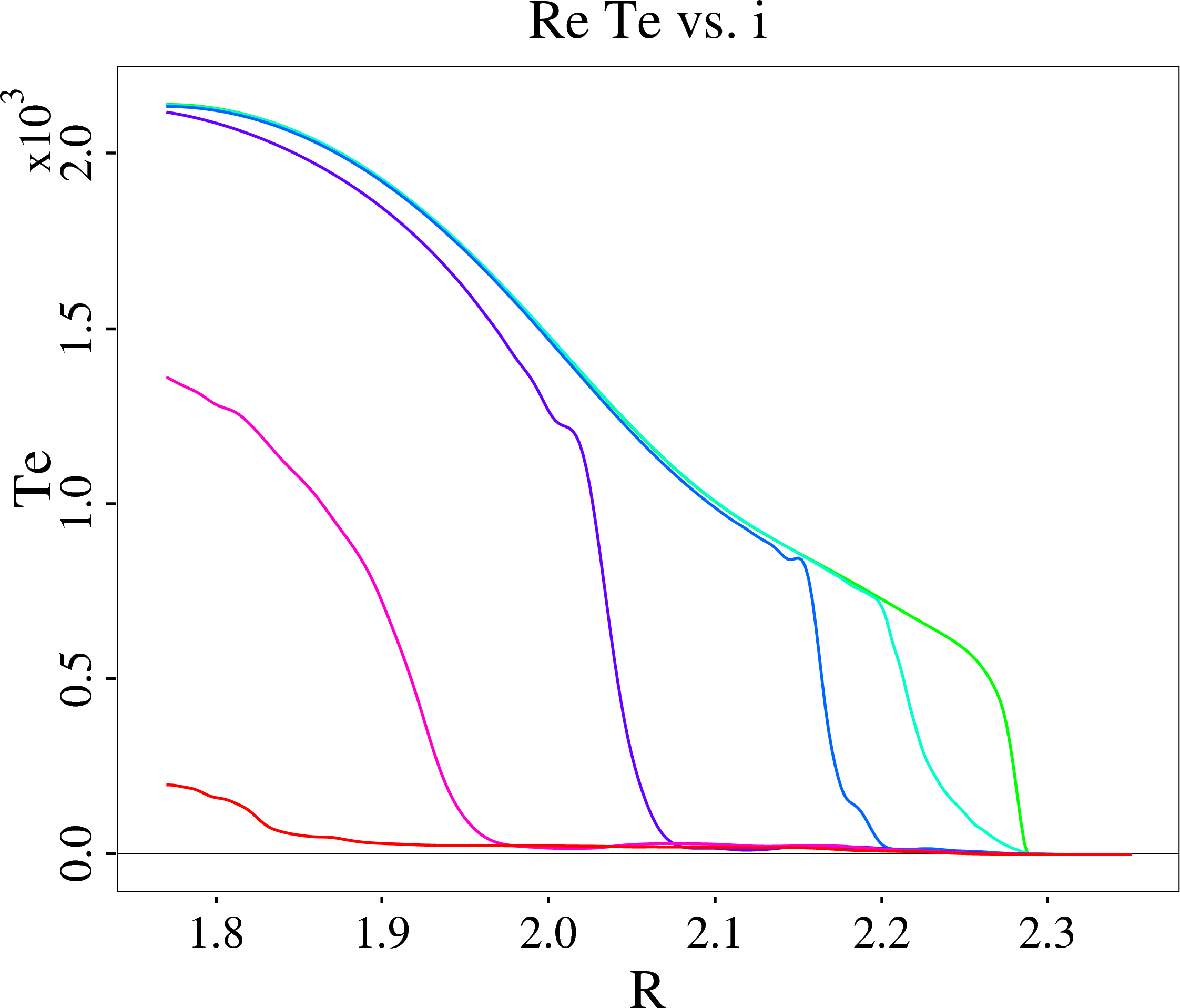}
            \caption*{$\phi=0^\circ$}
        \end{subfigure}
        \begin{subfigure}[c]{0.325\textwidth}
            \includegraphics[width=\textwidth,trim={0.0cm 0.0cm 0.0cm 2.8cm},clip]{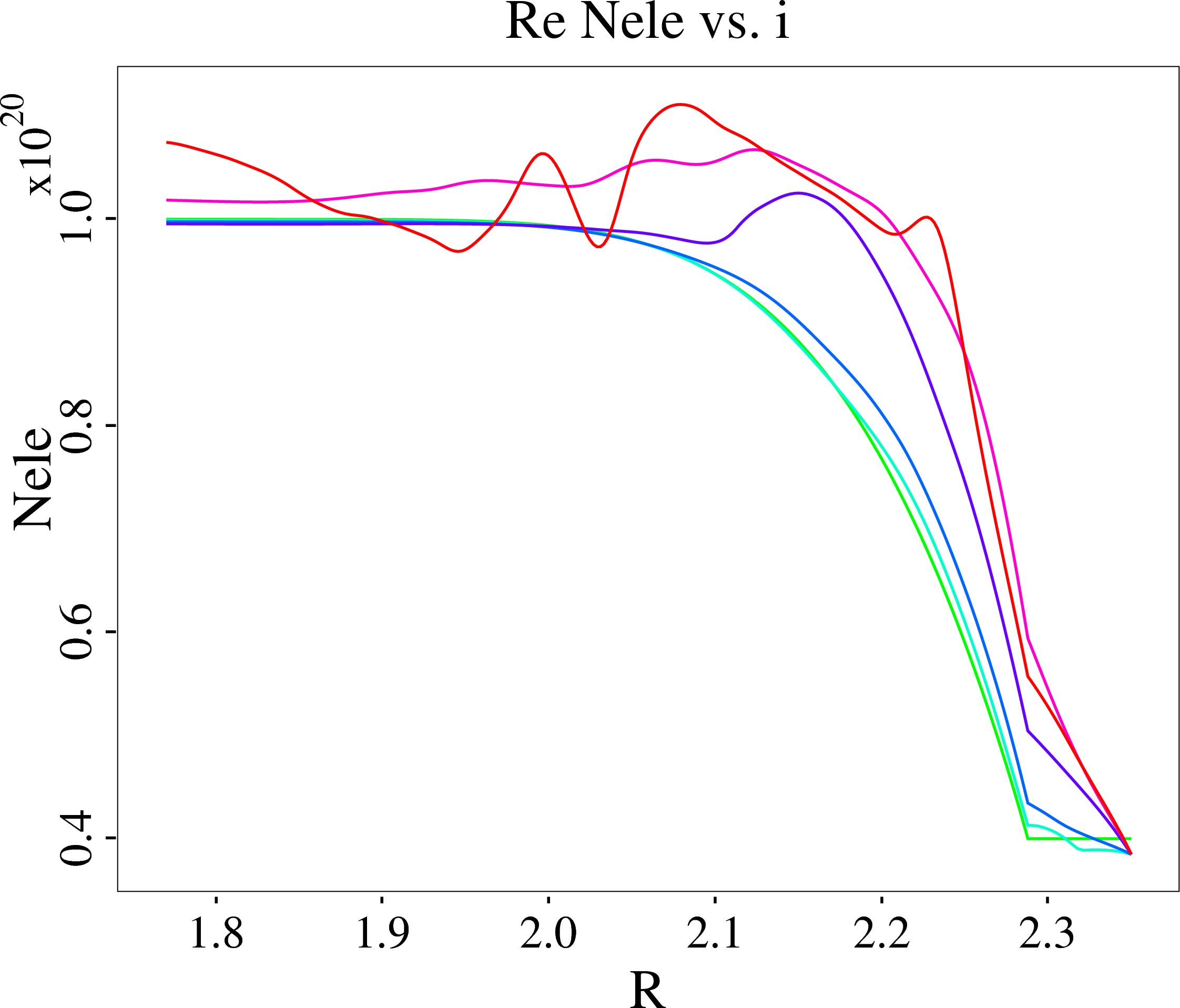}\\
            \includegraphics[width=\textwidth,trim={0.0cm 0.0cm 0.0cm 2.8cm},clip]{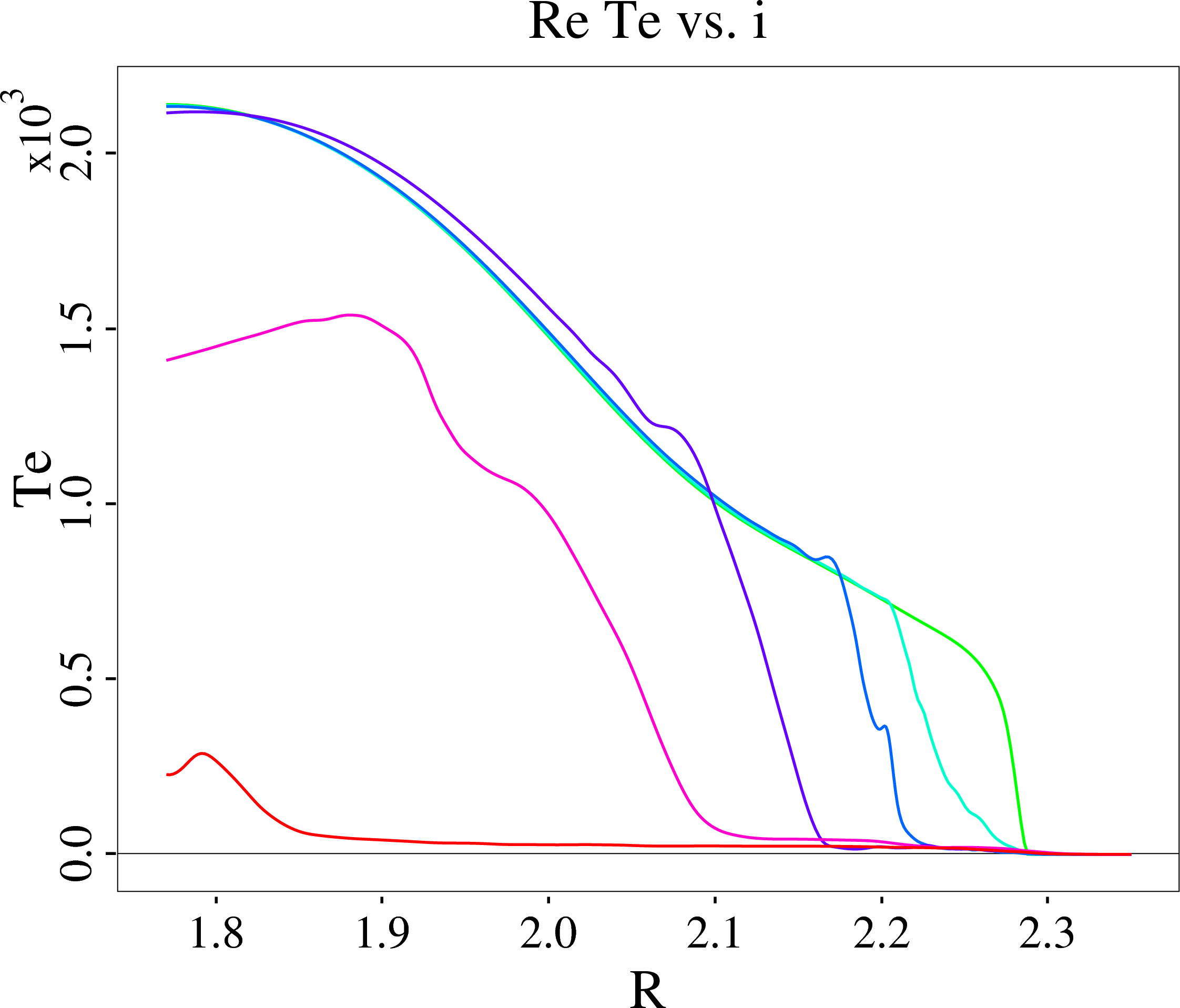}
            \caption*{$\phi=120^\circ$}
        \end{subfigure}
        \begin{subfigure}[c]{0.325\textwidth}
            \includegraphics[width=\textwidth,trim={0.0cm 0.0cm 0.0cm 2.8cm},clip]{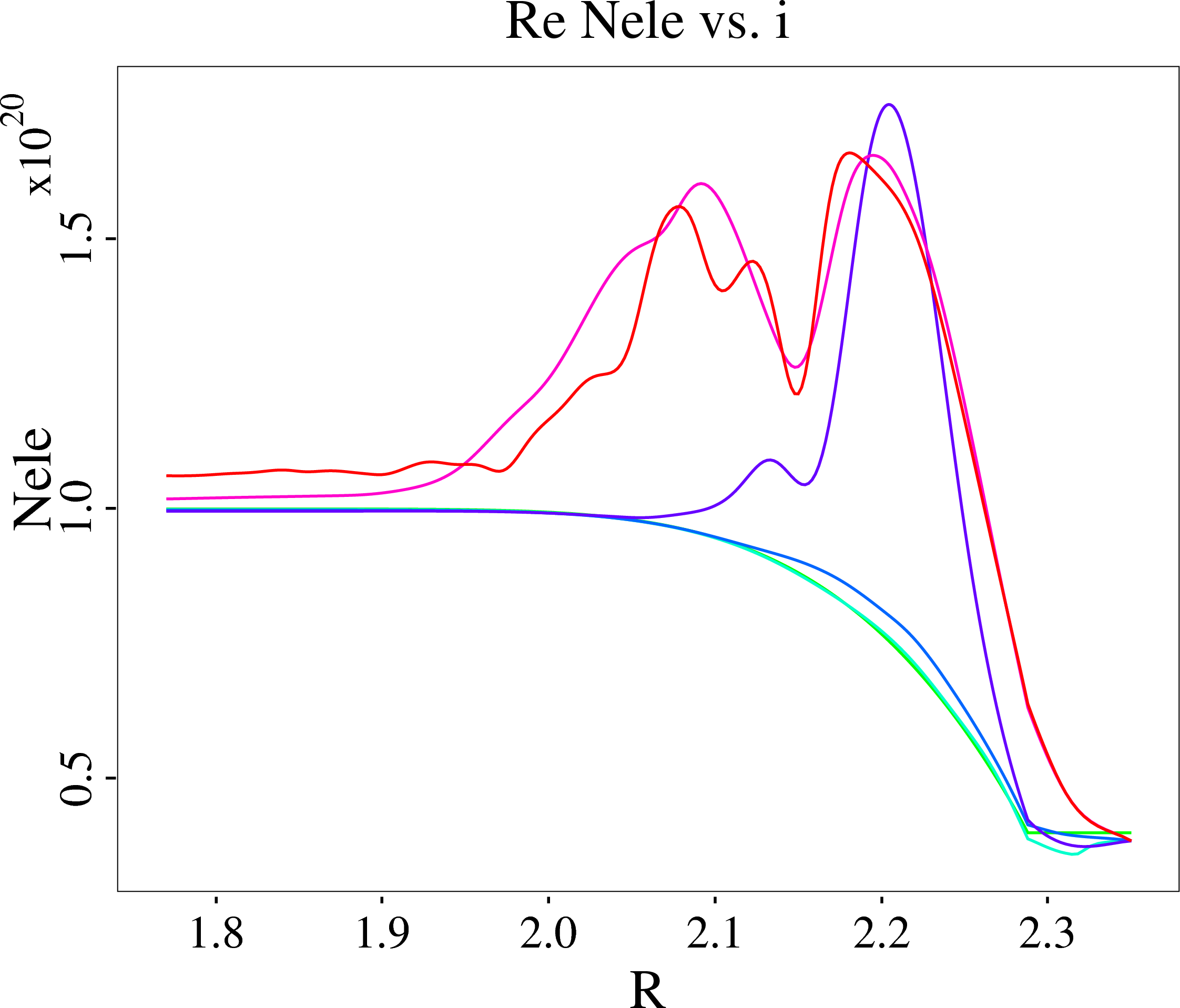}\\
            \includegraphics[width=\textwidth,trim={0.0cm 0.0cm 0.0cm 2.8cm},clip]{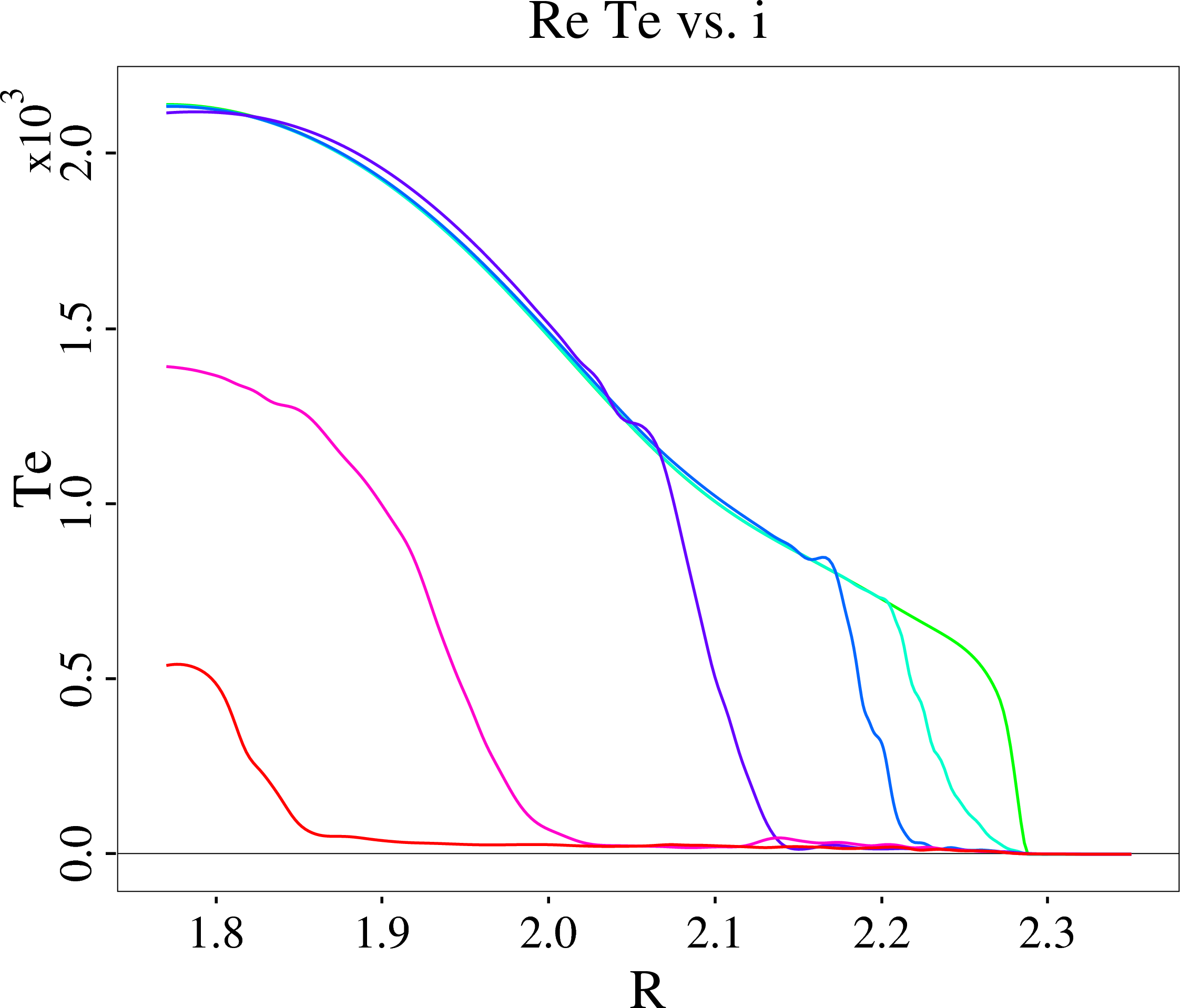}
            \caption*{$\phi=240^\circ$}
        \end{subfigure}
        \vspace{-0.2cm}
        \caption{NIMROD dual SPI dt=-0.4 midplane radial profiles of electron density(m$^{-3}$) and temperature(eV) at 
        t = [\clrlg{0.0}, \clrc{0.5}, \clrb{1.0}, \clrp{1.5}, \clrm{2.0}, \clrr{2.13}]ms.  Density profiles show variation
        with respect to the toroidal angle depending on their proximity to the injector.  Density profile at
        $\phi=240^\circ$ does not change much until \clrp{t=1.5ms}, after the delayed second injector
        fragments intercept the plasma.  The density profile at $\phi=120^\circ$ only changes marginally.
        Early temperature profiles look similar, a result of fast thermal conduction, but begin to show
        some differences beginning at \clrp{t=1.5ms} primarily reflecting the helical distortion of the plasma.
        }
        \label{fig:triprofs_dtn04}
    \end{figure}

Figure \ref{fig:triprofs_dtn04} shows NIMROD dual SPI dt=-0.4 midplane radial profiles of electron density and 
temperature at t = [\clrlg{0.0}, \clrc{0.5}, \clrb{1.0}, \clrp{1.5}, \clrm{2.0}, \clrr{2.13}]ms.  Density profiles 
show variation with respect to the toroidal angle depending on their proximity to the injector.  The density profile at 
$\phi=240^\circ$ does not change much until \clrp{t=1.5ms} (recall sec\ref{sec:dtn04}), after the delayed second injector
fragments intercept the plasma and begin ablating.  The density profile at $\phi=120^\circ$ change marginally, 
similar to the DIII-D profiles in figure \ref{fig:ne_p1}.  (note that the baseline density on axis is
1$\times$10$^{20}$m$^{-3}$ in the NIMROD simulations (inherited from 160606) while the DIII-D experiments are
0.7$\times$10$^{20}$m$^{-3}$.)

Early temperature profiles look similar at the three toroidal locations, a result of fast parallel thermal 
conduction.  They
begin to show some differences starting with \clrp{t=1.5ms} reflecting the helical distortion of the plasma resulting
from the dynamics induced by the thermal quench.  Comparison with the DIII-D temperatures profiles in figure
\ref{fig:te_p1} shows that the NIMROD simulations maintain more of a pedestal like structure through earlier
stages of the thermal quench than the DIII-D experiment.  This is likely due to the constant thermal conduction
used in the simulations.  Follow up simulations with temperature dependent thermal conduction will be checked.  
    
\subsection{Midplane Toroidal Array of Magnetic Probes}
\vspace{-0.5cm}
	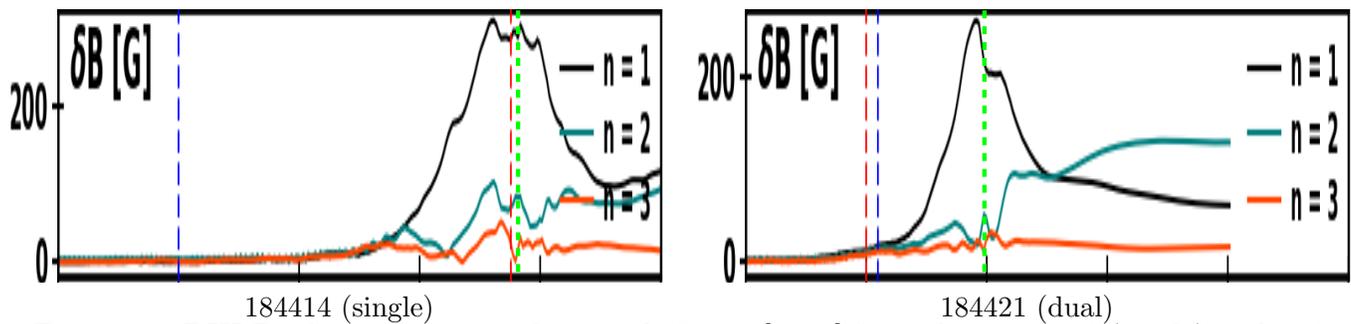
\begin{figure}[H]
		\hspace{-0.5cm}
		\begin{subfigure}[c]{0.500\textwidth}
			\includegraphics[width=\textwidth,height=0.15\textheight, trim={0.0cm 20.0cm 0.0cm 12.0cm},clip]{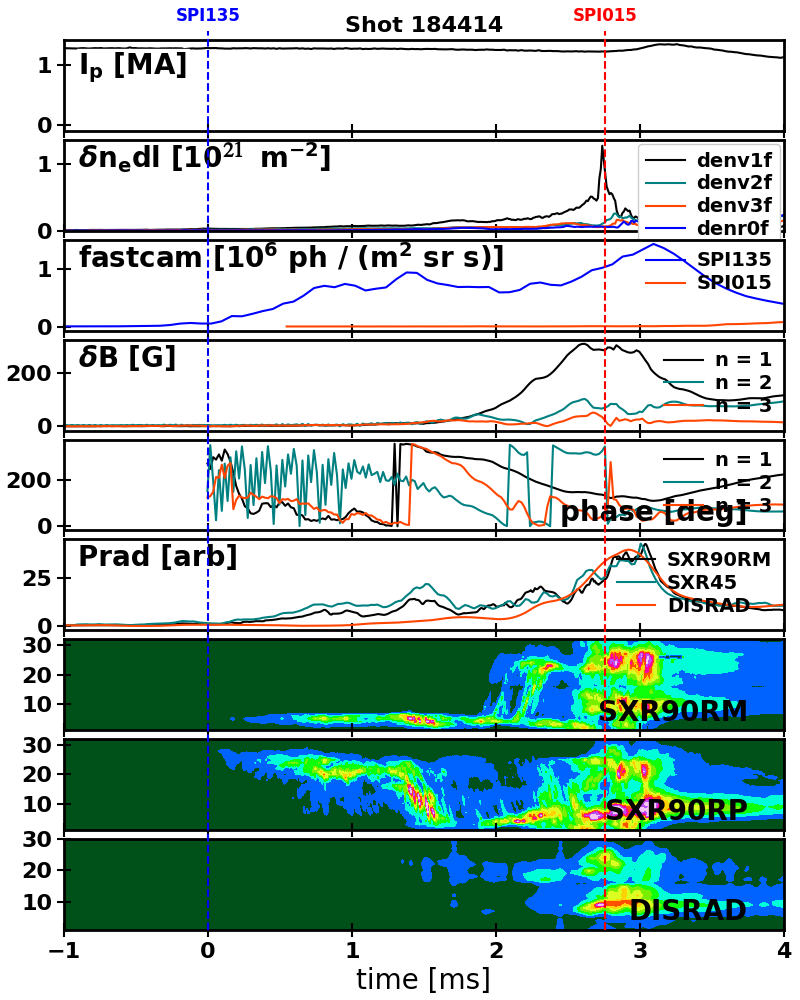}
			\vspace{-0.60cm}
			\caption*{184414 (single)}
			\begin{tikzpicture}[overlay]
				\draw[dashed, ultra thick, green] ( 6.82 ,1.1) -- ( 6.82 ,4.6);
			\end{tikzpicture}
		\end{subfigure}
		\hspace{0.0cm}
		\begin{subfigure}[c]{0.500\textwidth}
			\includegraphics[width=\textwidth,height=0.15\textheight, trim={0.0cm 20.0cm 0.0cm 12.0cm},clip]{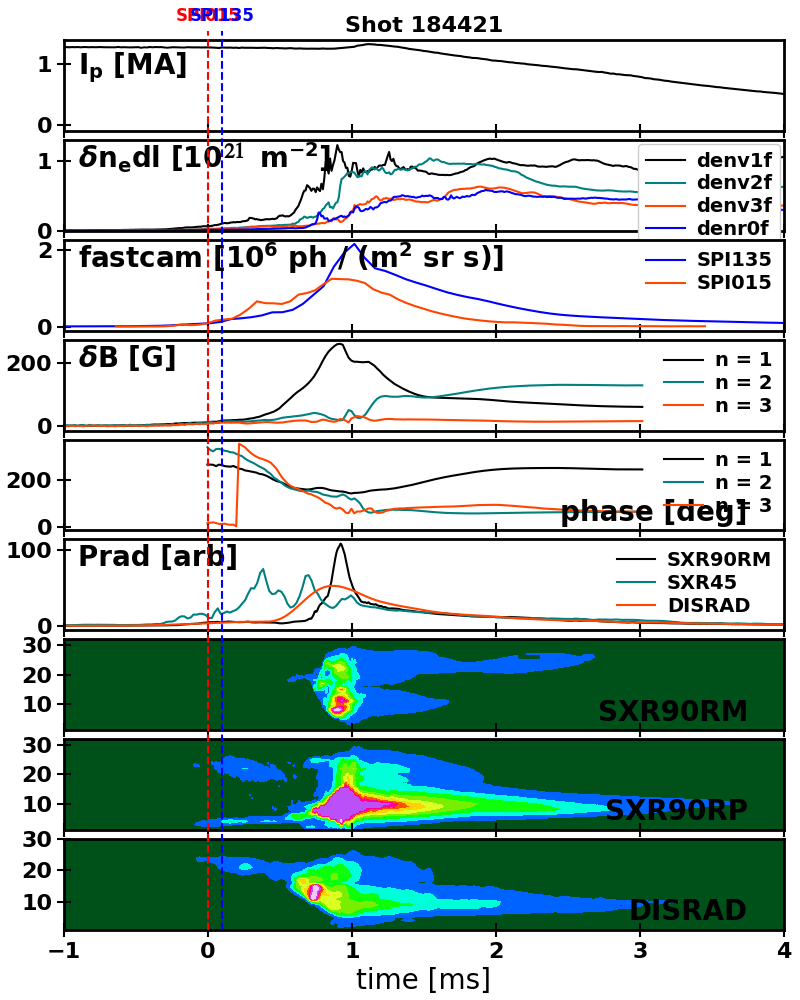}
			\vspace{-0.60cm}
			\caption*{184421 (dual)}
			\begin{tikzpicture}[overlay]
				\draw[dashed, ultra thick, green] ( 3.88 ,1.1) -- ( 3.88 ,4.6);
			\end{tikzpicture}
		\end{subfigure}
		\vspace{-0.8cm}
		\caption{DIII-D edge magnetic probe signals for n=[1,2,3] from shots 184414 (single) and 184421 (dual).  
		Red dashed line
		indicates arrival of fragments from SPI015 and blue dashed line indicates arrival of fragments from SPI135.  The
		green dashed line indicates the thermal quench time.}
		\label{fig:d3dbprobe}
	\end{figure}

	\begin{figure}[H]
		\begin{subfigure}[b]{0.5\textwidth}
		\includegraphics[width=\textwidth]{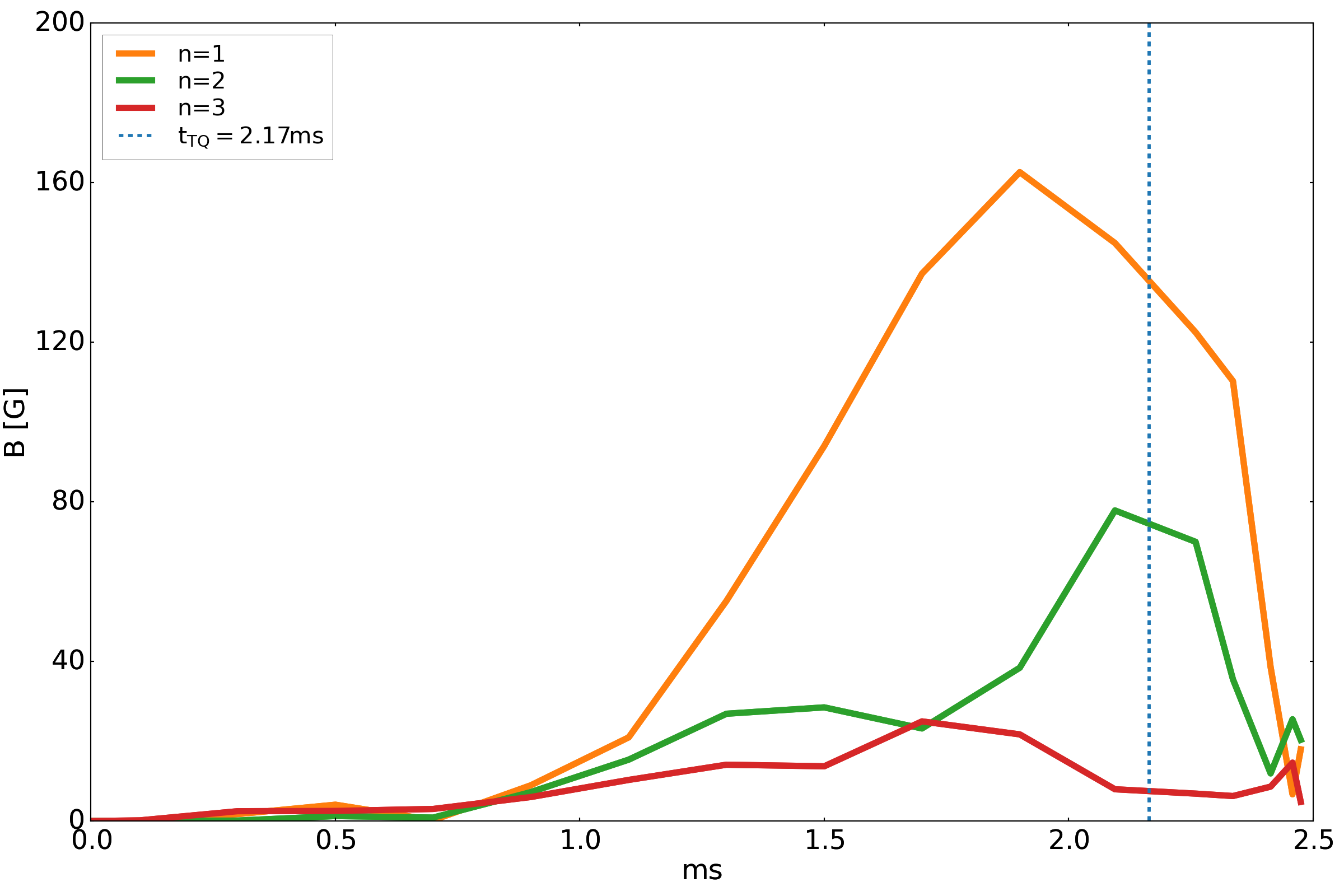}
		\vspace{-0.80cm}
			\caption{dt=+0.4ms}
		\end{subfigure}
		\begin{subfigure}[b]{0.5\textwidth}
		\includegraphics[width=\textwidth]{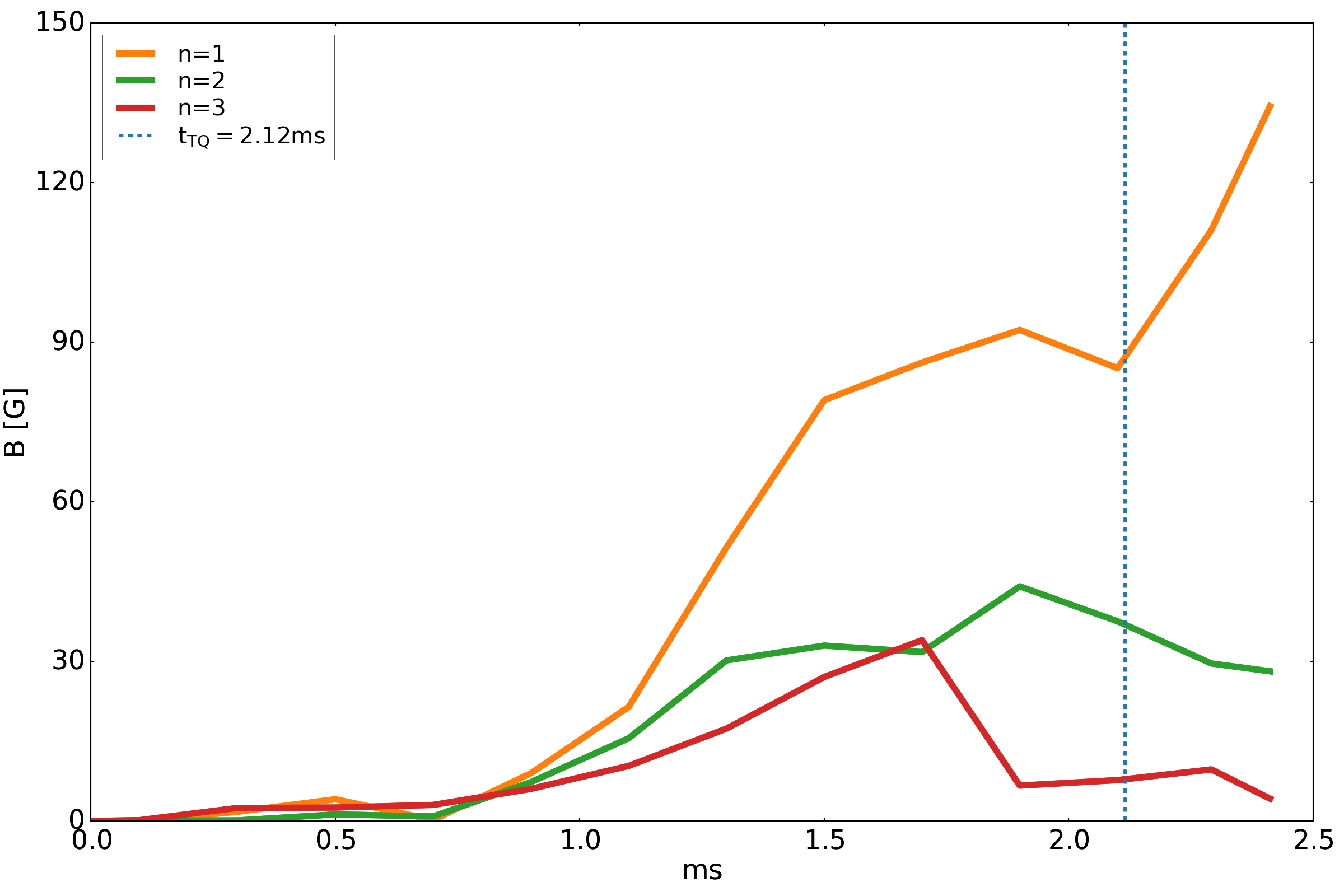}
		\vspace{-0.80cm}
			\caption{dt=-0.4ms}
		\end{subfigure}
		\\
		\begin{subfigure}[b]{0.5\textwidth}
		\includegraphics[width=\textwidth]{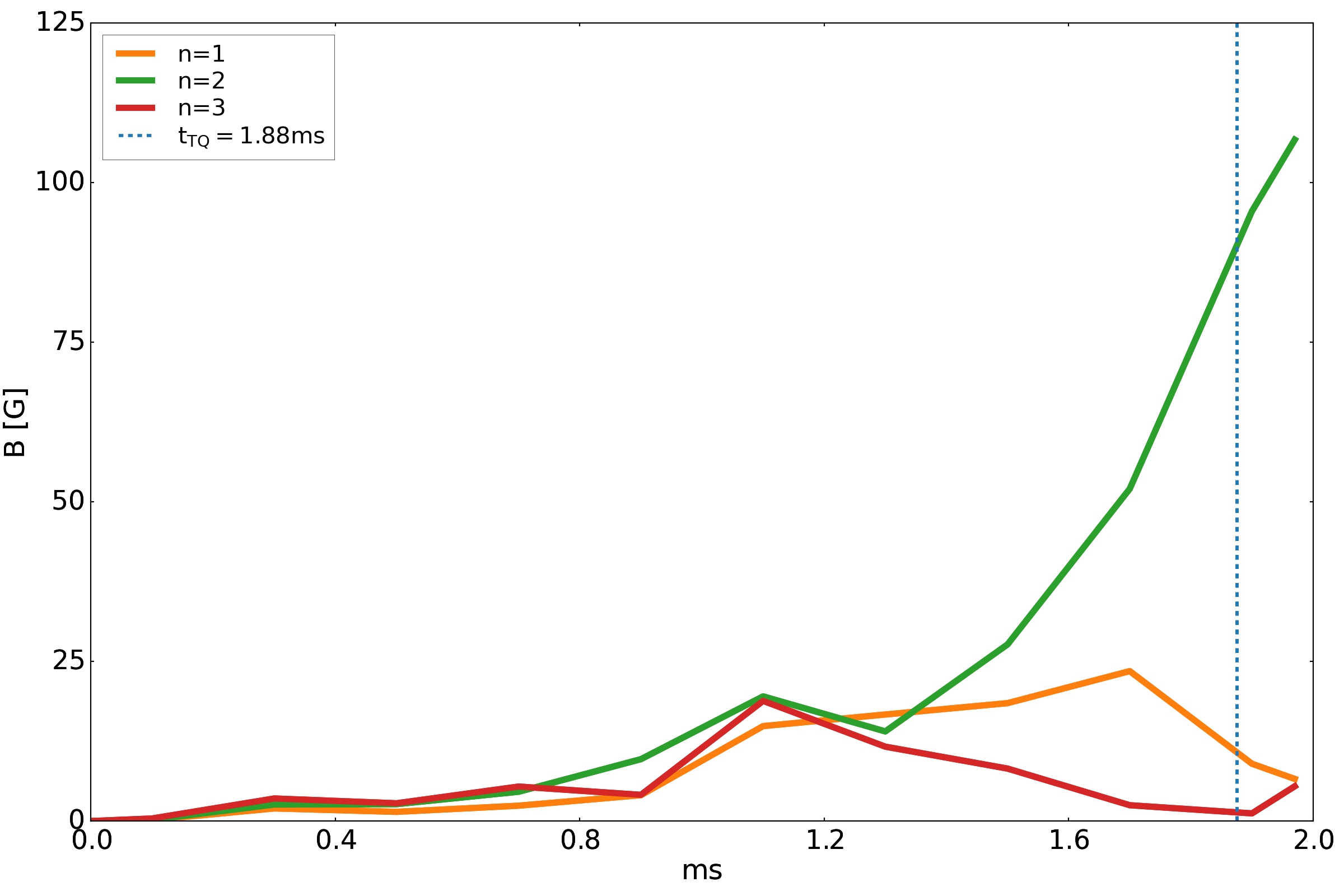}
		\vspace{-0.80cm}
			\caption{dt=0}
		\end{subfigure}
		\begin{subfigure}[b]{0.5\textwidth}
		\includegraphics[width=\textwidth]{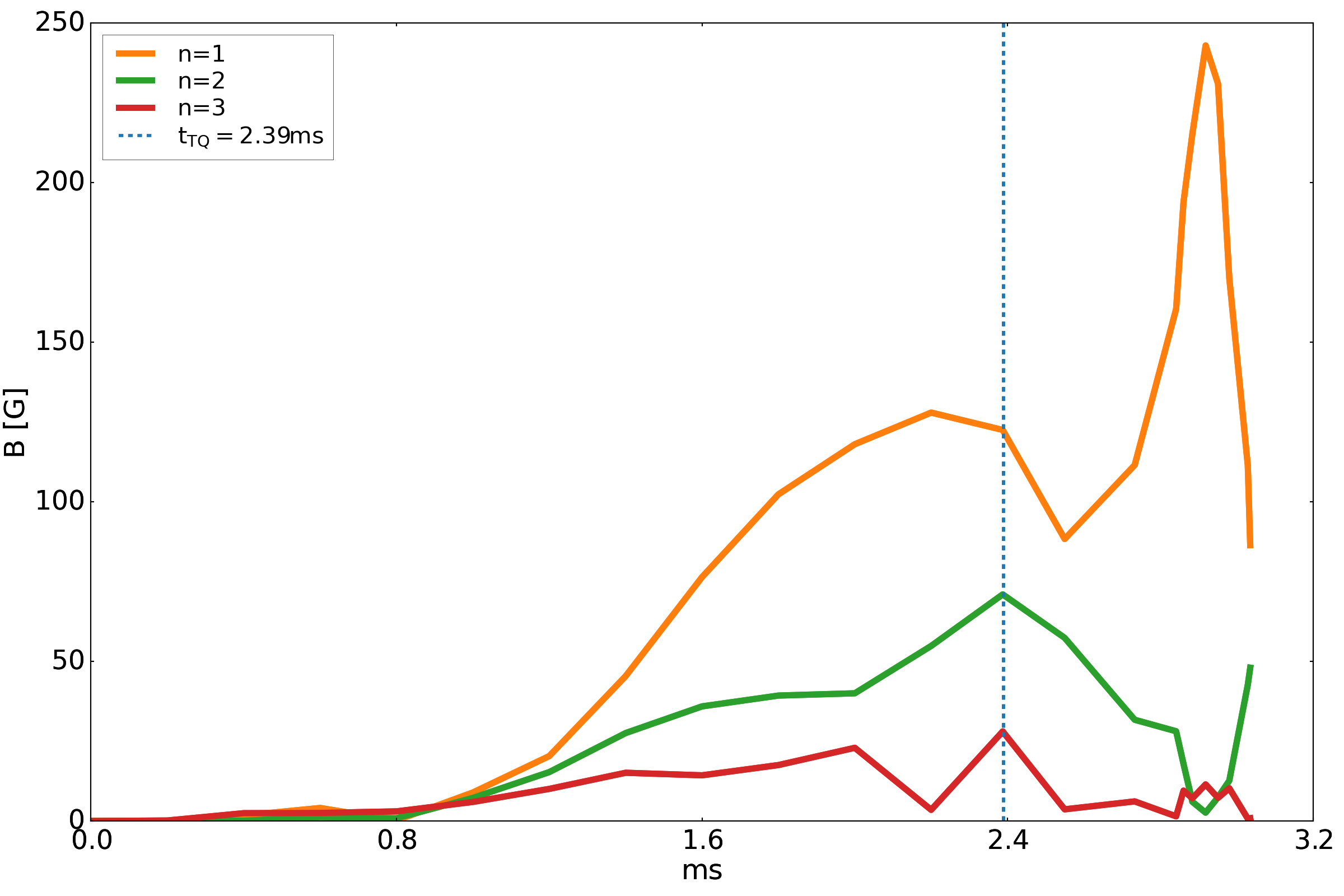}
		\vspace{-0.80cm}
			\caption{dt=-0.8ms}
		\label{fig:bprobe08}
		\end{subfigure}
		\vspace{-0.8cm}
		\caption{Synthetic edge magnetic probes located at the outboard midplane.
		Plotted are the field strength $|\delta\bvec{B}|$ for n=[1,2,3], in Gauss.
		Magnetic probe amplitudes show a similar decreasing trend to the magnetic energies in Figure
		\ref{fig:MEdt4}.  However, for dt=0.0ms, the n=1 signal is significantly reduced and the n=2 is
		more prominent.   }
		\label{fig:MagProbe}
	\end{figure}
	
The DIII-D experimental midplane toroidal array of magnetic probes in Figure \ref{fig:d3dbprobe} show a dominant 
n=1 signal accompanied by modest n=[2,3].  

Figure \ref{fig:MagProbe} shows the synthetic outboard midplane edge probes from the SPI simulations for delays
dt=[$\pm$0.4,0.0,-0.8]ms.  Plotted are the field strength $|\delta\bvec{B}|$ for n=[1,2,3].  Comparison of amplitudes
shows a factor of $\times$2 larger signal in the experiment than the simulation.  Magnetic probe amplitudes
show a similar decreasing trend to the magnetic energies in Figure \ref{fig:MEdt4}.  However, for dt=0.0ms, the n=1
signal is significantly reduced and the n=2 is more prominent.   This reduction may be particular to the simultaneous
injector dt=0.0ms and is likely unrealizable experimentally.  All other cases show a prominent n=1.

The dt=-0.8ms case shows a large n=1 corresponding to the current spike.  The dt=-0.4ms case also shows a rising n=1
corresponding to its rising current which seems on its way to a current spike.  The experimental magnetic probes in Figure
\ref{fig:d3dbprobe} show some increase in the n=1 amplitude in correlation with the current spike but not as prominent as
the simulation.  The experimental current spike is broader than that observed in the simulations.  

\subsection{Last Closed Flux Surface}
\begin{figure}[H]
	\centerline{
		\begin{subfigure}[c]{0.350\textwidth}
		\includegraphics[width=\textwidth]{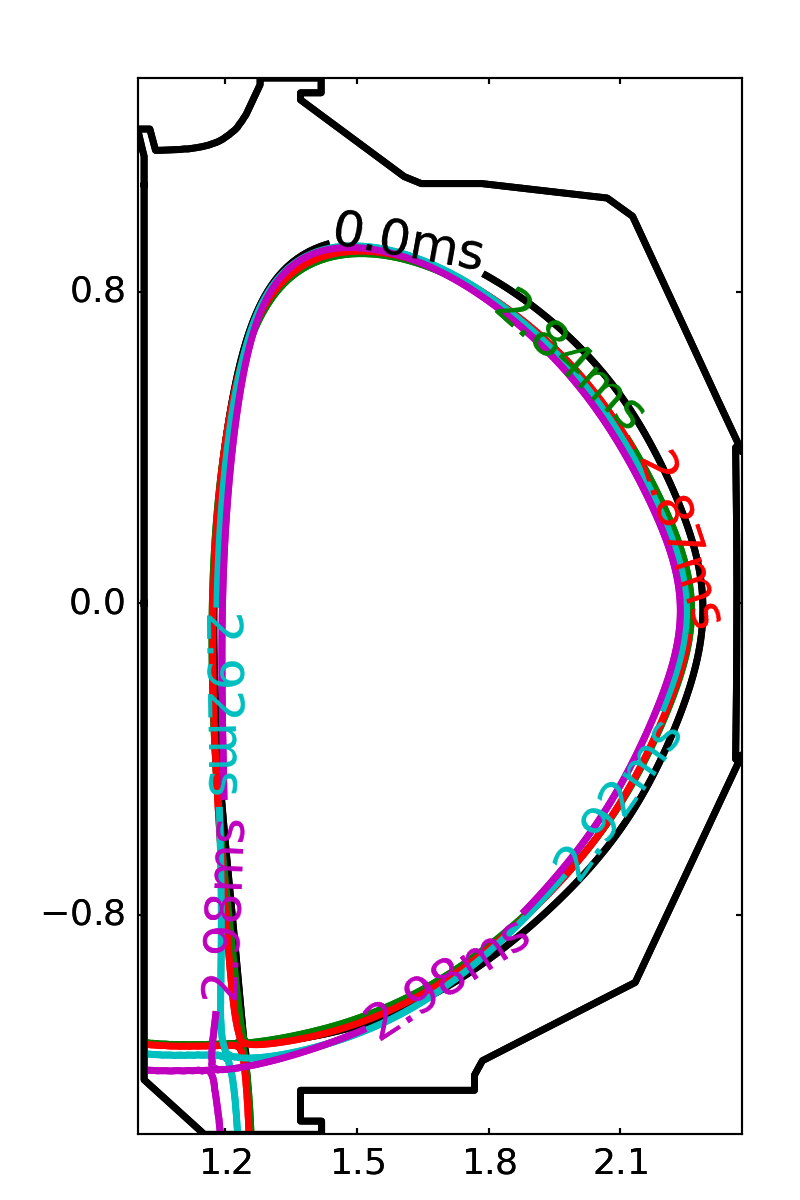}
		\vspace{1.40cm}
		\caption{NIMROD}
		\label{fig:lcfsdt08}
		\end{subfigure}
		\hspace{1.0cm}
		\begin{subfigure}[c]{0.680\textwidth}
		\includegraphics[width=\textwidth]{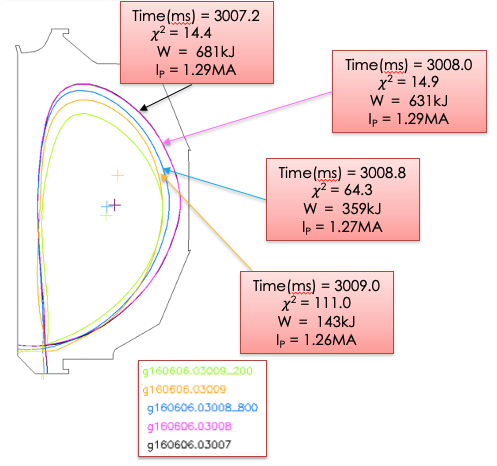}
		\vspace{-0.60cm}
		\caption{DIII-D/EFIT}
		\label{fig:lcfsd3d}
		\end{subfigure}
		}
		\caption{Comparison of last closed flux surface evolution during the thermal quench.  NIMROD lcfs show
		a deflection of the x-point corresponding to the current spike (fig \ref{fig:allddt08}).  The
		DIII-D/EFIT reconstructions show greater change in the poloidal cross section.
		}
		\label{fig:lcfs}
\end{figure}

Figure \ref{fig:lcfs} shows a comparison of a sequence of last closed flux surface reconstructed from NIMROD SPI
simulations with dt=-0.8ms and a series of EFIT reconstructions of shot 160606\cite{Shiraki2016}, the canonical single
SPI experiment.  

The NIMROD lcfs were reconstructed from the flux surfaces computed from the axisymmetric fields and correspond to times
indicated by the colored dots along the plasma current trace in Figure \ref{fig:allddt08}.  These axisymmetric lcfs do
not change much until after the thermal quench, during the current spike.  Close inspection shows that during the
current spike (t=[2.92,2.98]ms) the x-point is deflected.  A similar observation regarding the current spike and x-point
were first made by V.~Izzo\cite{VIzzo:2021} in NIMROD Dispursive Shell simulations.  Otherwise, only a modest decrease
in the cross section is observed throughout the entire thermal quench.

The EFIT reconstructions from t=[3007.0-3009.2]ms show much more change in the cross section of the poloidal flux during
the thermal quench.  This difference may be due to reduced thermal conduction model.  The slow evolution may be an
indication that the constant parallel thermal conduction of 1$\times$10$^9$m/s$^2$ may be too low maintaining too
high of a core temperature.  The single temperature model may be another limit.
Additional factors may be the non-axisymmetric contributions to the EFIT data absent in the NIMROD lcfs.
However, there is an absence of significant n>0 magnetic energies early in the simulations.
Further investigations are ongoing.

\subsection{Toroidal Asymmetry}
\begin{figure}[H]
		\centerline{
		\begin{subfigure}[b]{0.245\textwidth}
			\includegraphics[width=\textwidth,trim={0.0cm 0.0cm 0.0cm 2.0cm},clip]{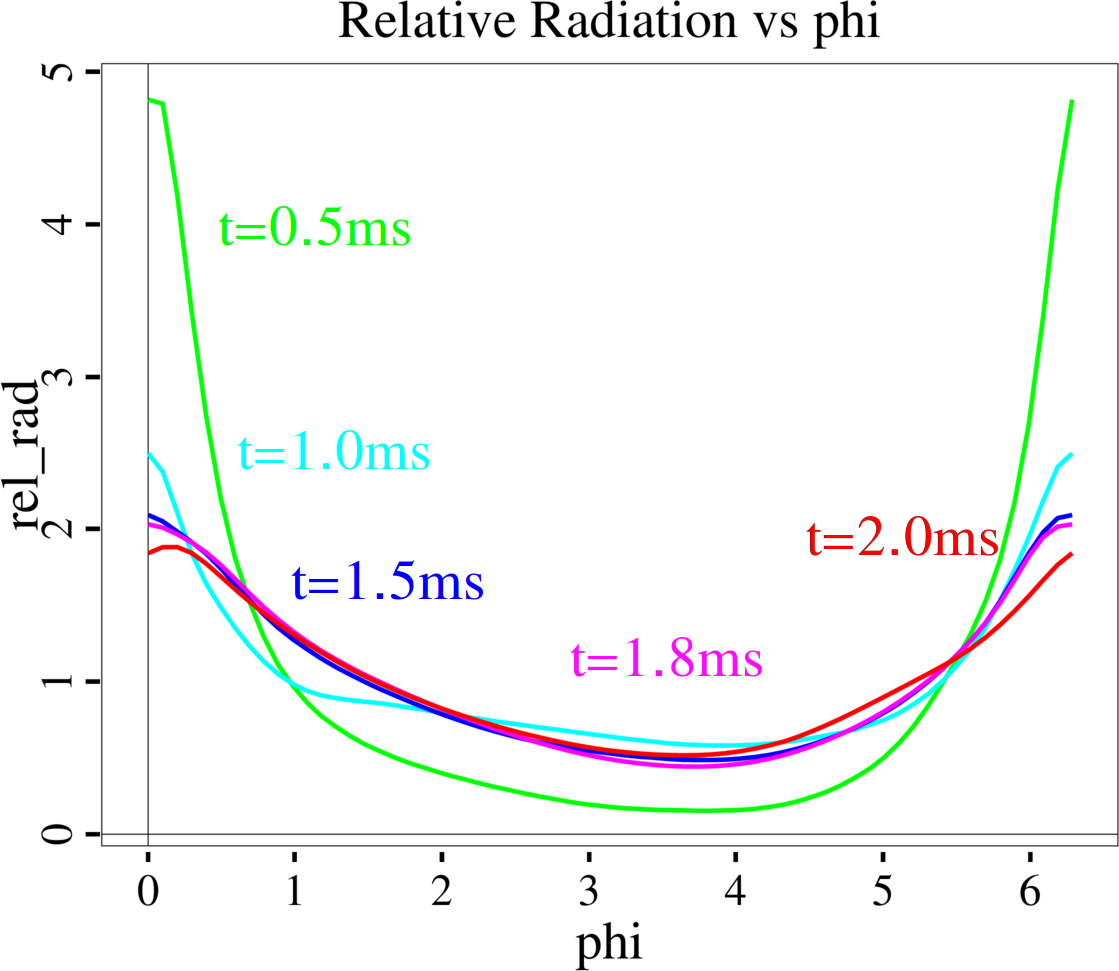}
			\vspace{-0.60cm}
			\caption{double load}
			\label{fig:torasymsd}
		\end{subfigure}
		\hspace{0.0cm}
		\begin{subfigure}[b]{0.245\textwidth}
			\includegraphics[width=\textwidth,trim={0.0cm 0.0cm 0.0cm 2.0cm},clip]{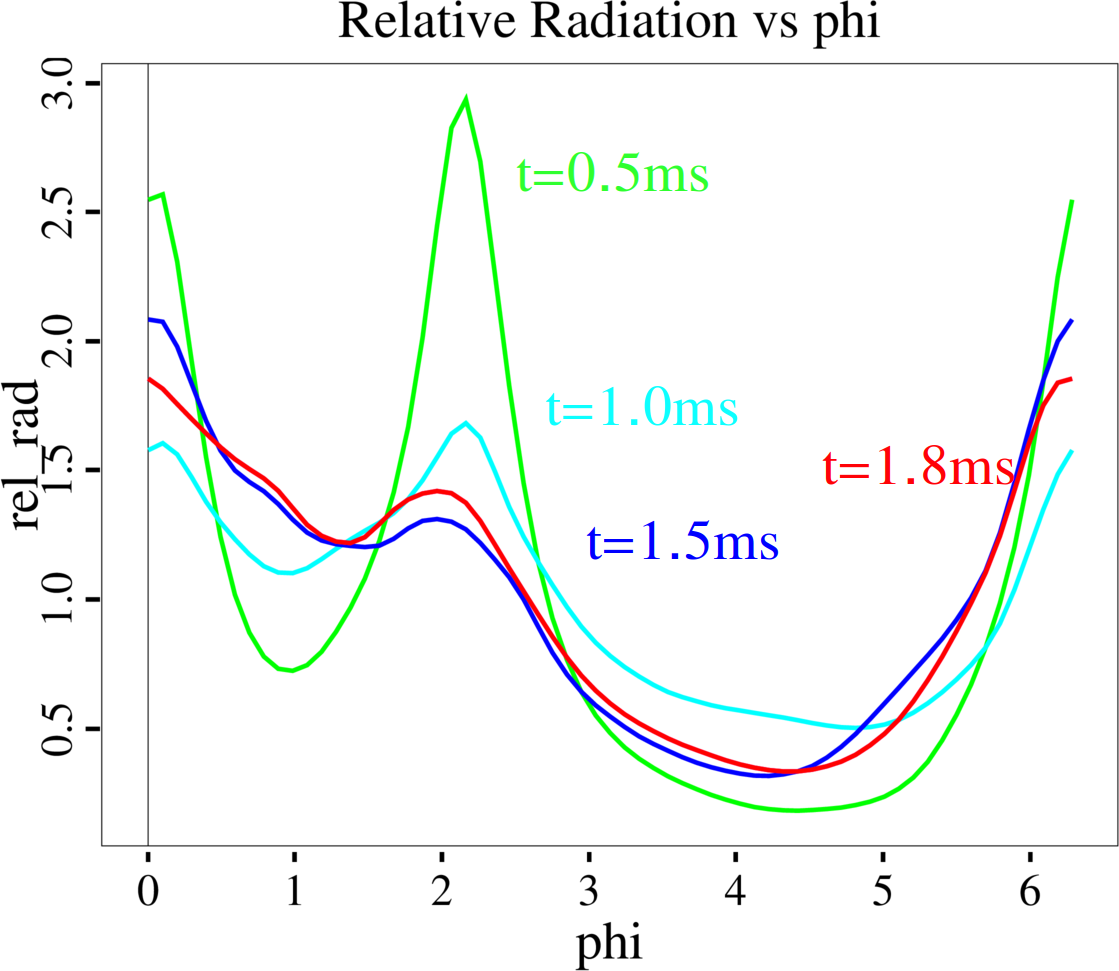}
			\vspace{-0.60cm}
			\caption{\clrlg{dt=0.0ms}}
			\label{fig:torasymdl}
		\end{subfigure}
		\hspace{0.0cm}
		\begin{subfigure}[b]{0.245\textwidth}
			\includegraphics[width=\textwidth,trim={0.0cm 0.0cm 0.0cm 2.0cm},clip]{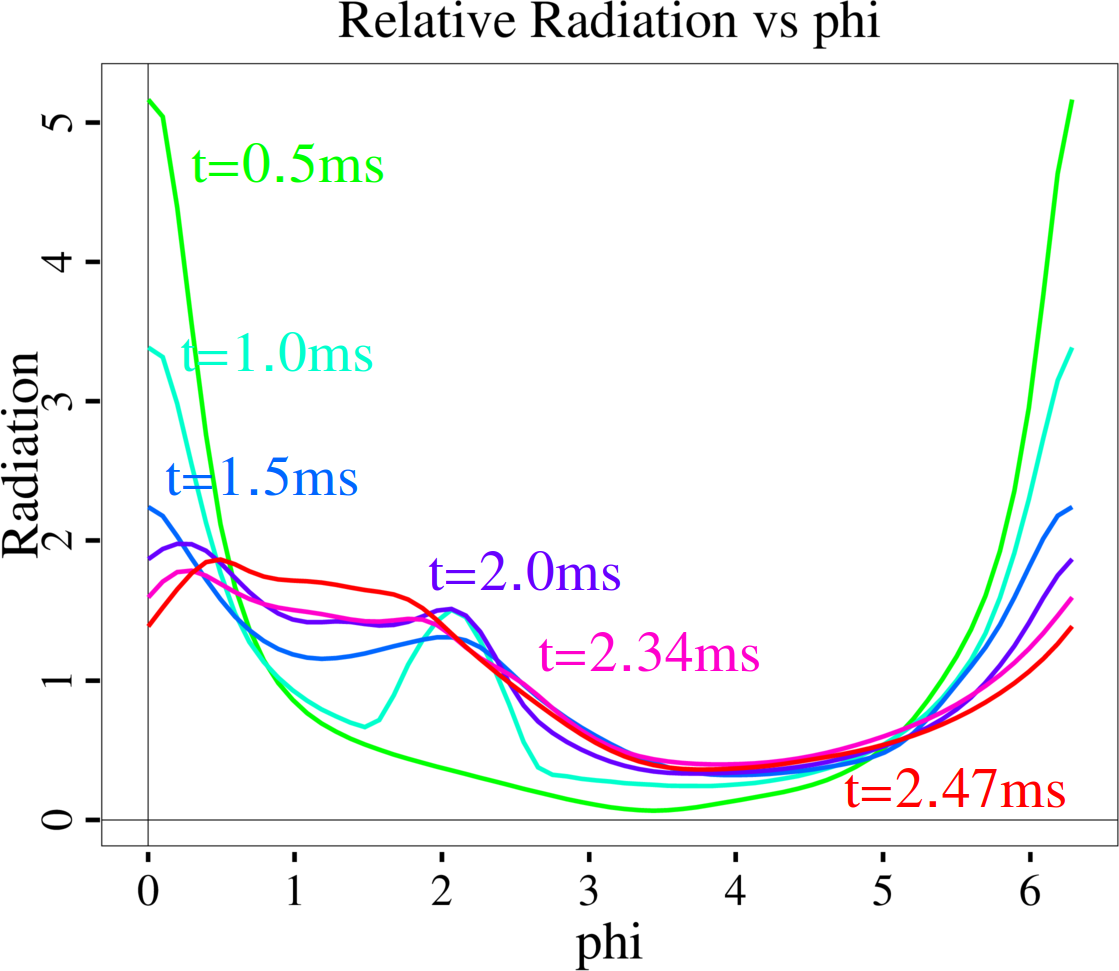}
			\vspace{-0.60cm}
			\caption{\clrb{dt=+0.4ms}}
			\label{fig:torasymp4}
		\end{subfigure}
		\hspace{0.0cm}
		\begin{subfigure}[b]{0.245\textwidth}
			\includegraphics[width=\textwidth,trim={0.0cm 0.0cm 0.0cm 2.0cm},clip]{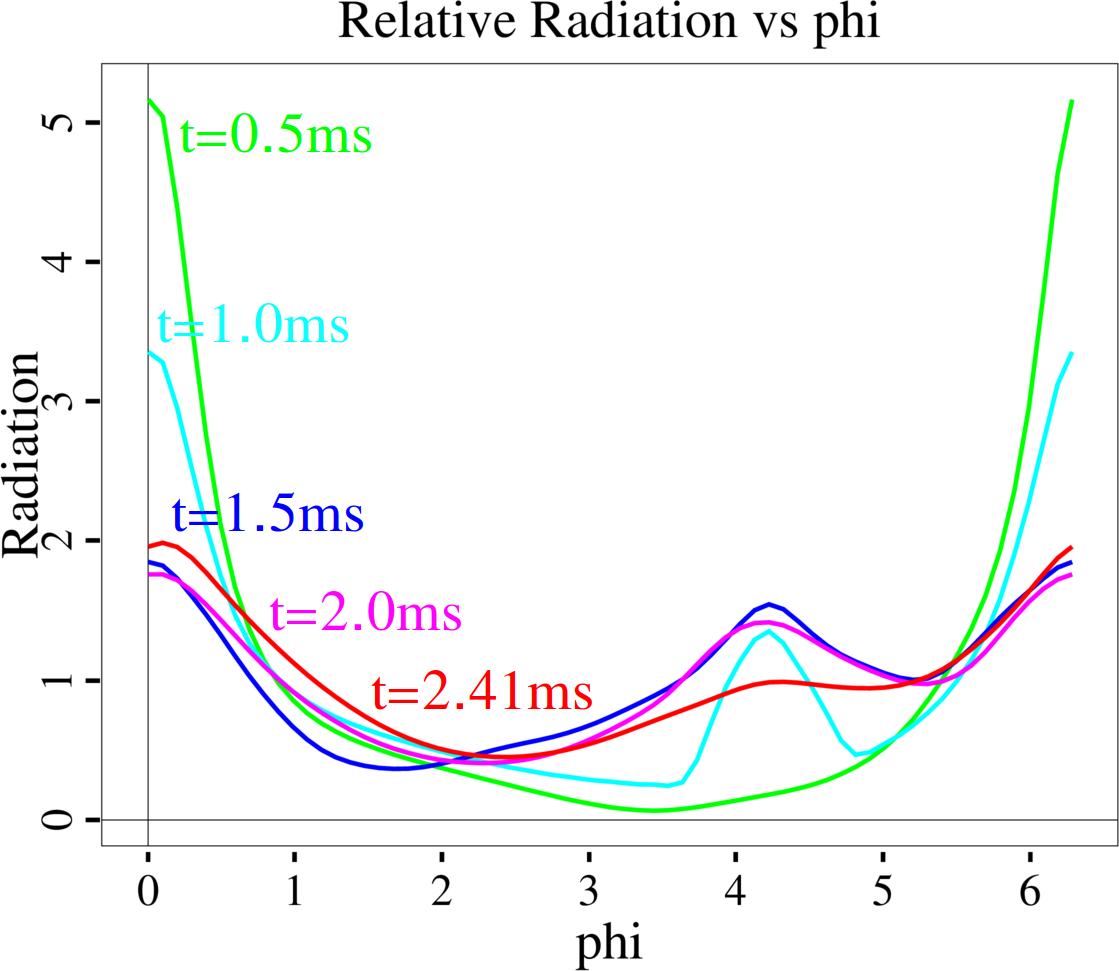}
			\vspace{-0.60cm}
			\caption{\clrr{dt=-0.4ms}}
			\label{fig:torasymn4}
		\end{subfigure}
		}
		\caption{Toroidal radiation asymmetry plot for single SPI double load and dual SPI 
		\clrlg{dt=0.0ms}, \clrb{dt=+0.4ms} and \clrr{dt=-0.4ms}.  The dual SPI radiation asymmetries show
		that during the highest radiation intensity (t$\simeq$[\clrb{1.5}-\clrr{2.4}]ms) the radiation
		peaking is broadened.} 
		\label{fig:torasym}
	\end{figure}

Figures \ref{fig:torasym} show the toroidal radiation asymmetry ($A(\phi)=R(\phi)/R_{average}$) at several times during
single SPI double load and dual SPI \clrlg{dt=0.0ms},the \clrb{dt=+0.4ms} and \clrr{dt=-0.4ms} simulation.
Initially (t=0.5ms) we can see that radiation is peaked around $\phi=0^\circ$ as the leading injector dominates early
on, except for the \clrlg{dt=0.0ms} where there is a clear dual peak.  For the dual SPI cases with delays, as the
second injector fragments encounter the plasma, the radiation asymmetry can be seen to decrease, with reduced peaks near
the injector locations.  In particular, during the most intense radiation peak (t$\simeq$[\clrb{1.5}-\clrr{2.4}]ms) the
toroidal radiation asymmetry is reduced regardless of the sign of the delay.  For all three dual SPI cases, the
second radiation peak during the most intense radiation is not very prominent and could easily be confused with the
single SPI case, particularly in situations of low resolution.

We note that the toroidal width of the radiation follows the width of the toroidal deposition
(gaussian 3\% half-width at half max results in features of $\sim$12\% width), which is limited by the
spatial resolution.  However, recall that the simulations do not include plasma rotation.  As was mentioned, even a modest
rotation frequency of a few 100Hz can rotate the plasma enough to broaden the toroidal deposition, thus broadening the
radiation.

\begin{figure}[H]
		\centerline{
		\includegraphics[width=0.7\textwidth,trim={0.0cm 0.0cm 0.0cm 0.0cm},clip]{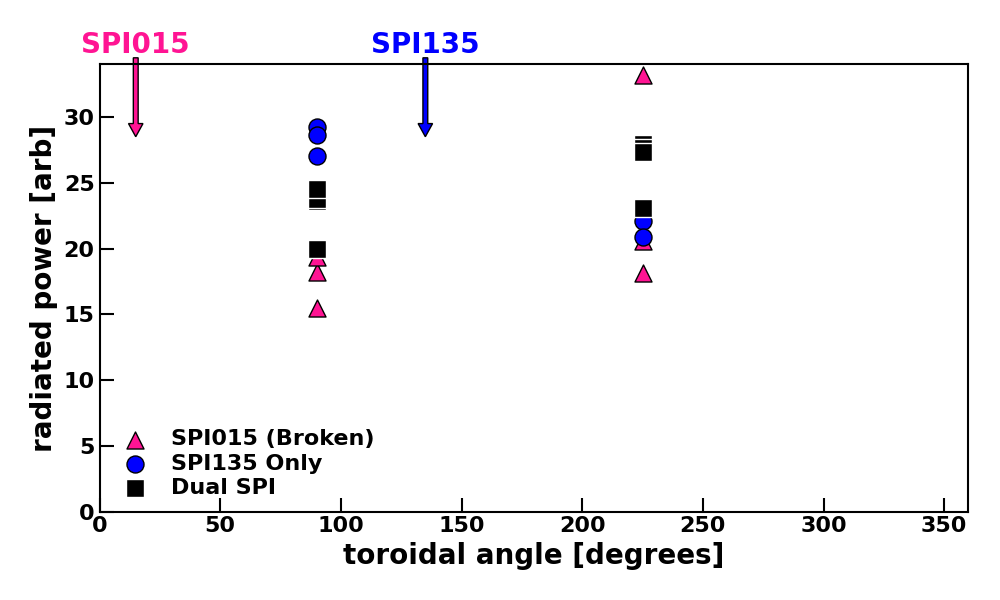}
			}
			\vspace{-0.5cm}
			\caption{DIII-D local toroidal radiation measurements hint at reduced peaking with dual SPI
			compared to single SPI.  However the lack of coverage makes definite conclusions difficult.}
			\label{fig:d3dradasym}
	\end{figure}

The reduced toroidal radiation asymmetry is also indicated in experiments as shown in Figure \ref{fig:d3dradasym}.
Plotted are radiation measurements at two toroidal locations for three different SPI shots: single SPI at
$\phi=15^\circ$, single SPI at $\phi=135^\circ$ and both injectors.  The radiation peaks near the respective
injectors for each of the single SPI shots.  The dual SPI shots shows a reduced toroidal peaking similar to
the simulations.

Additional evidence of decreased toroidal radiation asymmetry is indicated in the fast camera signals in Figures
\ref{fig:d3dexsum}.  The single SPI (184414) fast camera plot shows light corresponding to SPI135 is much larger
than SPI015, indicating the radiation is toroidally localized around SPI135.  The dual SPI (184421) shows both fast
cameras, SPI135 and SPI015, record comparable amounts of light, indicating the radiation is more toroidally spread,
similar to what is observed in the simulations.

\section{Conclusion}
This report presents NIMROD dual SPI simulations validated against DIII-D experiments\cite{herfindal2021}.  We show that
thermal quench improves as the time delay is decreased from dt=+0.4ms to dt=-0.4ms.  This linear trend is attributed to
the handedness of the tokamak plasma and a preferred direction of evolution of quenching plasma column.  As the plasma
quenches, MHD motion of the plasma column may intercept the incoming fragments of the second injector plume, causing
additional ablation and improving the thermal quench efficiency.  

Simulations show a strong correlation between reduced magnetic mode activity and improved thermal quench efficiency;
the quiescent plasma quenches best.  This is not too surprising since an unstable plasma is often a stochastic plasma
that allows for high conductive losses.  Analysis shows that for the dt=-0.4ms case, the delayed fragment ablation of
the second injector plume dampens the growing n=1 resulting in a less stochastic plasma and a higher thermal quench
efficiency,  even higher than the simultaneous dt=0.0 case.  This particular case does not necessarily suggest a 
particular strategy for dual SPI.  Rather, it demonstrates that understanding and utilizing MHD dynamics excited by
the SPI is critical to an efficient mitigation technology.

Similar to the experiment, the simulations show that the acceptable delay between injectors is a small time window that
is a fraction of the thermal quench time, dt at most 50\%$\tau_{TQ}$, more likely dt<25\%$\tau_{TQ}$. 
Experimental control and timing of the injectors and
shattered fragments is not precise enough to systematically test this. 

Gaps remain.  Repeatability and precise control of the SPI experiments, particularly the dual injector experiments,
remains elusive.  Uncertainties and variations in velocity and composition of the fragment plume make comparisons
challenging.  Limited diagnostics appropriate for disruption mitigation experiments also limit the ability to
validate NIMROD results.  The high computational cost limits the number of runs to a handful and so scans in 
parameter space are shallow.  

\newpage

\section*{Acknowledgments and Disclaimers}
This work supported in part by US DOE contract DE-SC0020299 and the Center for Tokamak Transient Simulation (CTTS)
SciDAC (DE-SC0018109), and in collaboration with work under Awards DE-FG02-95ER54309, and DE-FC02-04ER54698.  This work
is supported by the ITER Organization under Contract Number IO/IA/20/4300002130 and is contributing to the ITER
Organization's Disruption Mitigation Task Force work program. The views and opinions expressed herein do not necessarily
reflect those of the ITER Organization.  This research used resources of the National Energy Research Scientific
Computing Center (NERSC), a U.S. Department of Energy Office of Science User Facility located at Lawrence Berkeley
National Laboratory, operated under Contract No.  DE-AC02-05CH11231 using NERSC award FES-ERCAP0018139.

This report was prepared as an account of work sponsored by an agency of the United States Government. Neither the
United States Government nor any agency thereof, nor any of their employees, makes any warranty, express or implied, or
assumes any legal liability or responsibility for the accuracy, completeness, or usefulness of any information,
apparatus, product, or process disclosed, or represents that its use would not infringe privately owned rights.
Reference herein to any specific commercial product, process, or service by trade name, trademark, manufacturer, or
otherwise, does not necessarily constitute or imply its endorsement, recommendation, or favoring by the United States
Government or any agency thereof. The views and opinions of authors expressed herein do not necessarily state or reflect
those of the United States Government or any agency thereof.

\newpage
\bibliographystyle{unsrt}
\bibliography{ref}

\newpage
\appendix

\section{DIII-D Dual SPI Shot 184414 and 184421}
\vspace{-0.5cm}
\begin{figure}[H]
	\hspace{-1.25cm}
		\begin{subfigure}[b]{0.55\textwidth}
		\includegraphics[width=\textwidth]{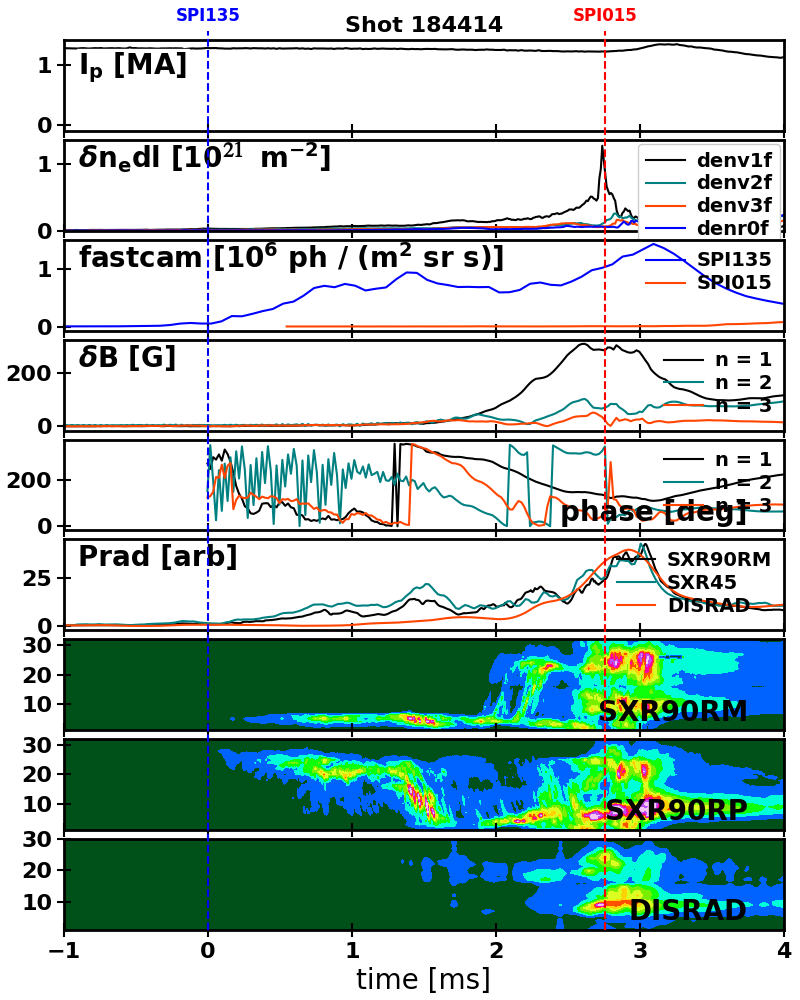}
		\vspace{-0.50cm}
			\caption{184414 - single}
		\label{fig:184414sum}
		\end{subfigure}
		\hspace{-0.25cm}
		\begin{subfigure}[b]{0.55\textwidth}
		\includegraphics[width=\textwidth]{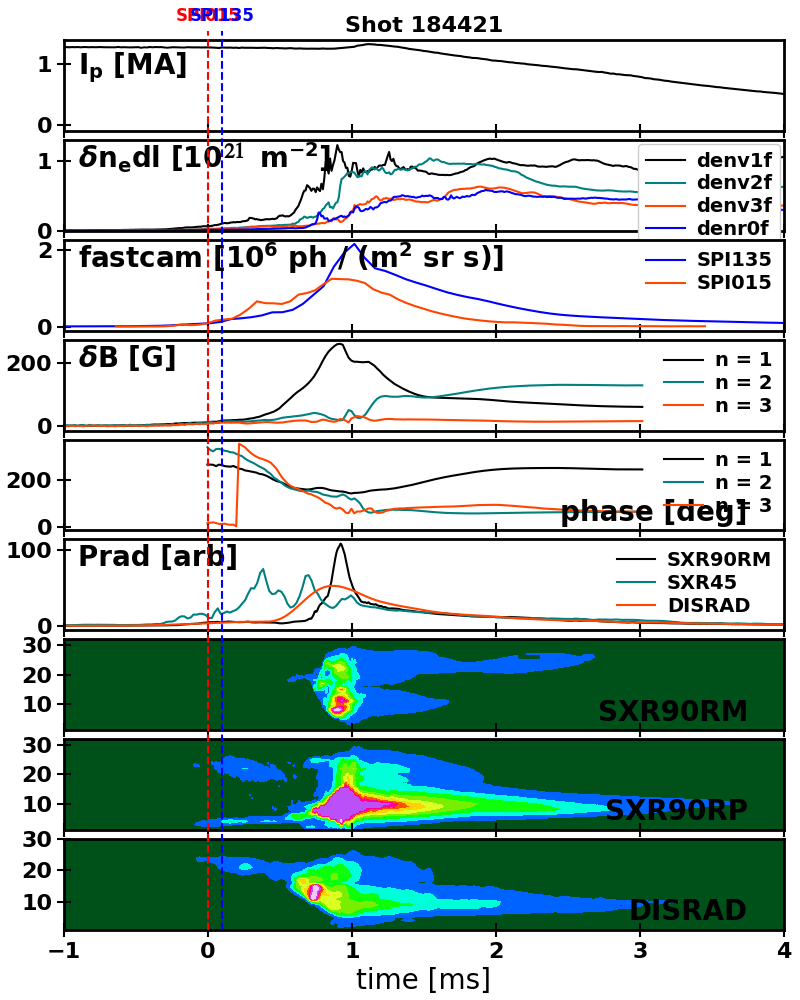}
		\vspace{-0.50cm}
			\caption{184421 - dual}
		\label{fig:184421sum}
		\end{subfigure}
		\vspace{-0.40cm}
		\caption{DIII-D SPI experiment thermal quench summary for shots 184414 and 184421 showing the plasma
		current, interferometer line integrated electron density from 3 vertical chords and a radial midplane
		chord, integrated light from fast cameras at 15$^\circ$ and 135$^\circ$, midplane edge B-probe signals
		for n=[1,2,3], the integrated radiate power from the soft x-rays arrays at 90$^\circ$(SXR90RM) and 45$^\circ$(SXR45) and 225$^\circ$(DISRAD)
		and contours plots from the soft x-ray diode arrays, where the vertical axis corresponds to the array channel.
		The arrival times of the respective SPI fragments is indicated by the vertical dashed lines.
			}
		\label{fig:d3dexsum}
\end{figure}
Figure \ref{fig:d3dexsum} show the SPI quench experiment summary for shots 184414 and 184421 from
DIII-D\cite{herfindal2021}.  Plotted are the plasma current, interferometer line integrated electron density from 3
vertical chords and a radial midplane chord, integrated light from fast cameras at 15$^\circ$ and 135$^\circ$, midplane
edge B-probe signals for n=[1,2,3], the radiate power from the soft x-rays at 90$^\circ$ and 45$^\circ$ and
225$^\circ$(DISRAD), and contour plots from the soft x-ray diode arrays, where the vertical axis corresponds to the
array channel.  Note that SXR90RM shows an inverted poloidal view (bottom up view, channels numbered top to bottom) to
SXR90RP and DISRAD (top down view, channels numbered bottom to top).  The arrival times of the respective SPI fragments
is indicated by the vertical dashed lines in red(SPI015) and blue(SPI135).

These two shots represent a single SPI (184414) and dual SPI (184421) SPI thermal quench.  Both cases show a
minimal loss of plasma current before the thermal quench and current spike.  This is also observed in the NIMROD plasma
currents of the SPI simulations shown in Figure \ref{fig:MEdt4} (note the scaling indicated in the legend).  The dual
SPI case shows a faster thermal quench than the single SPI.  Both cases show that the majority of the
radiation is emitted during the latter half of the thermal quench.  Both cases show increasing magnetic activity as the
quench proceeds; the n=1 mode is particularly prominent.  The integrated light from the fast cameras and the soft x-ray
arrays show that the dual SPI radiates more than the single SPI.

\newpage
\section{Experimental Layout}
\label{sec:explayo}
\vspace{-1.2cm}
	\begin{figure}[H]
		\begin{subfigure}[c]{0.70\textwidth}
			\includegraphics[width=\textwidth]{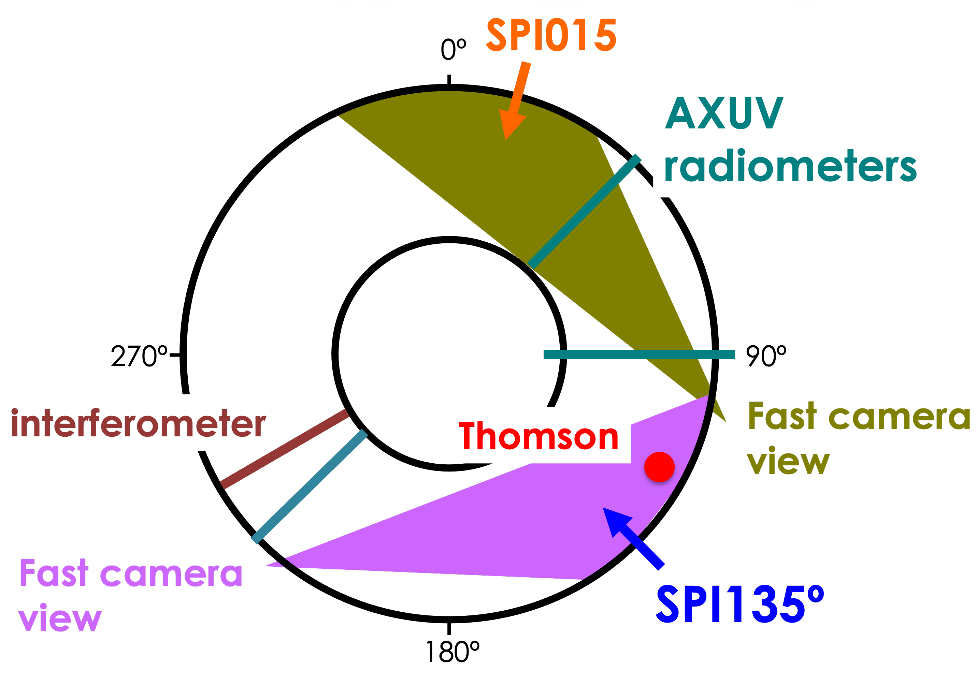}
			\vspace{-0.10cm}
			\caption*{injectors and diagnostics}
		\end{subfigure}
		\hspace{-0.30cm}
		\begin{subfigure}[c]{0.30\textwidth}
			\includegraphics[width=\textwidth]{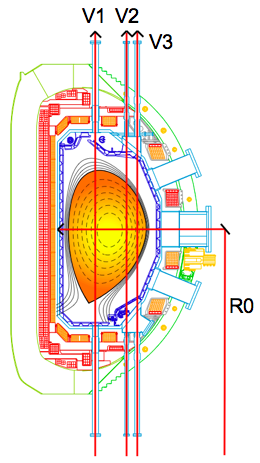}
			\vspace{-0.30cm}
			\caption*{C02 interferometers}
		\end{subfigure}
	\end{figure}
\vspace{-0.75cm}
	\begin{figure}[H]
	\centerline{
		\includegraphics[width=0.7\textwidth]{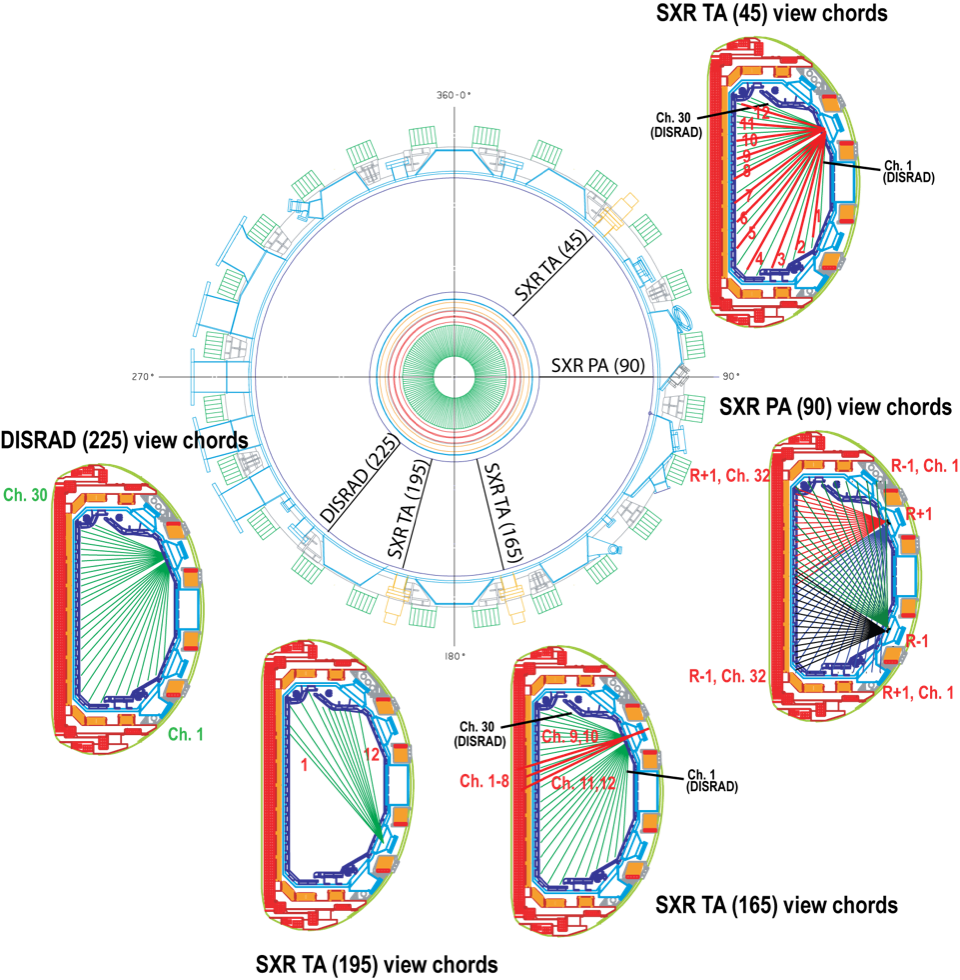}
		}
		\vspace{-0.10cm}
		\caption*{soft x-ray arrays}
	\end{figure}

\newpage
\section{\href{https://www.youtube.com/playlist?list=PLoStj5o1MeTu2gIDiWVWtvVUe3FsVEdni}{Visualizations}}
\label{sec:viz}
Images are active links that will bring up VisIt\cite{VisIt2012} generated visualizations on YouTube.

\subsection{Delay dt=0.0}
	\begin{figure}[H]
	\centerline{
		\href{https://youtu.be/4ZbGe0s62fk}
		{\includegraphics[width=1.10\textwidth]{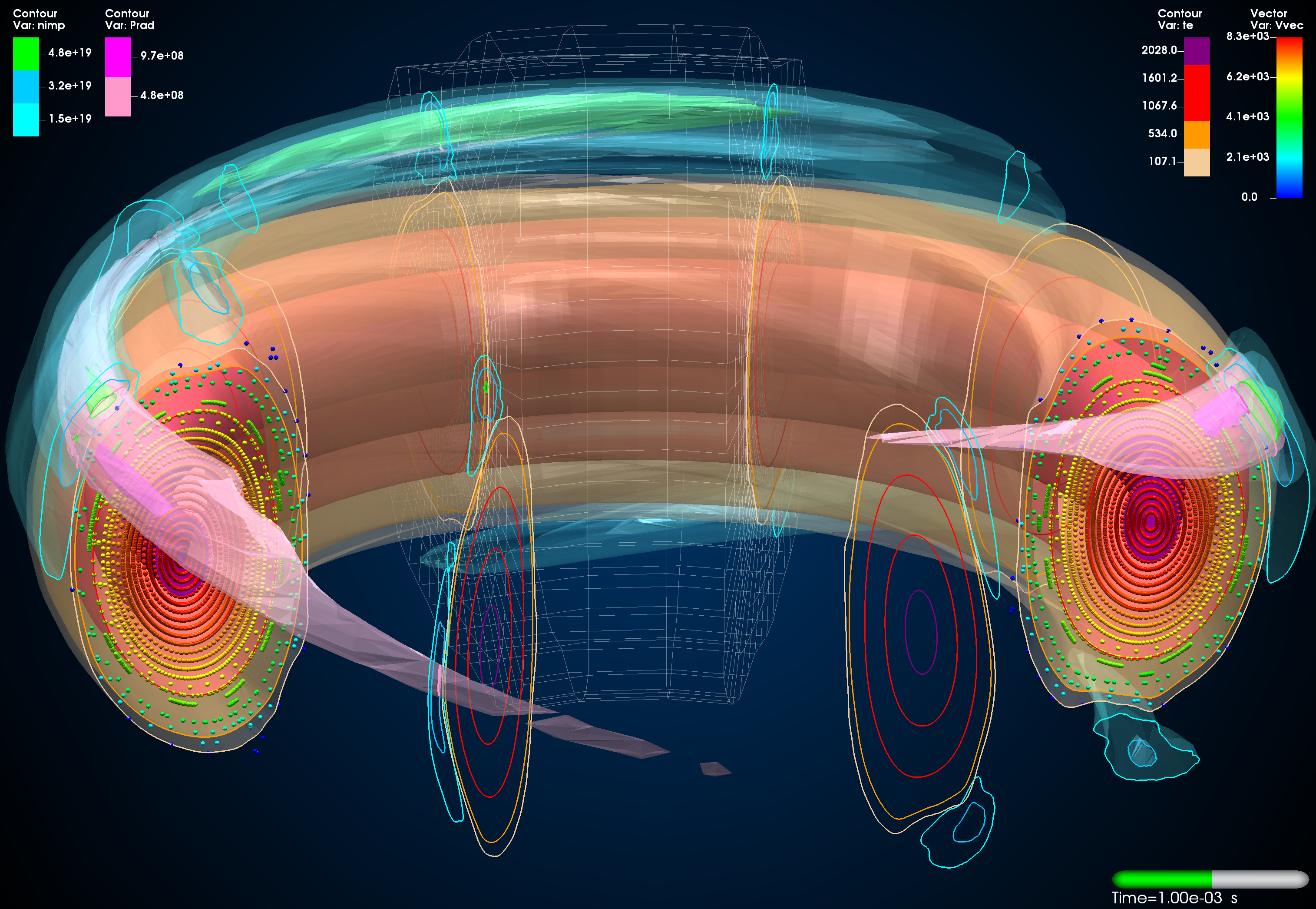}}
		}
	\end{figure}
	Simultaneous dual SPI.  3D visualization shows impurity density in blues and greens, radiation
	in pink, and the main isosurface shows temperature.  Vector arrows show flow but are relatively small at this
	time and not very visible.  Legends are in the upper corners. A 120$^\circ$ section of temperature and impurity
	isosurfaces have been removed for an internal view.  The exposed poloidal planes correspond to the dual SPI injection planes.
	Magnetic field puncture point are shown at each injector plane.  Poloidal cross-sections every 60$^\circ$ show
	contours of temperature, impurity density, and radiation.  t=1.0ms is near the end where both fragment plumes still interact
	symmetrically with the plasma.

	2D cross sections at $\phi$=[0$^\circ$,120$^\circ$,240$^\circ$] in the next Figure show contours and contour
	lines of the impurity density, radiation and temperature at t=[0.2,1.1,1.5]ms.  Contour lines of the temperature
	are proxies for magnetic flux surfaces.  At t=0.2ms both fragments plumes at 0$^\circ$ and 120$^\circ$ are
	ablating equally.  By t=1.1ms, MHD causes the flux surfaces to divot at 0$^\circ$ and bulge at 120$^\circ$
	promoting ablation at 120$^\circ$ and reducing it at 0$^\circ$.  By t=1.5ms, the helical distortion is evident.

	\begin{figure}
		\href{https://youtu.be/bDXCbaVzwAQ}
		{
	\centerline{
		\includegraphics[width=0.9\textwidth]{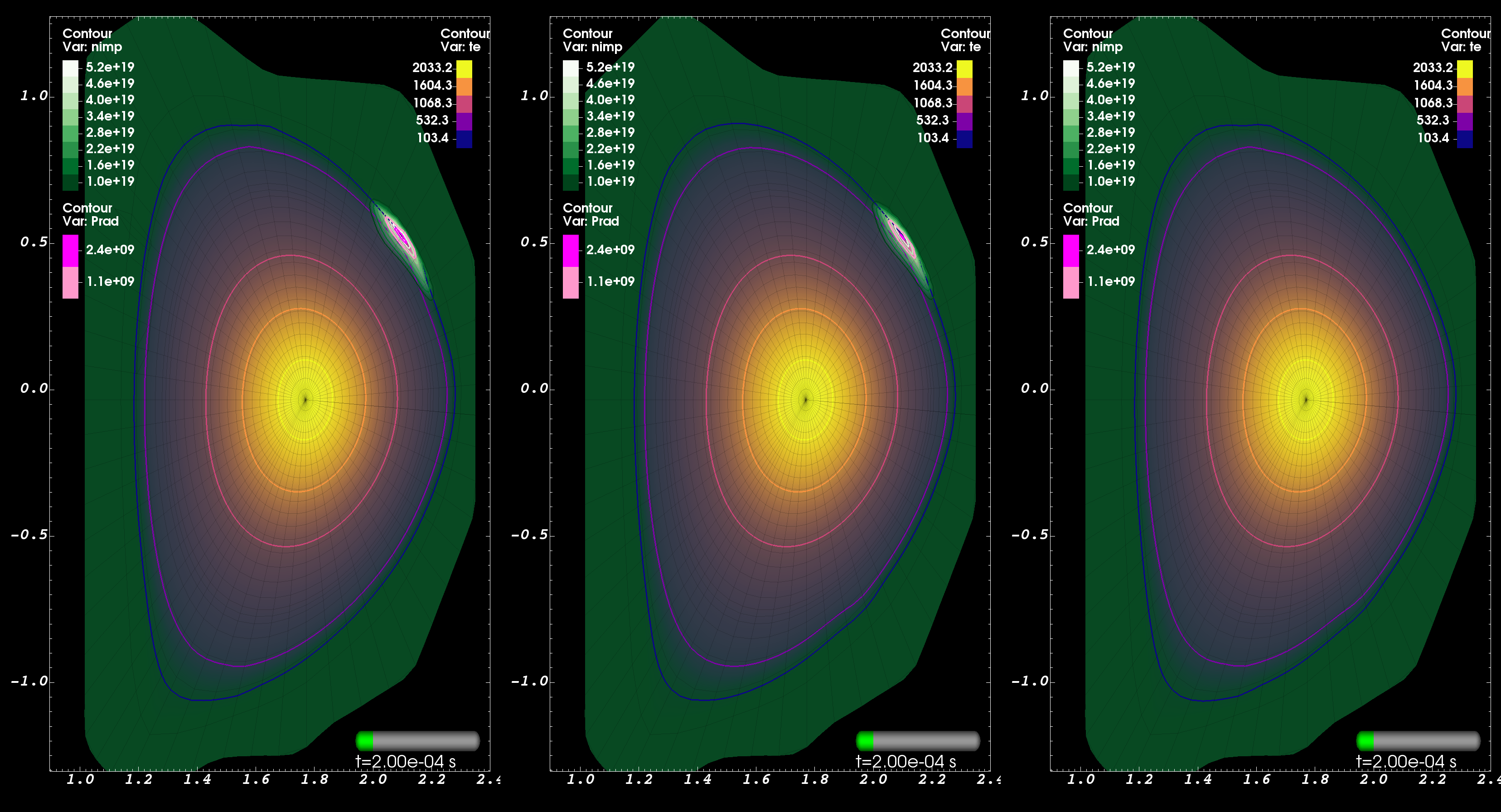}
		}
	\centerline{
		\includegraphics[width=0.9\textwidth]{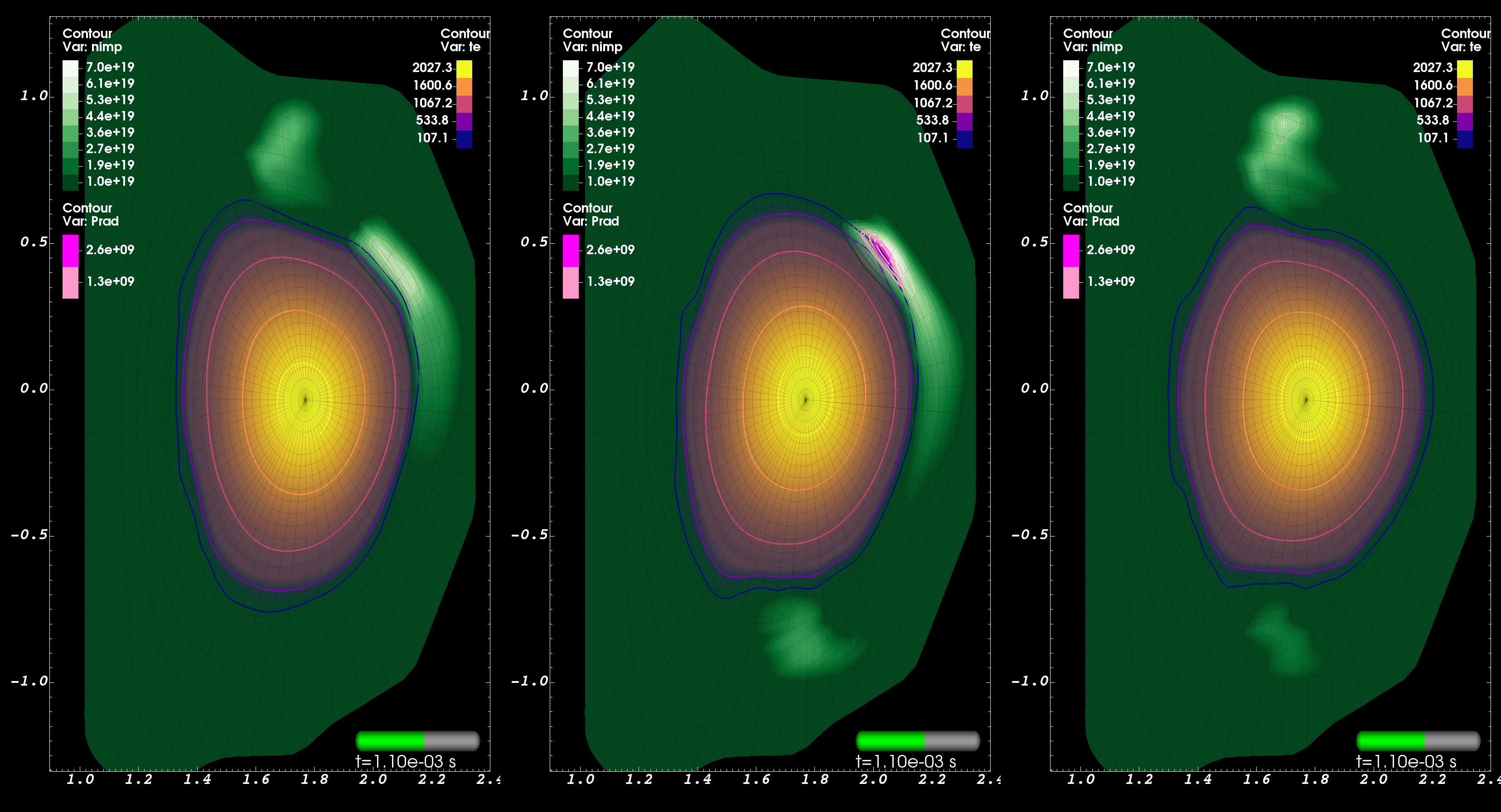}
		}
	\centerline{
		\includegraphics[width=0.9\textwidth]{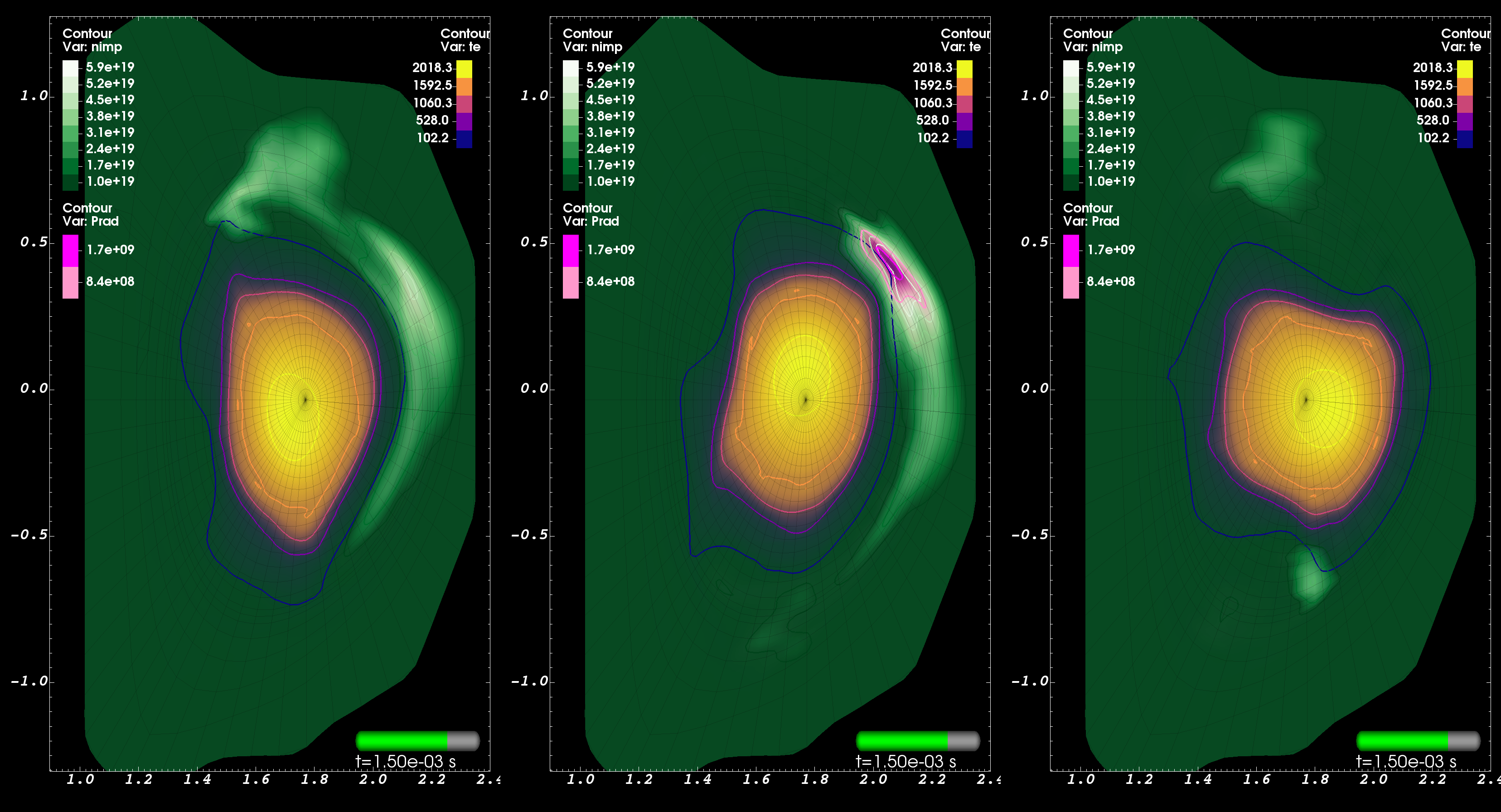}
		}
		}
		\label{fig:xscdt0}
	\end{figure}

\subsection{Delay dt=-0.4ms}
\label{sec:vizdtn04}
	\begin{figure}[H]
	\centerline{
		\href{https://youtu.be/yVQtO40OAdU}
		{\includegraphics[width=1.10\textwidth]{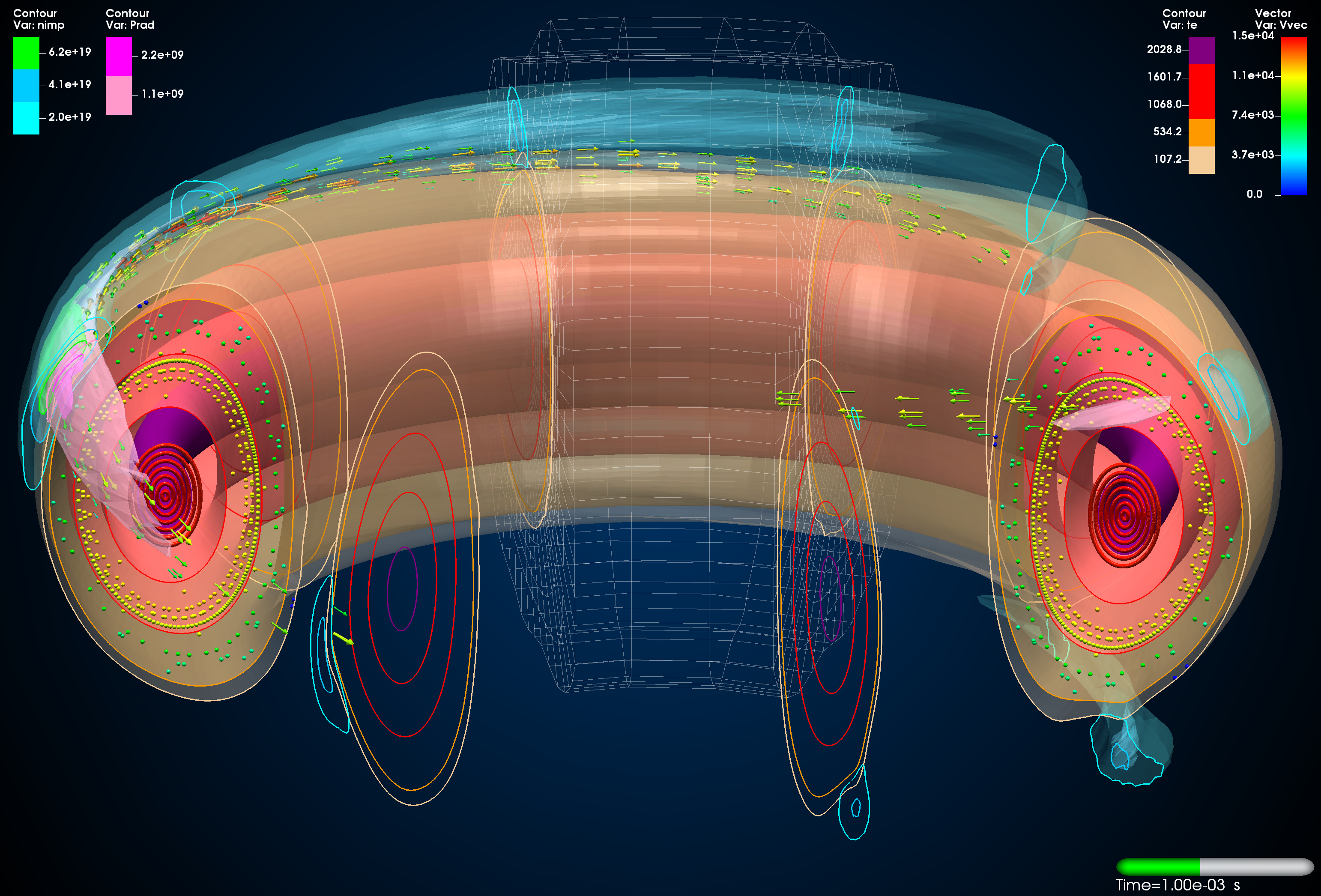}}
		}
	\end{figure}
	Dual SPI delay dt=-0.4ms.  3D visualization at t=1.0ms show onset of plasma interacting with second
	injector at 120$^\circ$.  Prior to this, ablation was dominated by the first injector at 0$^\circ$.

	2D cross sections at $\phi$=[0$^\circ$,120$^\circ$,240$^\circ$] at t=[0.2,1.1,1.5]ms.  At t=0.2ms only the first
	injector fragments at 0$^\circ$ are ablating.  At t=1.1ms, MHD causes the flux surfaces to flex and distort and
	promotes ablation at 120$^\circ$ while maintaining it at 0$^\circ$.  At t=1.5ms, both injector fragments are
	still ablating, in contrast to the dt=0.0 case, and the helical distortion is visibly smaller.

	\begin{figure}
		\href{https://youtu.be/KXN6r-mq_40}
		{
	\centerline{
		\includegraphics[width=0.9\textwidth]{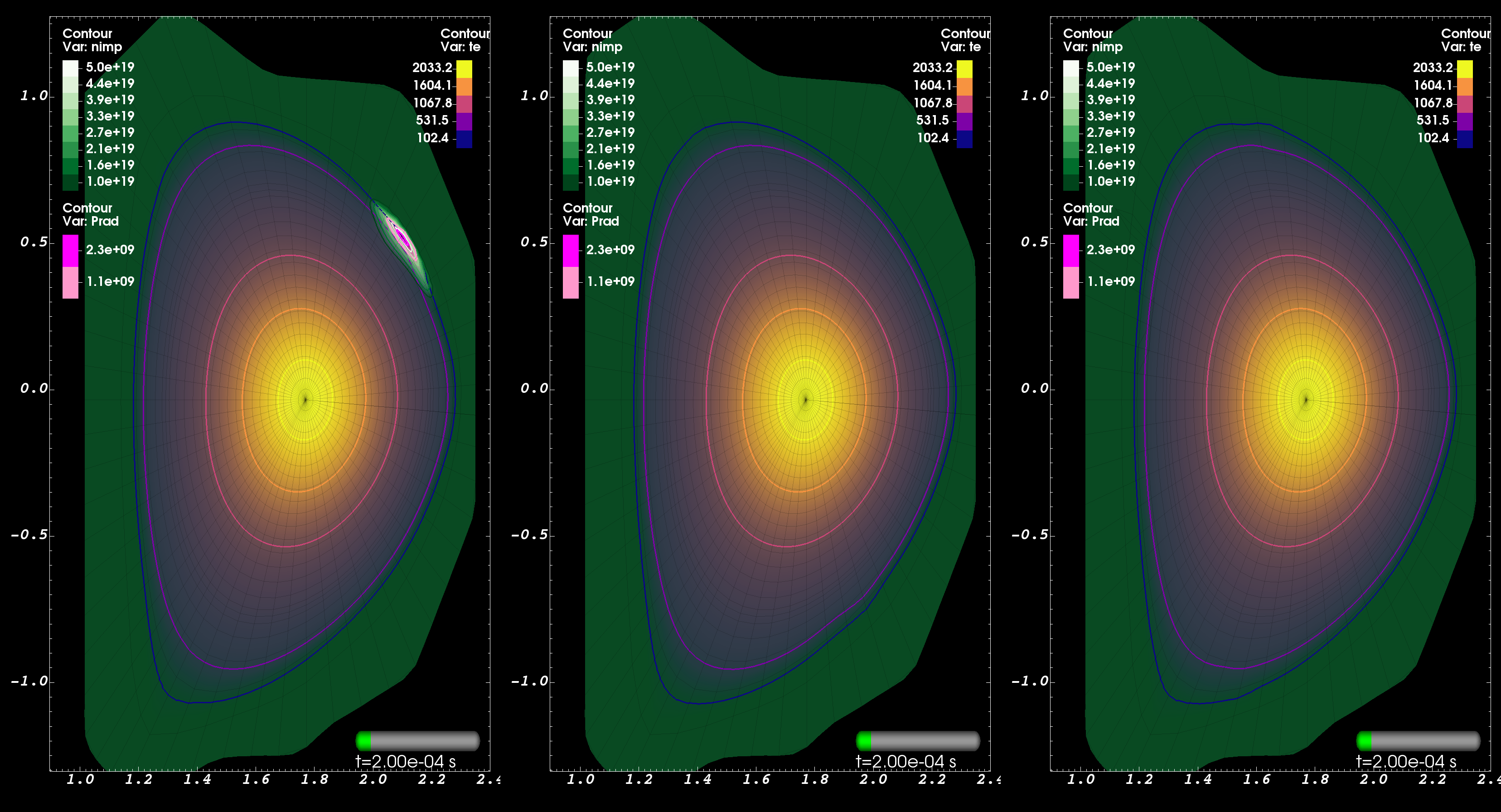}
		}
	\centerline{
		\includegraphics[width=0.9\textwidth]{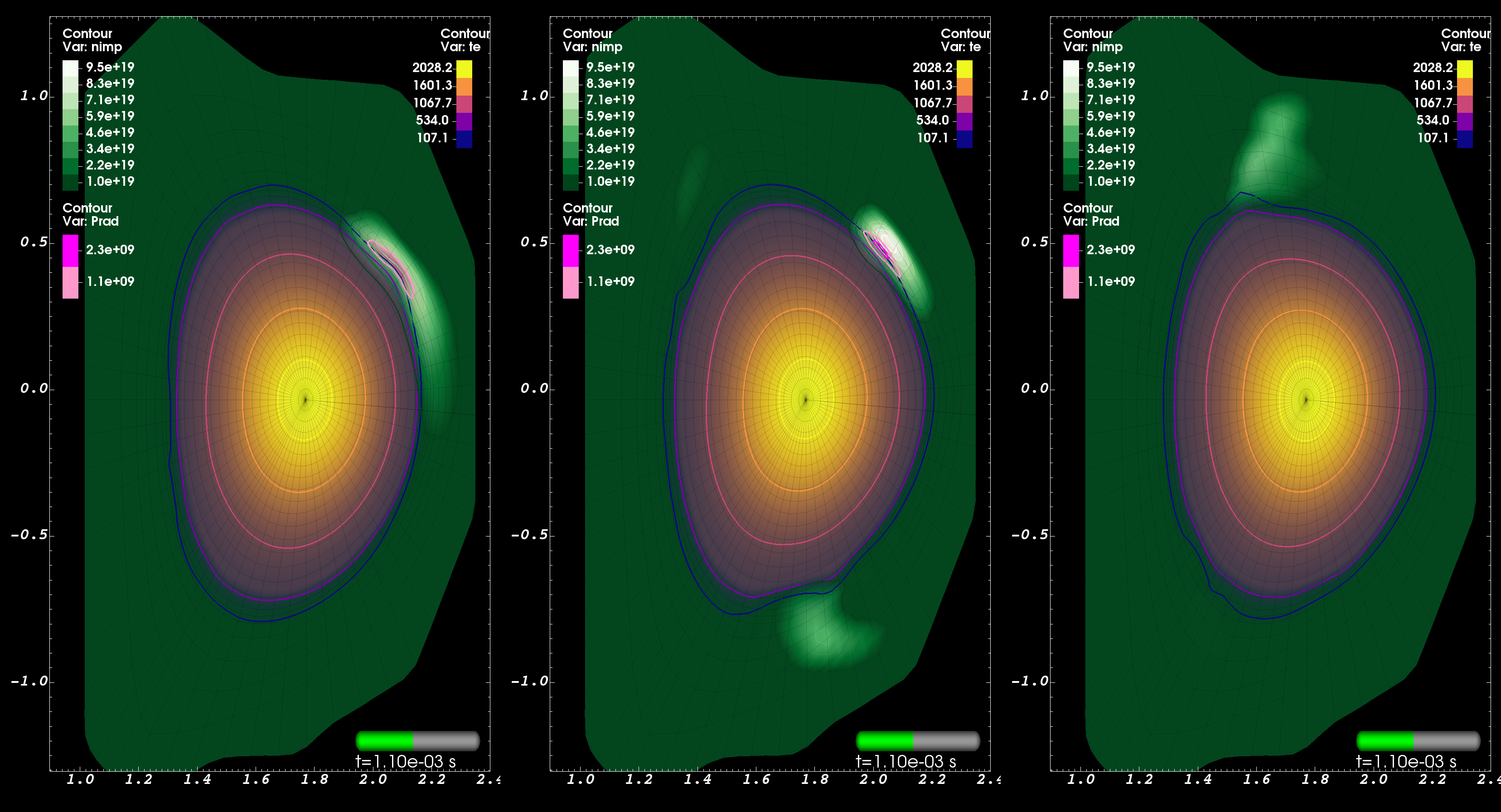}
		}
	\centerline{
		\includegraphics[width=0.9\textwidth]{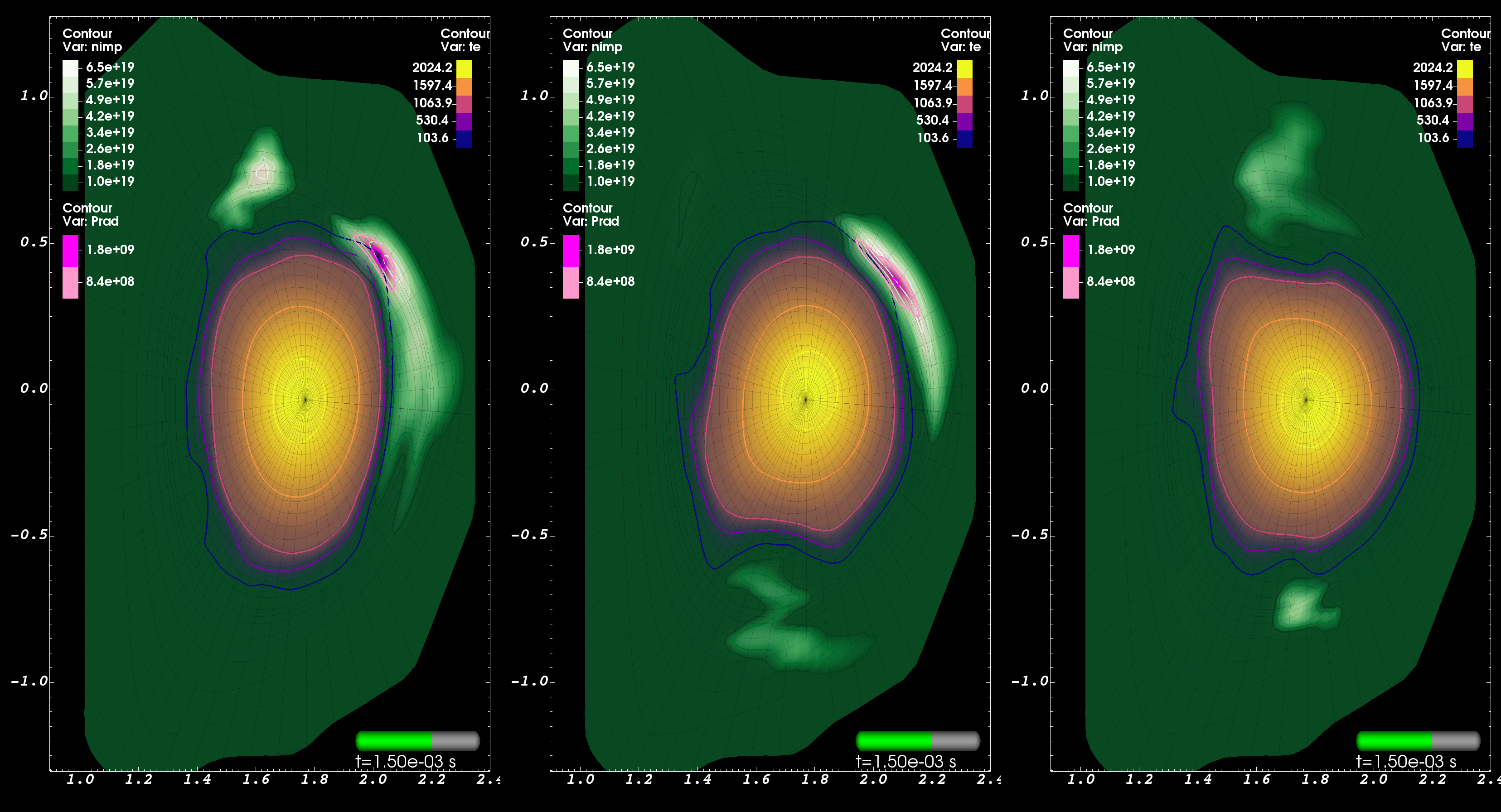}
		}
		}
		\label{fig:xscdtn04}
	\end{figure}
	
\newpage
\section{Increasing $\bten{\chi}$ and Fragment Plume Inventory and Velocity}
\label{sec:follow}
Follow up single injector simulations are presented to test the impact of increasing the thermal conduction, fragment plume
inventory, and fragment plume velocity.

	\begin{figure}[H]
		\begin{subfigure}[b]{0.5\textwidth}
			\includegraphics[width=\textwidth,trim={0.0cm 3.5cm 0.0cm 3.5cm},clip]{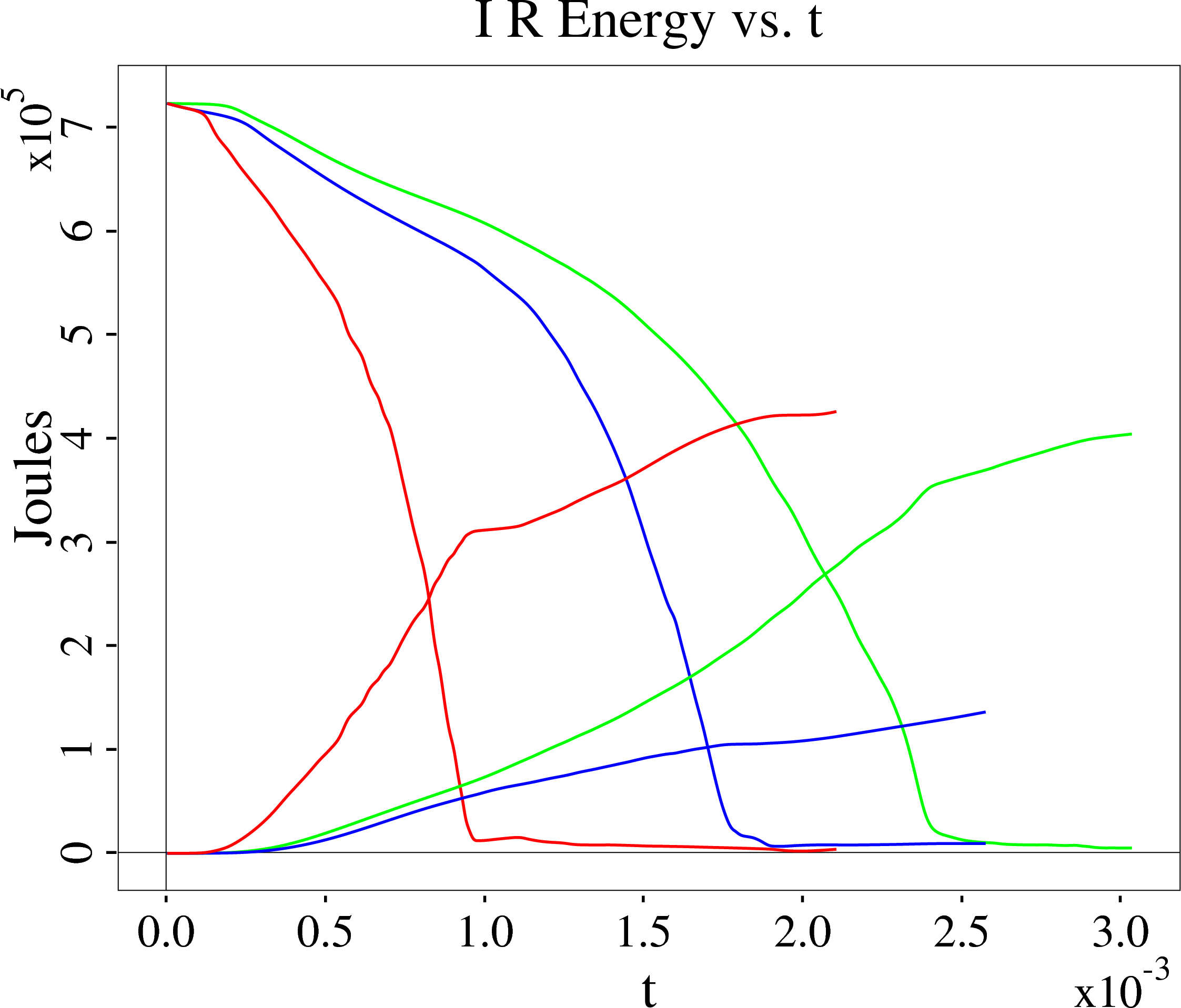}
			\vspace{-0.60cm}
			\caption*{thermal and radiated energy}
			\label{fig:te_follow}
		\end{subfigure}
		\hspace{0.25cm}
		\begin{subfigure}[b]{0.5\textwidth}
			\includegraphics[width=\textwidth,trim={0.0cm 3.5cm 0.0cm 3.5cm},clip]{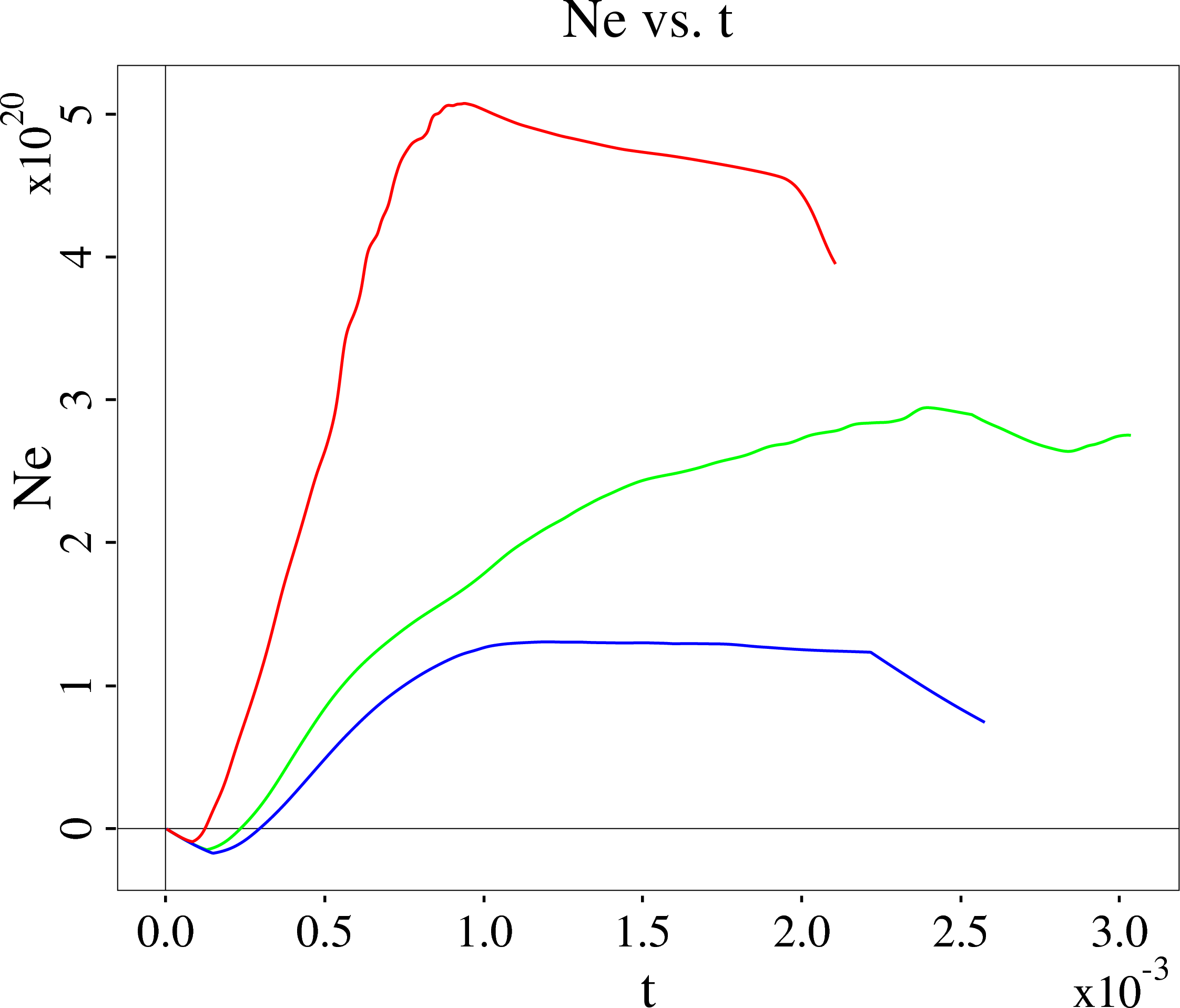}
			\vspace{-0.60cm}
			\caption*{electron population}
			\label{fig:Ne_follow}
		\end{subfigure}
		\caption{Single injector thermal and radiated energy and electron population compares \clrlg{\bf dt=-0.8ms} with
		two follow up case: \clrb{\bf $\bten{\chi}$ increased} and \clrr{\bf $\bten{\chi}$ increased, plume inventory
		increase, fragment velocity increased}.  \clrb{\bf Increasing thermal conduction} accelerates the thermal loss
		and reduces the radiated energy.  \clrr{\bf Increasing the velocity and plume inventory} brings the radiated 
		energy back to the \clrlg{\bf base case}.  The electron population does not increase dramatically.}
		\label{fig:followup}
	\end{figure}
	
Figure \ref{fig:followup} compares the thermal and radiated energy and the electron population for three single SPI
simulations: the base case \clrlg{\bf dt=-0.8ms} (see sec.\ref{sec:dt08}), a case where \clrb{\bf thermal conduction is 
increased} - $\chi_\perp$=0.2m$^2$/s to 5.0m$^2$/s and $\chi_\parallel$=1$\times$10$^{9}$ to 1$\times$10$^{10}$ - 
and a case where \clrr{\bf thermal conduction is increased} as in the previous case  and 
\clrr{\bf the fragment plume inventory} is increased by a factor of $\times$40 and \clrr{\bf velocities are increased}
from 120m/s to 200m/s.

\clrb{\bf Increasing $\bten{\chi}$ only} results, unsurprisingly, in an acceleration of the thermal loss and
reduction in radiation and electron population.  

With the same \clrr{\bf increase in $\bten{\chi}$ and increase in fragment plume inventory and velocity}, the radiated 
energy increases significantly over the \clrb{\bf increased $\bten{\chi}$ only} case.  The radiated energy is almost
the same as the \clrlg{\bf base case} but the thermal quench time is decreased by almost a factor of $\times$2.5 due 
to the faster fragments.

Additionally,we see an increase of the electron population.  However, despite the factor of $\times$40 increase, 
the electron population is not much more than the dual SPI dt=$\pm$0.4ms case (see fig.\ref{fig:Ne_dt_comp}).  
We conclude that the ablation is limited by the thermal inventory of a flux surface not the entire plasma.  
In particular,  the early phase of the thermal quench proceeds from one flux surface to the next, like peeling
an onion.  Once the thermal inventory of a flux surface is exhausted there is no more thermal energy to 
ablate any further and
the fragments must advance to the next flux layer to ablate more;
due to magnetic confinement, the thermal transport is too slow to significantly replenish the quenched region. 
We see this reflected in the temperature profiles in figure \ref{fig:triprofs_dtn04}, the core remains mostly unperturbed
while the outer regions are quenched.  

Due to this limit in ablation, the total ablation for the \clrr{\bf increase in $\bten{\chi}$ and $\times$40 
increase in fragment plume inventory 
(and velocity)} at the end of the thermal quench is $\simeq$2\% of the total inventory.  That is only 2\% of the injected
material is assimilated, leaving 98\% to pass through the vessel.  

As a concrete example of why we reduce the simulation plume fragment inventory consider that at
t=1.0ms (end of the thermal quench), the fragment plume with a $\Delta v$=100m/s is 20cm long.  2\% total 
ablation means that only 0.4cm 
of 20cm long plume has ablated away.  A second fragment plume with a time delay of only dt=0.1ms trails behind the lead
plume by 2.0cm.  At the end of the thermal quench, the second fragment plume still has a 1.6cm gap to close before 
contact with the plasma, i.e. even with a small delay of dt=0.1ms, this dual injector scenario will look like a 
single injector case because the second injector can not close the gap unless the plasma itself moves towards it.
As yet, we cannot reconcile this reduction in inventory with the experiment but require it in order to observe
any interaction with the second injector.  

\end{document}

%% file: macros.tex
\newcommand{\bvec}[1]{\mathbf{#1}}
\newcommand{\bten}[1]{\mathbf{\underline{#1}}}

\newcommand{\clrr}[1]{\textcolor{red}{#1}}

\newcommand{\clrlg}[1]{\textcolor{green!100!black!95}{#1}}
\newcommand{\clrb}[1]{\textcolor{blue}{#1}}

\newcommand{\clrm}[1]{\textcolor{magenta}{#1}}
\newcommand{\clrp}[1]{\textcolor{violet}{#1}}
\newcommand{\clrc}[1]{\textcolor{cyan}{#1}}

\def\grapprox{\mathrel{\rlap{\lower3.5pt\hbox{$\mathchar"218$}}\raise 
1pt \hbox{$\mathchar"13E$}}}
\def\lsapprox{\mathrel{\rlap{\lower3.5pt\hbox{$\mathchar"218$}}\raise 
1pt \hbox{$\mathchar"13C$}}}